\documentclass[oldversion]{aa}
\usepackage{graphicx}
\usepackage{txfonts}

\newcommand{\mi}{\mbox{$M_{I}$}}

\newcommand{\Mbol}{\mbox{$M_{\rm bol}$}}

\newcommand{\Msun}{\mbox{$M_{\odot}$}}
\newcommand{\Teff}{\mbox{$T_{\rm eff}$}}

\newcommand{\comment}[1]{}
\newcommand{\beq}{\begin{equation}}
\newcommand{\eeq}{\end{equation}}
\newcommand{\beqa}{\begin{eqnarray}}
\newcommand{\eeqa}{\end{eqnarray}}
        \def\smallskip{\vskip 2pt}

\begin{document}
\title{Evolution of asymptotic giant branch stars} 
\subtitle{I. Updated synthetic TP-AGB models and their basic calibration}
\author{Paola Marigo$^1$ \and 
 L\'eo~Girardi$^2$}
\institute{
 Dipartimento di Astronomia, Universit\`a di Padova,
	Vicolo dell'Osservatorio 3, I-35122 Padova, Italy \\ {\tt(e-mail:
	  paola.marigo@unipd.it)} \and
 Osservatorio Astronomico di Padova -- INAF, 
	Vicolo dell'Osservatorio 5, I-35122 Padova, Italy \\{\tt(e-mail:
	  leo.girardi@oapd.inaf.it)}}


\date{Received December 2006 / Accepted March 2007}

\abstract{
We present new {{\bf} synthetic} models of the TP-AGB evolution. 
They are computed for 7
values of initial metal content ($Z$ from 0.0001 to 0.03) and for
initial masses between 0.5 and 5.0 $M_\odot$, thus extending the low-
and intermediate-mass tracks of Girardi et al. (2000) until the 
beginning of the post-AGB phase. The calculations are performed by
means of a synthetic code that incorporates many recent
improvements, among which we mention: (1) the use of detailed and
revised analytical relations to describe the evolution of 
quiescent luminosity, inter-pulse period, third
dredge-up, hot bottom burning, pulse cycle luminosity variations,
etc.; (2) the use of variable molecular opacities -- i.e. opacities
consistent with the changing photospheric chemical composition -- in
the integration of a complete envelope model, instead of the standard
choice of scaled-solar opacities; (3) the use of formalisms for the 
mass-loss rates derived from pulsating dust-driven wind models of C- and
O-rich AGB stars; and (4) the switching of pulsation modes between
the first overtone and the fundamental one along the evolution, which has
consequences in terms of the history of mass loss. It follows that, 
in addition to the time evolution on the HR diagram, 
the new models predict in a consistent fashion also variations 
in surface chemical compositions,
pulsation modes and periods, and mass-loss rates. The onset and
efficiency of the third dredge-up process are calibrated in order to
reproduce basic observables like the carbon star luminosity functions
in the Magellanic Clouds, and TP-AGB lifetimes (star counts) in
Magellanic Cloud clusters.
In this paper, we describe in detail the model ingredients, basic
properties, and calibration. Particular emphasis is put in
illustrating the effects of using variable molecular opacities. 
Forthcoming
papers will present the theoretical isochrones 
and chemical yields 
derived from these
tracks, and additional tests performed with the aid of a complete population
synthesis code.
}

\authorrunning{P. Marigo \& L. Girardi}
\titlerunning{Updated synthetic TP-AGB models}
\maketitle

\section{Introduction}
\label{intro}

Owing to their large luminosities and cool photospheres, thermally
pulsing asymptotic giant branch (TP-AGB) stars are among the most
remarkable objects in near- and mid-infrared surveys of nearby
galaxies. The TP-AGB is also one of the most intriguing phases of
stellar evolution, marked by the development of high mass loss rates,
the presence of recurrent luminosity and temperature excursions, the
rich nucleosynthesis, and the long-period variability. Several ``third
dredge-up'' episodes -- followed by hot-bottom burning (HBB) in the
most massive AGB stars -- cause the surface pollution with He and CNO
elements of primary origin, which become directly observable first via
the changes in the spectral features of AGB stars themselves, and
later in the emission line spectrum of the subsequent planetary
nebulae phase. In the context of galaxy models, AGB stars are crucial
contributors to both their integrated spectra and their chemical
enrichment.

Computations of the TP-AGB phase by means of complete evolutionary
codes are very demanding in terms of computational time, and in most
cases they fail to predict basic observational facts such as the
conversion from M to C spectral types in AGB stars with initial masses
and luminosities as low as $M_{\rm i}\sim1.5~\Msun$ and $M_{\rm
bol}\sim-3.5$ (see Herwig 2005 for a review, and Stancliffe et
al. 2005 for a recent exception).
Such difficulties derive from the complex structure of these stars, as
well as the uncertainties in the modelling of convective dredge-up and
mass-loss processes.  In order to provide extended grids of TP-AGB
models that reproduce basic observational constraints, the only viable
alternative so far is the use of the so-called synthetic codes, in
which the stellar evolution is described by means of simplified
relations derived from complete stellar models, while convective
dredge-up and mass-loss are tuned by means of a few adjustable
parameters.

Purpose of this paper is to describe new improved models of the TP-AGB
phase of low- and intermediate mass stars, whose computation has been
motivated by a series of recent improvements in the area:

1) Starting from the work by Wagenhuber \& Groenewegen (1998), recent
theoretical works (Karakas et al. 2002, hereafter K02; Izzard et
al. 2004, hereafter I04) have significantly detailed and extended the
range of validity of analytical relations describing the evolution of
TP-AGB stars -- e.g., the core-mass luminosity relation, inter-pulse
period, pulse cycle variations, dredge-up efficiency, etc. The
theoretical modelling of the mass-loss phenomenon has also
significantly improved (Willson 2000; Winters et al. 2003; H\"ofner et
al. 2003, and references therein). Moreover, Marigo (2002) has
recently demonstrated that the low-temperature opacities used in
present-day TP-AGB calculations are in substantial error, due to the
non-consideration of the dramatic changes in molecular concentrations
that occur in the outer envelope of TP-AGB stars as their chemical
composition is altered by third dredge-up episodes; the consequences
of using consistent molecular opacities turn-out to be remarkable when
stars enter into the C-rich phase of their evolution (see Marigo 2002;
Marigo et al. 2003).  All these theoretical improvements should allow
a better description of many aspects of the TP-AGB evolution, and
also a better understanding of their dependence on the stellar mass
and initial chemical composition.

2) Present-day near-infrared cameras are enormously increasing our
knowledge of the AGB populations of Local Group galaxies. DENIS
and 2MASS have provided photometry in the red and near-infrared for
the complete sample of non-obscured AGB stars in the Magellanic
Clouds, revealing striking features such as the ``red tail'' of carbon
stars (Cioni et al. 1999; Nikolaev \& Weinberg 2000). Dedicated C star
surveys (e.g. Battinelli \& Demers 2005ab, and references therein), as
well as future near-IR surveys using UKIDSS and VISTA, will soon
provide complete TP-AGB samples for a large subset of the Local
Group. On the other hand, infrared surveys (IRAS, ISO, Spitzer) are
revealing the samples of dust-enshrouded, optically obscured, TP-AGB
stars in the Magellanic Clouds (Loup et al. 1999; Cioni et al. 2003;
van Loon et al. 2005; Blum et al. 2006). Interpreting these large
photometric databases in terms of galaxy properties (star formation
rate, age-metallicity relation, density profiles) requires the use of
suitable TP-AGB models. An example of the potentialities of
present-day data-sets are given by Cioni et al. (2006ab), who detect
variations of the mean stellar metallicity and star formation rate
across the LMC and
SMC galaxies, using only the near-IR properties of their AGB stars as
compared to those predicted by theoretical models.

3) Micro-lensing surveys in the Magellanic Clouds and in the Galactic
Bulge (MACHO, OGLE) have provided a huge amount of high-quality
optical data for long-period variables; these data are nicely
complemented with near-IR photometry from DENIS and
2MASS. Particularly striking has been the discovery of several different
sequences in the luminosity-period plan, four of which represent the
various pulsation modes of AGB and upper-RGB stars (Wood et al. 1999;
Soszynski et al. 2006; Groenewegen 2004; Ita et al. 2004; Fraser et
al. 2005). The data clearly indicate that TP-AGB stars start pulsating
as small-amplitude semi-regular variables (first and second overtone,
and mixed modes) and then later in the evolution become high-amplitude
Mira variables (fundamental mode). Since it is generally believed that
the pulsation period correlates with the mass loss rate of such stars,
this may have important evolutionary implications that still have to
be explored.

Therefore, we aim at producing TP-AGB models -- and their derivatives,
like isochrones, luminosity functions, synthetic colour-magnitude
diagrams, etc. -- including the above-mentioned improvements in the
analytical relations, molecular opacities, and mass-loss prescription,
and consistently predicting variations in surface chemical
compositions, pulsation modes and periods, and mass-loss rates. Like
in any other set of synthetic models, uncertain parameters are to be
calibrated in order to reproduce basic observables like the carbon
star luminosity functions in the Magellanic Clouds, and TP-AGB
lifetimes (star counts) in star clusters. In a second phase, we intend
to check also relatively new observables like the period distributions
and relative numbers of stars in the several sequences of LPVs, and
the properties of dust-enshrouded AGB stars with high mass-loss.

The present paper is organised as follows. In Sect.~\ref{sec_syntagb}
we describe model ingredients and computations.
Sect.~\ref{sect_modpre} illustrates the basic model predictions, with
special emphasis put on the distinctive features found in the present
calculations. Section~\ref{sec_calibr} together with the Appendix 
detail the calibration of
model parameters, based on selected observational constraints.
Finally, in Sect.~\ref{sec_end}
we briefly recall the major achievements of this work, 
and the next steps to be followed. 

\section{Synthetic AGB evolution}
\label{sec_syntagb}

A very detailed code for the synthetic evolution of TP-AGB stars has
been developed by Marigo et al. (1996, 1998) and Marigo (2001 and
references therein).  It couples the use of updated analytical
relations (e.g., the core mass-luminosity relation, the interpulse
duration), with a parametric description of the third dredge-up
episodes, and numerical integrations of a complete envelope
model. Over the last years, several additional improvements have been
introduced, the most relevant of which are briefly recalled:
\begin{itemize}
\item  Development of a  consistent method for dealing with the
over-luminosity effect caused by hydrogen burning at the bottom of the
convective envelope and its related nucleosynthesis (HBB; see Marigo
1998).
\item Adoption of more physically-sound dredge-up parameters, linked 
to the minimum post-flash temperature at the bottom of the convective
envelope, $T_{\rm b}^{\rm dred}$. The two free parameters associated
with the possible occurrence -- described by $T_{\rm b}^{\rm dred}$
--, and the efficiency -- described by $\lambda$ -- of the third
dredge-up were calibrated by fitting the carbon stars luminosity
functions in both Magellanic Clouds (Marigo et al. 1999).
\item Calculation of molecular opacities properly coupled to the actual
chemical composition of the envelope ($\kappa_{\rm var}$), in place of
the common and inappropriate choice of low-temperature opacity tables
($\kappa_{\rm fix}$) strictly valid for solar-scaled chemical mixtures
(Marigo 2002; Marigo et al. 2003).
\end{itemize}

The present study stands on these previous works, adopting additional
improvements to be described below. For a general description on how
synthetic TP-AGB models work, the reader is referred to Groenewegen \&
Marigo (2003).

\subsection{Initial conditions}
\label{ssec_initial}
The initial conditions at the first thermal pulse -- namely total
mass, luminosity, core mass, and surface chemical composition as
affected by the first and second dredge-ups -- are taken from Girardi
et al. (2000) for the initial metallicities $Z=0.0004$, $0.001$,
$0.004$, $0.008$, $0.019$, and $0.03$. For $Z=0.0001$, we use an
additional set of tracks computed by Girardi (2001, unpublished; see
{\tt http://pleiadi.oapd.inaf.it}) using the same input physics as
Girardi et al. (2000). Without entering in details, suffice it to
recall that these tracks are computed with OPAL (Iglesias \& Rogers
1993, and references therein) and Alexander \& Ferguson (1994)
opacities, and include a moderate amount of convective core
overshooting\footnote{{{\bf} Namely, in these models the overshooting
parameter $\Lambda_{\rm c}$ (see Alongi et al. 1993) 
is set to 0.5 for initial masses
$\mi\ge1.5$~\Msun, which implies an overshooting region extending for
about 0.25 pressure scale heigths {\em above} the classical
Schwarzschild border. For $\mi<1.5$~\Msun, $\Lambda_{\rm c}$ is
gradually reduced until it becomes null for $\mi\le1.0$~\Msun}}.  As
a consequence of overshooting, the upper mass limit for the
development of the AGB phase is located close to 5~\Msun\ (instead of
the $\sim8$~\Msun\ found in classical models) for all values of $Z$,
and is slightly uncertain because the initial phase of carbon burning
has not been followed in detail (see section 3.1 in Girardi et
al. 2000). Therefore, $M_{\rm i}=5$~\Msun\ is the largest initial mass
to be considered in this work.

Finally, we remark that throughout this paper, except when otherwise
stated,  the term initial mass is equivalent to the stellar mass at
the first thermal pulse, as the pre-AGB evolution in Girardi et
al. (2000) is followed at constant mass. The effect of mass loss
on lower-mass models while evolving on the red giant branch 
will be considered a posteriori in the construction of stellar
isochrones (Marigo \& Girardi, in prep.).

\subsection{Luminosity and core-mass growth}
\label{ssec_lum}
\begin{figure*}[!tbp]  
\begin{minipage}{0.49\textwidth}
	\resizebox{\hsize}{!}{\includegraphics{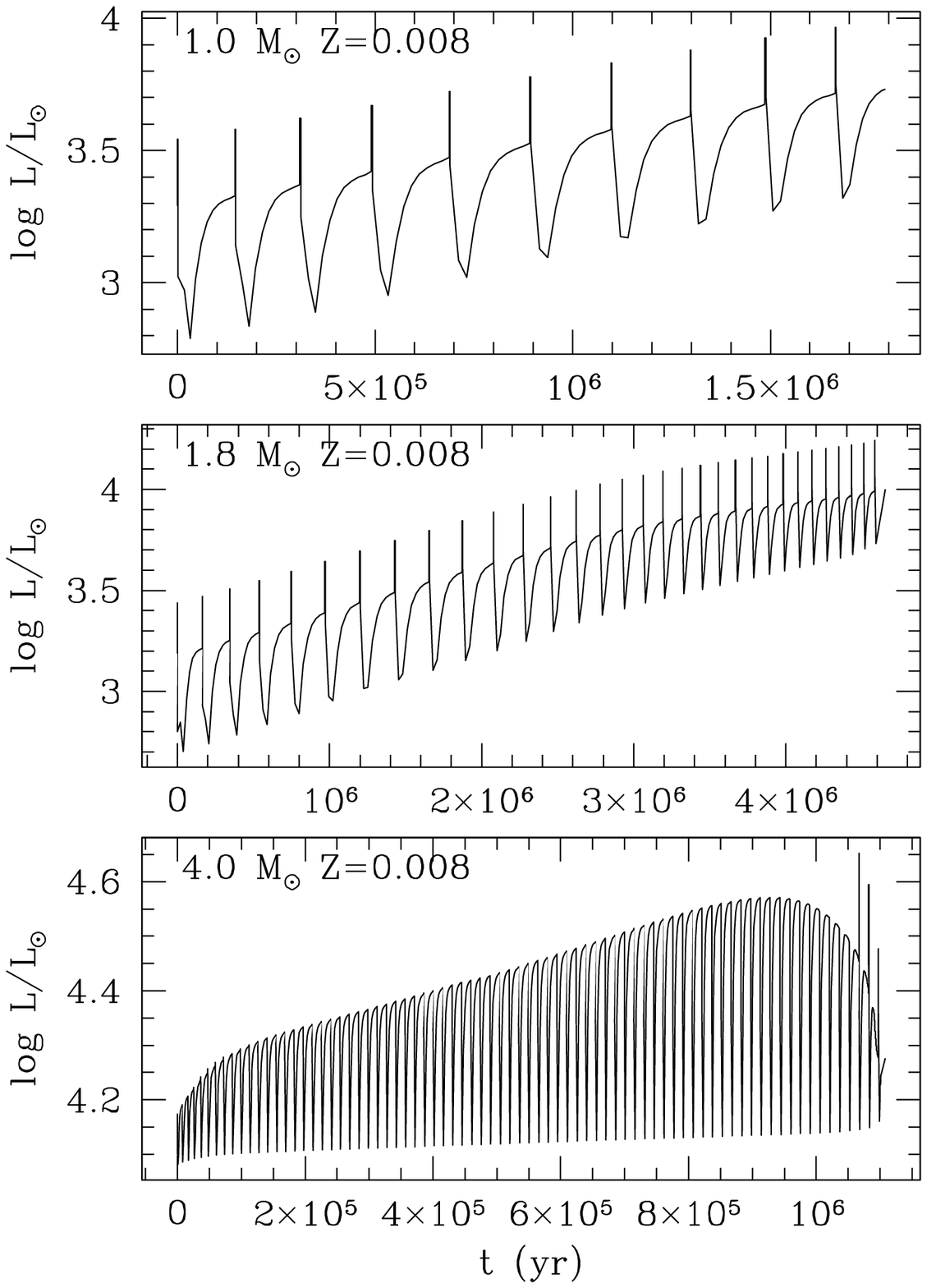}}
\end{minipage} 
\hfill
\begin{minipage}{0.49\textwidth}
	\resizebox{\hsize}{!}{\includegraphics{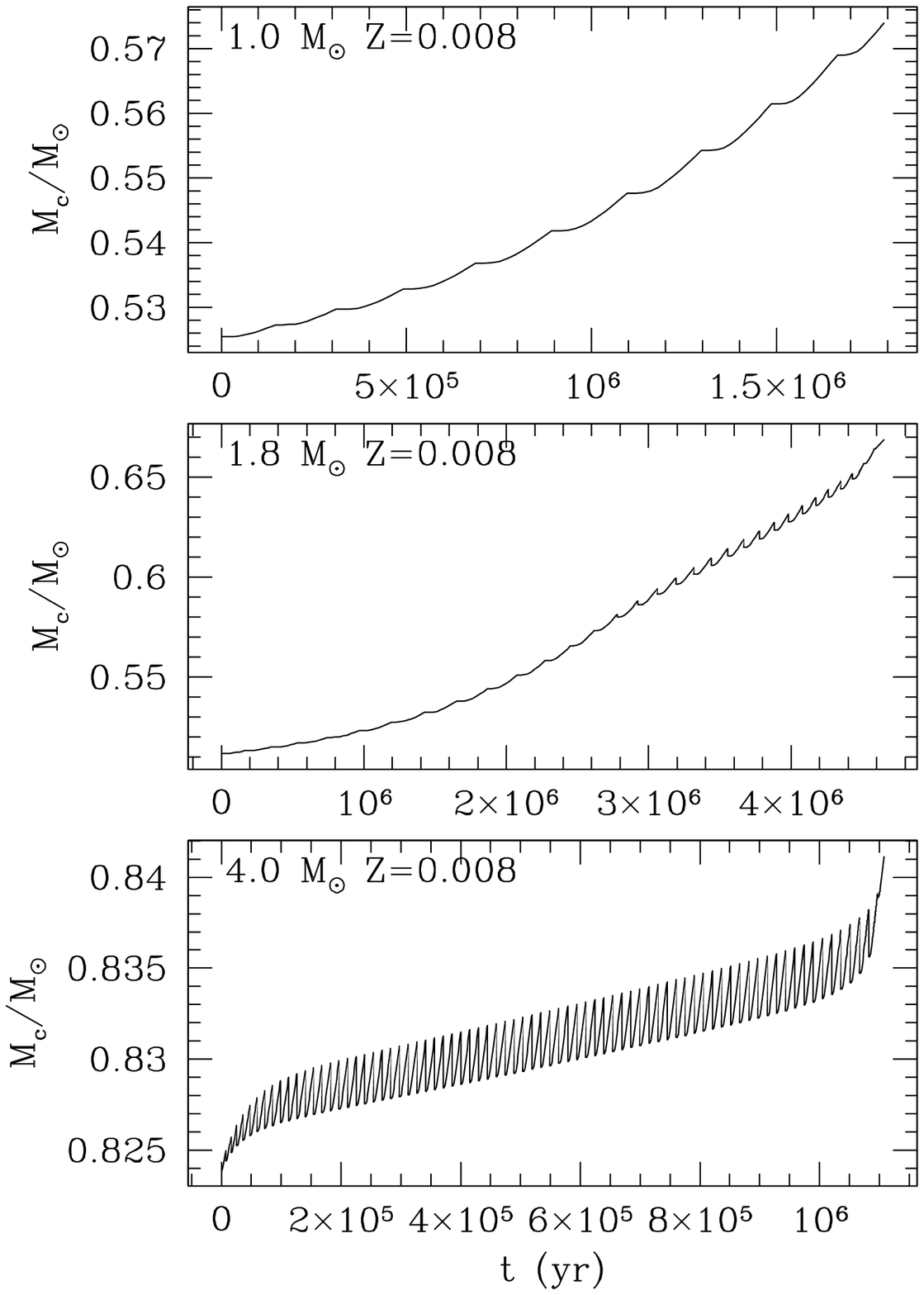}}
\end{minipage} 
\caption{Left panels: Evolution of surface luminosity 
along the TP-AGB phase of two representative models with different
initial masses with $Z=0.008$, according to the analytic formalism by
Wagenhuber \& Groenewegen (1998). Note the high detail in the
description of the luminosity variations driven by thermal pulses
(i.e. peaks and dips), as well as of the overluminosity caused by the
occurrence of HBB in the $4.0\, M_{\odot}$ model.  Right panels:
Evolution of the core mass. The saw-like trend reflects the occurrence
of dredge-up events at thermal pulses.  Note that the $4.0\,
M_{\odot}$ model is characterised by very deep dredge-up for most of
its evolution ($\lambda\approx 1$).}
\label{fig_lum_mc_z008}
\end{figure*}  
\begin{figure*}[!tbp]  
\begin{minipage}{0.49\textwidth}
	\resizebox{\hsize}{!}{\includegraphics{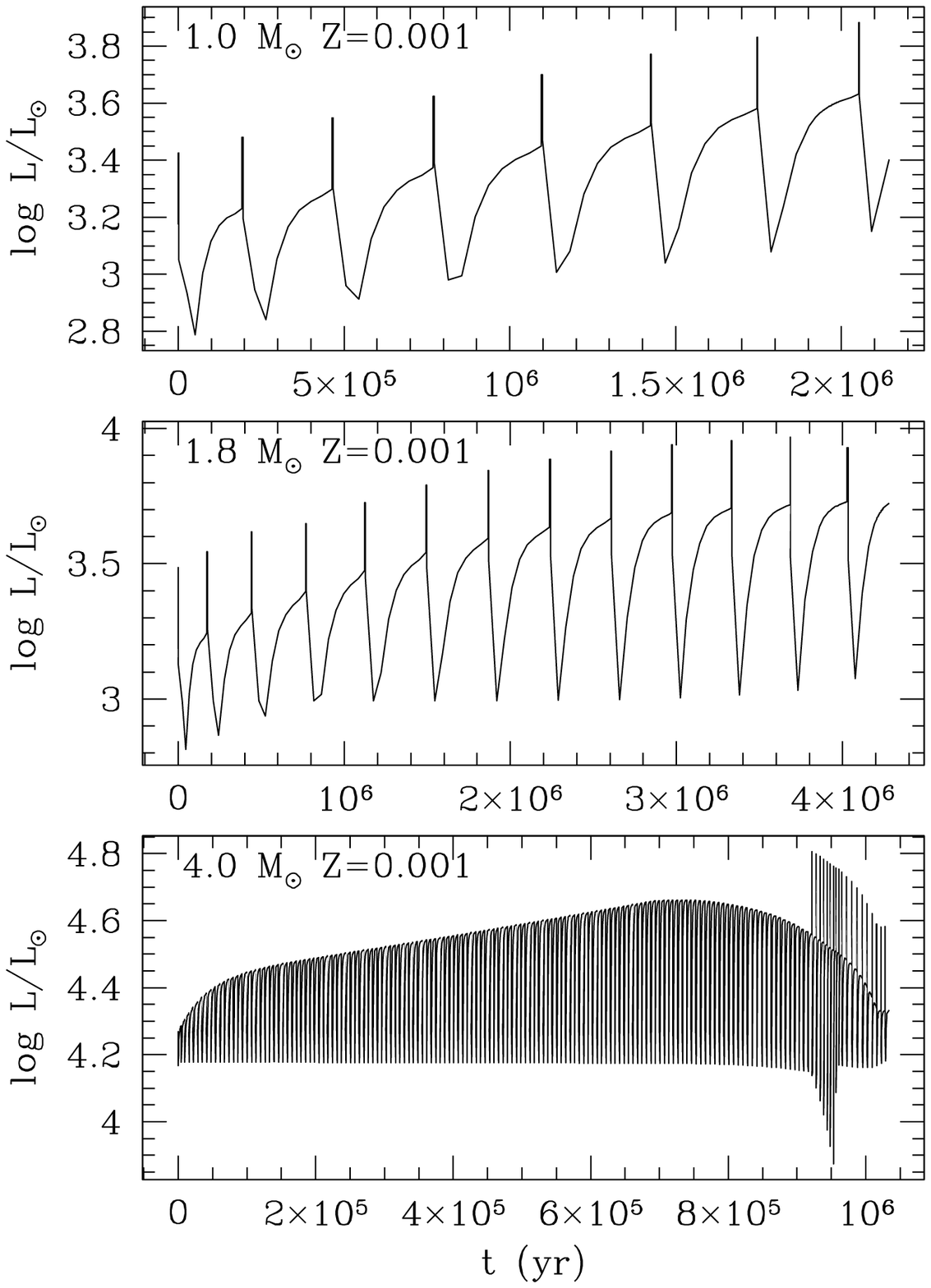}}
\end{minipage} 
\hfill
\begin{minipage}{0.49\textwidth}
	\resizebox{\hsize}{!}{\includegraphics{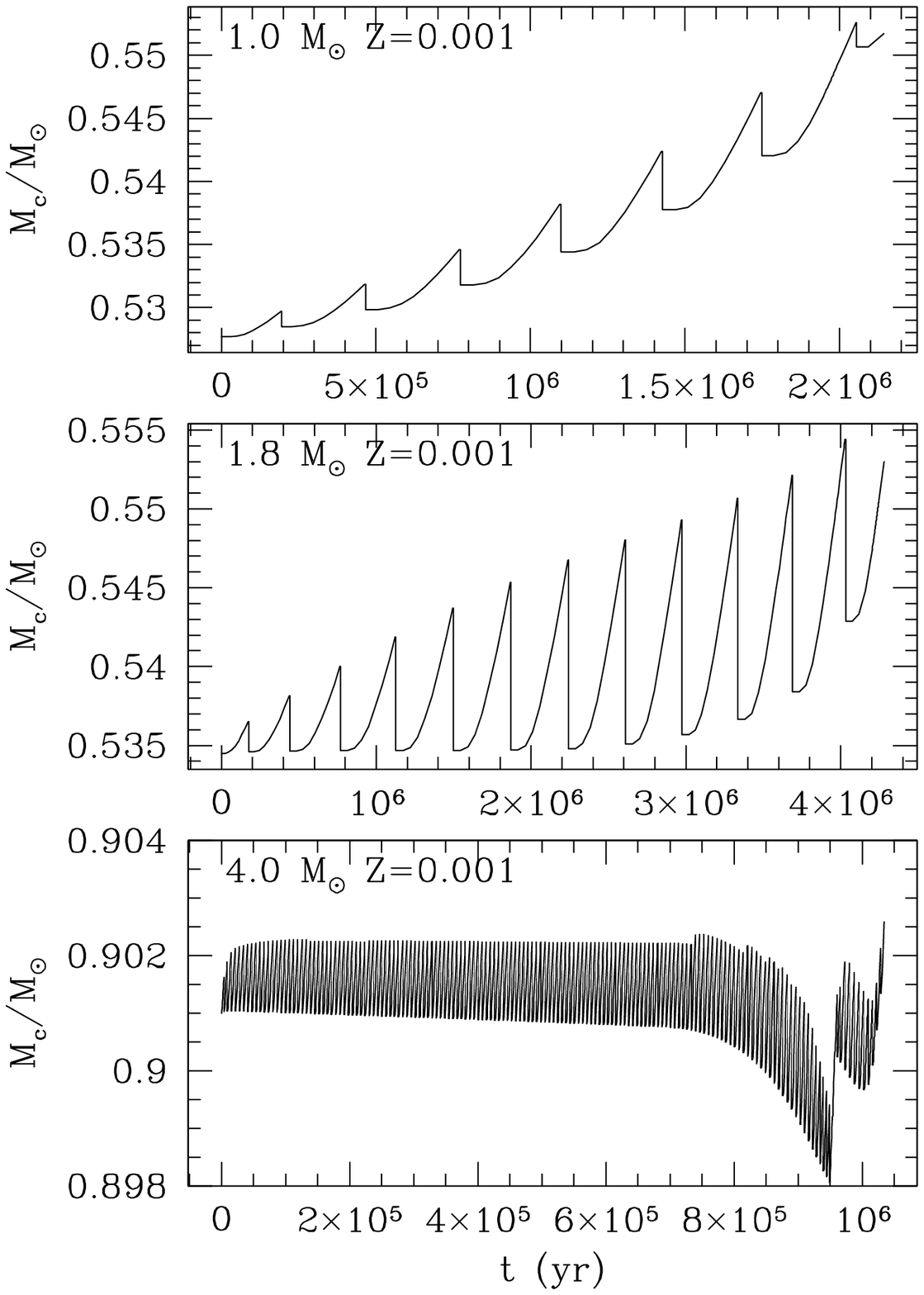}}
\end{minipage} 
\caption{The same as in Fig.~\ref{fig_lum_mc_z008}, but for 
initial metallicity $Z=0.001$.}
\label{fig_lum_mc_z001}
\end{figure*}  

Important improvements have been introduced in the present
calculations to follow (i) the evolution of the stellar luminosity
$L(t)$ and, (ii) the growth rate of the core mass $d M_{\rm c}/d t =
L_{\rm H}\; Q_{\rm H}/X({\rm H})$, where $L_{\rm H}$ is the rate of
energy generation due to radiative shell H-burning, $Q_{\rm H}$ is the
emitted energy per unit mass of H burnt into He, and $X({\rm H})$ is
the H abundance in the stellar envelope.
 
To this aim we need first to choose a core mass-luminosity ($M_{\rm
c}-L$) relation, among the various available in the literature.  This
key-ingredient of any synthetic AGB model is usually derived from
complete AGB evolutionary calculations by fitting the surface
luminosity at the stage of the quiescent pre-flash maximum as a
function of the core mass, and possibly including the effect of other
less influent parameters (metallicity, envelope mass, initial
conditions at the first thermal pulse).

For lower-mass models not experiencing HBB, we adopt the core
mass-luminosity ($M_{\rm c}-L$) relationship provided by Wagenhuber \&
Groenewegen (1988; their equation 5), that includes important effects
like dependence on metallicity and the peculiar behaviour owing to the
first thermal pulses.
 
The deviation from the $M_{\rm c}-L$ relation due to the
overluminosity produced by HBB in more massive models (i.e. with
$M_{\rm i} > 3.5-4.5\, M_{\odot}$, depending on metallicity) is
estimated as an additional term according to I04 (their equation 32).
This correction depends on metallicity and it takes into account 
the increase in the core degeneracy during the TP-AGB evolution.  This
means that even when $M_{\rm c}$ remains almost constant in massive
AGB stars with very deep dredge-up ($\lambda \approx 1$), the surface
luminosity still increases because of the increasing core degeneracy.

It is important to remark that the $M_{\rm c}-L$ relation is strictly
valid under the quiescent conditions just before a thermal pulse,
while significant deviations (i.e. the so-called rapid dip, rapid peak
and slow dip) take place during the interpulse period due to the
complex interplay between the occurrence of the flash and the
consequent structural and thermal readjustment across the star.
Compared to our previous studies -- where a very simplified
description of the rapid peak and slow dip was adopted -- a crucial
implementation in our synthetic code is that, with the aid of the
formalism proposed by Wagenhuber \& Groenewegen (1988) we now follow
in detail the evolution of both $L(t)$ and $L_{\rm H}(t)$, which vary
significantly during thermal pulse cycles.

Figures~\ref{fig_lum_mc_z008} and \ref{fig_lum_mc_z001} provide 
examples of the detailed description of the luminosity variations
driven by thermal pulses (left panels) and the evolution of the core
mass (right panels) in models of different masses and
metallicities. These results are in close agremeent with full
calculations of the TP-AGB phase (see e.g., figures 4-9 in Vassiliadis
\& Wood 1993).  We note that, except for the final stages of intense
mass loss, the most massive models ($M_{\rm i}= 4 M_{\odot}$) do not
show the rapid luminosity peaks because of their large envelope mass,
and that they are also affected by HBB while evolving in luminosity
well above the $M_{\rm c}-L$ relation.

It is worth remarking that the growth of the core mass due to shell
H-burning during the inter-pulses period may be followed by a
quasi-instantaneous reduction due to convective dredge-up at thermal
pulses. Such recursive increase-recession of the core mass gives
origin to the saw-like behaviour displayed by all models in the right
panels of Figs.~\ref{fig_lum_mc_z008} and \ref{fig_lum_mc_z001}
(except for the $1.0 M_{\odot},\, Z=0.008$ case which does not present
dredge-up). It is already clear from these figures that 
convective dredge-up is predicted to be more efficient in models
of larger mass and lower metallicity.
 
\begin{table*}[!tbp]   
\caption{Predicted properties of synthetic TP-AGB models with 
$M_{\rm i}=1.8 M_{\odot},\, Z=0.008$ under various physical
assumptions, namely: detailed description of the luminosity evolution
($L(t)$) or use of the quiescent core mass-luminosity relation
($M_{\rm c}-L$ rel.); adoption of molecular opacities for solar-scaled
chemical mixtures ($\kappa_{\rm fix}$) or coupled with the current
envelope composition ($\kappa_{\rm var}$); inclusion or not of the
third dredge-up.  The other table entries correspond to predicted
stellar lifetimes (in Myr) covering the whole TP-AGB phase ($\tau_{\rm
TP-AGB}$), the M-type phase ($\tau_{\rm M}$) with C/O$\,<1$ and $\log
L/L_{\odot} > 3.3$, the C-type phase ($\tau_{\rm C}$) with C/O$\,>1$;
final stellar mass ($M_{\rm f}$) in solar units and surface C/O ratio;
effective temperature (in K) at the onset of the superwind regime of
mass loss.}
\begin{tabular}{cccccccccc}
\hline
model & luminosity & molecular opacity & $3^{\rm rd}$dredge-up & 
$\tau_{\rm TP-AGB}$ & $\tau_{\rm M}$ & $\tau_{\rm C}$ & $M_{\rm f}$ & 
(C/O)$_{\rm f}$ & $T_{\rm eff}(\dot{M}_{\rm SW})$\\
 & & & & (Myr) & (Myr) & (Myr) & ($M_{\odot}$) &  & (K) \\
\hline
A &  $L(t)$ & $\kappa_{\rm var}$ & yes & 
4.65 & 2.98 & 1.33 & 0.67 & 3.04 & 2630\\
B &  $M_{\rm c}-L$ rel. & $\kappa_{\rm var}$ & yes & 
3.09 & 1.85 & 0.89 & 0.66 & 2.84 & 2630 \\
C &  $M_{\rm c}-L$ rel. & $\kappa_{\rm fix}$ & no &
3.18 & 2.83 & 0 & 0.73 & 0.34 & 3311 \\
D &  $L(t)$ & $\kappa_{\rm fix}$ & yes &
5.91 & 2.96 & 2.62 & 0.76 & 5.79 & 3162 \\
\hline
\end{tabular}
\label{tab_mod}
\end{table*}

Table~\ref{tab_mod} presents a few relevant quantities predicted by
four test synthetic TP-AGB calculations (A-B-C-D models), carried out
with varying input prescriptions. Model A corresponds to our reference
case, being obtained with the ``best'' combination of physical
assumptions at our disposal. It turns out that the results obtained by
simply assuming the $M_{\rm c}-L$ relation all over the entire
interpulse period (model B) are quite different from those found with
a more realistic description of the pulse cycle (model A).  
In fact, despite the two calculations end
up with almost the same final mass, the duration of the whole TP-AGB
evolution, as well as the M- and C-type phases are significantly
shorter in model B (by $\approx 30-40\, \%$) than in model A. Model C,
in which the third dregde-up is suppressed, predicts a sizeable
reduction of the TP-AGB lifetime as well, while the C-type phase is
completely missed.  This simple experiment tells us that, for
instance, the predicted counts of TP-AGB stars in a galaxy via a
population-synthesis scheme may be affected by a substantial error if
the underlying TP-AGB tracks are computed using a simplified
description of the luminosity evolution, i.e. if the $M_{\rm c}-L$
relation is assumed anytime.

\begin{figure}[!tbp]  
\resizebox{\hsize}{!}{\includegraphics{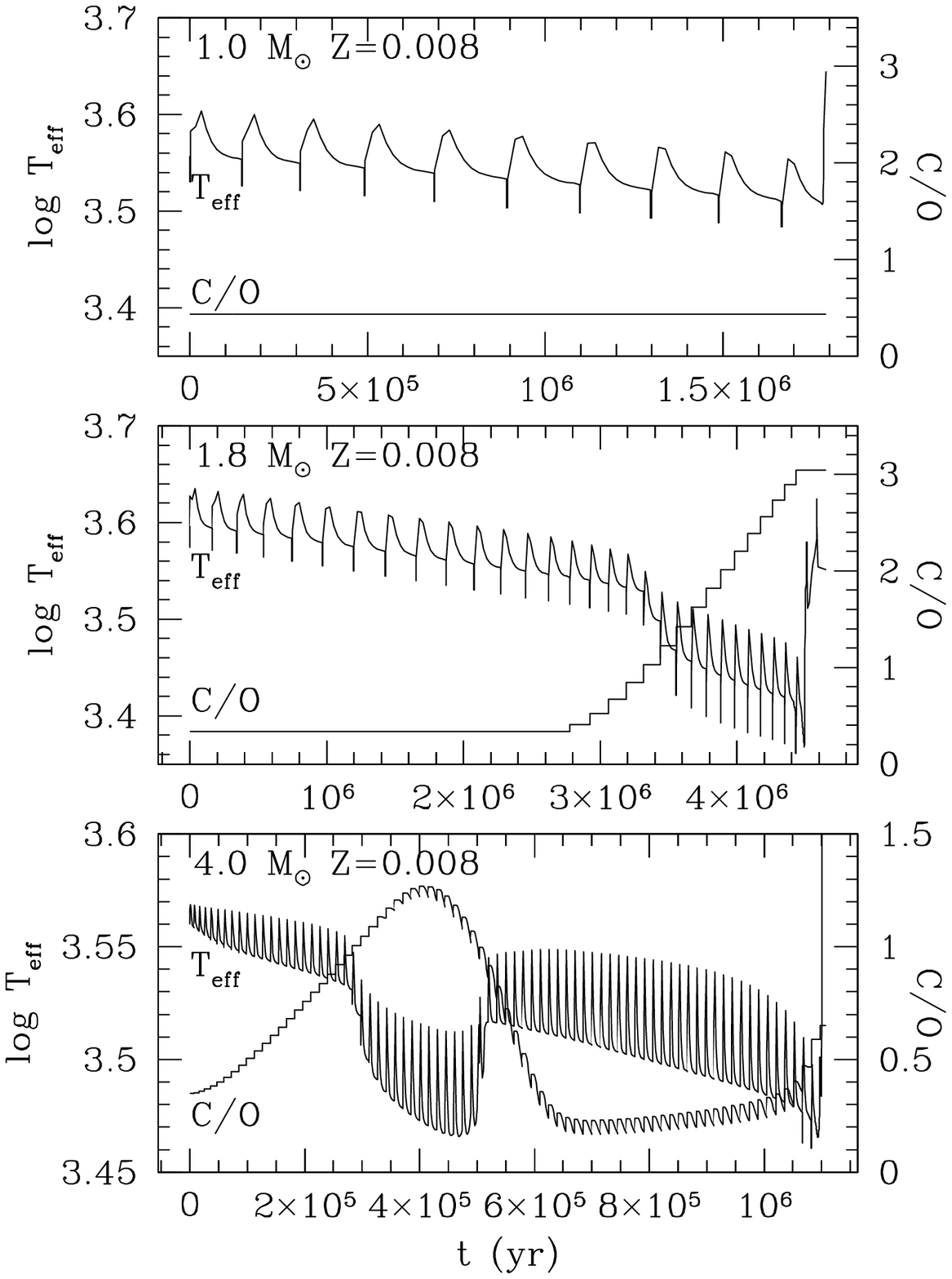}}
\caption{Evolution of effective temperature and photosperic C/O ratio
for the same models as in Fig.~\ref{fig_lum_mc_z008}. Note the
specular behaviour compared to that of the luminosity, and the
remarkable average decrease in $T_{\rm eff}$ as soon as C/O increases
from below to above unity. The reverse trend, i.e. an increase in
$T_{\rm eff}$, occurs instead in the $4.0\, M_{\odot}$ model due to
the subsequent reconversion from C/O$\,>1$ to C/O$\,<1$ caused by HBB.}
\label{fig_teff_z008}
\end{figure}

\begin{figure}[!tbp]  
\resizebox{\hsize}{!}{\includegraphics{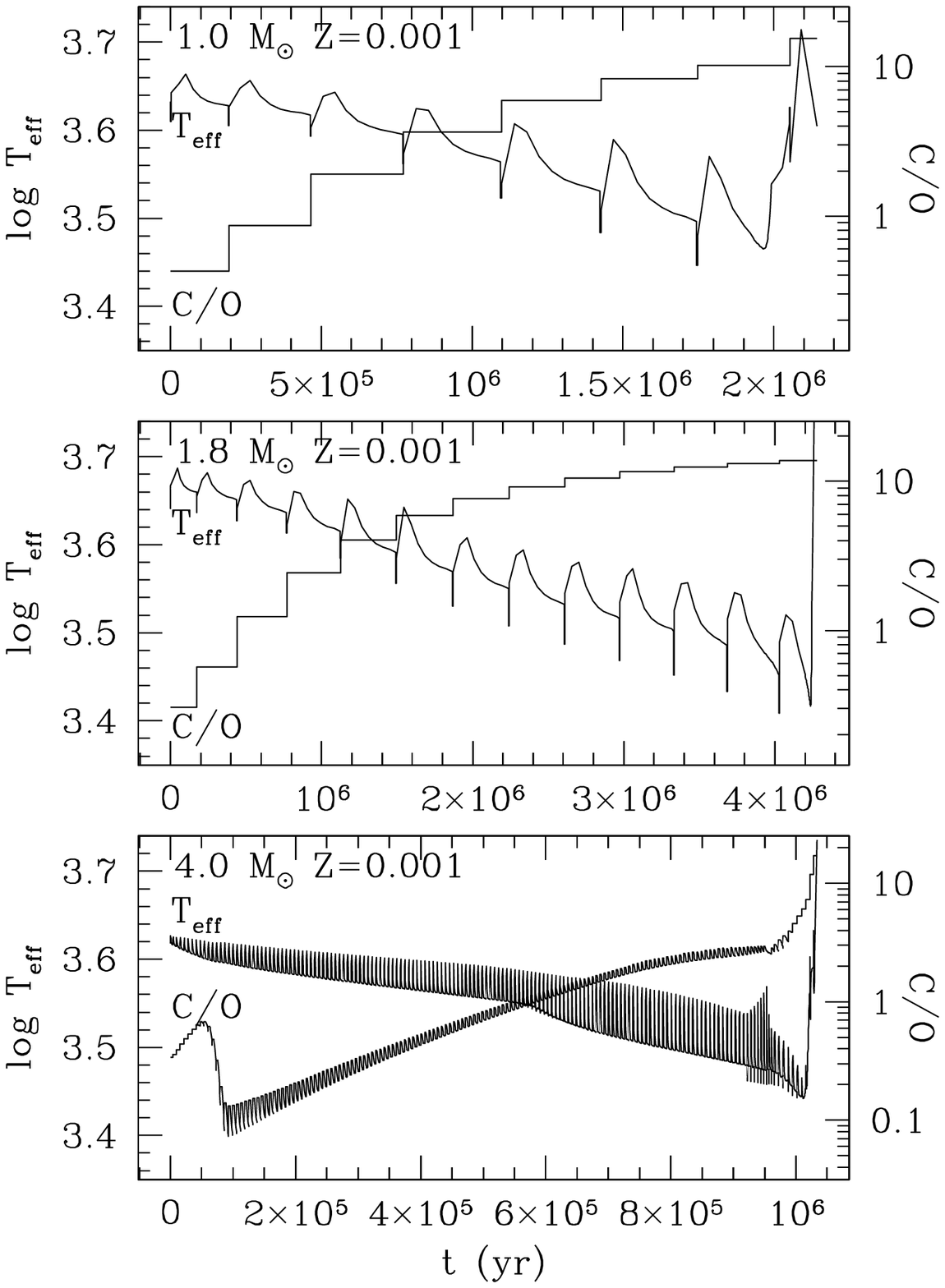}}
\caption{The same as in Fig.~\ref{fig_teff_z008}, but for initial metallicity
$Z=0.001$.}
\label{fig_teff_z001}
\end{figure}

\begin{figure}[!tbp]  
\resizebox{\hsize}{!}{\includegraphics{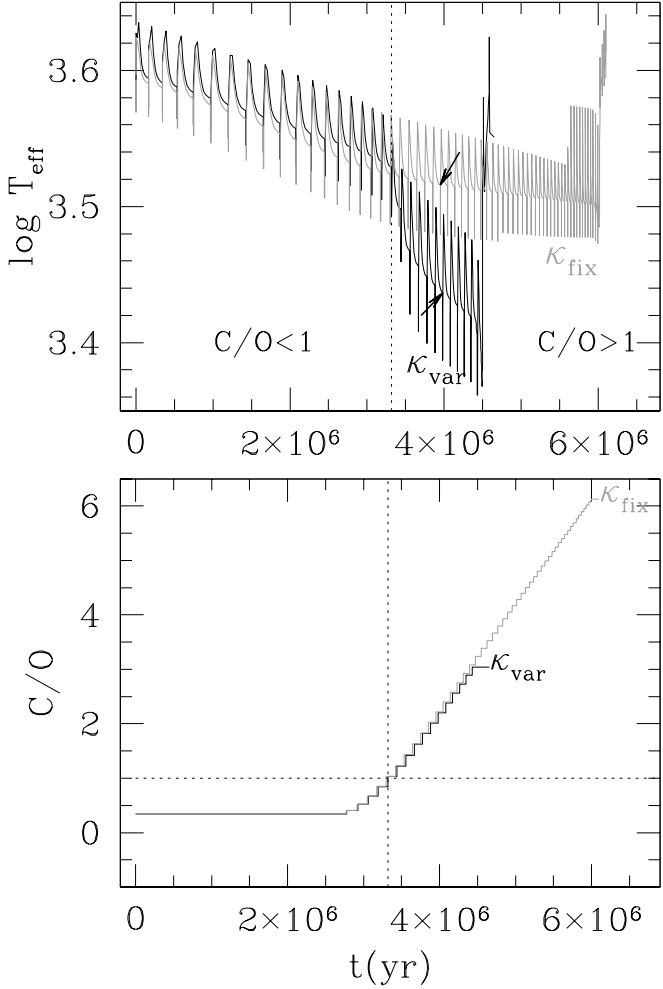}}
\caption{Predicted behaviour of the effective temperature 
(top panel) and photospheric C/O ratio (bottom panel) over the entire
TP-AGB evolution of a model with $M_{\rm i}=1.8 M_{\odot},\, Z=0.008$,
adopting either fixed solar-scaled (model A in Table~\ref{tab_mod};
gray line) or variable (model D in Table~\ref{tab_mod}; black line)
molecular opacities. The two $T_{\rm eff}$ curves suddenly separate as
soon as the models becomes carbon-rich as a consequence of the third
dredge-up.  Arrows indicate the quiescent stage just preceeding the
$24^{\rm th}$ thermal pulse in both models, whose photospheric
structures are shown in Fig.~\ref{fig_envelope}.}
\label{fig_teff_compare}
\end{figure}

\begin{figure}[!tbp]  
\resizebox{\hsize}{!}{\includegraphics{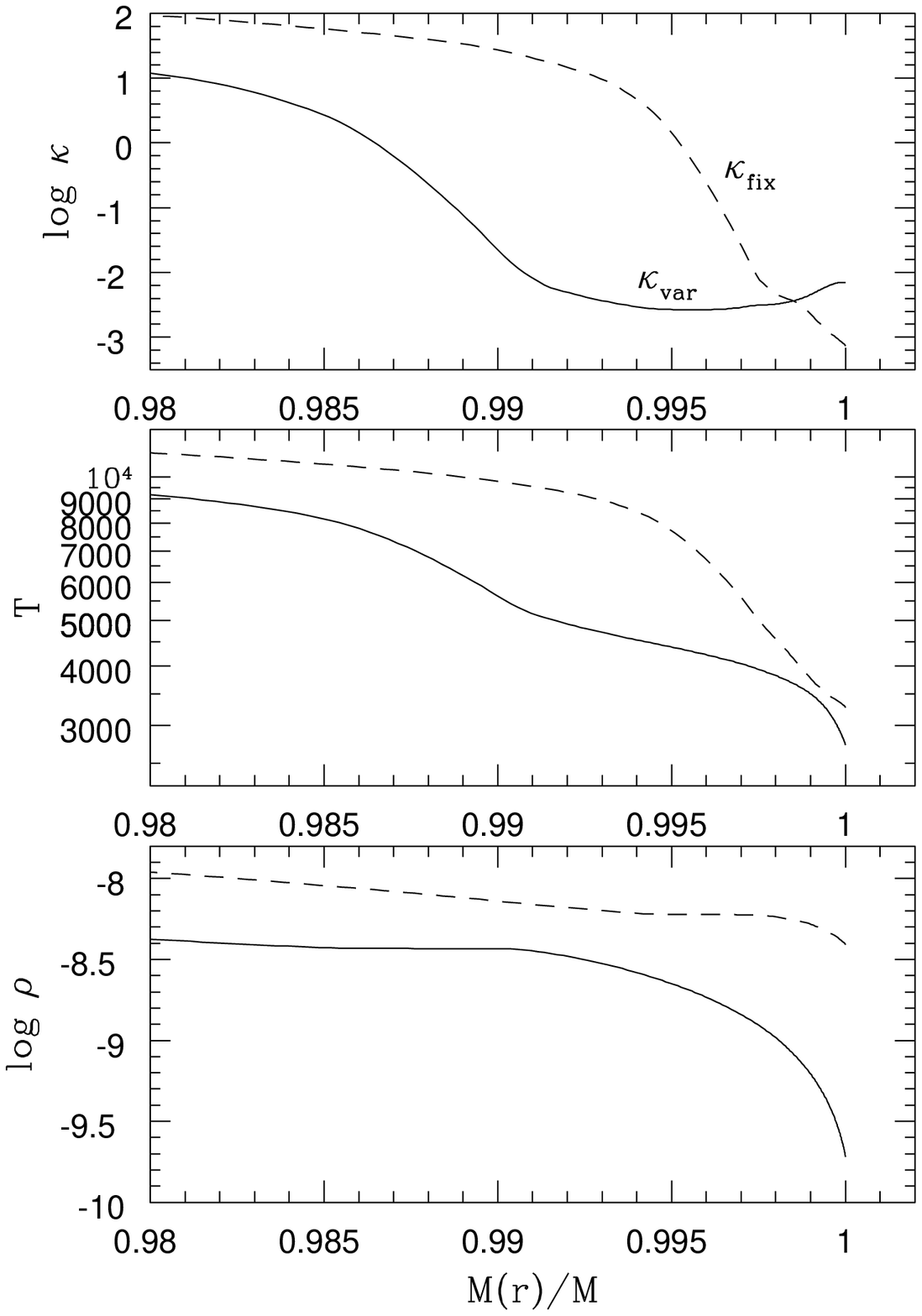}}
\caption{Structure of the surface layers  
of a model with $M=1.776 M_{\odot},\, M_{\rm c}=0.632M_{\odot},\,\log
L=3.926 L_{\odot}$ and C/O$=2.560$, as a function of the mass
coordinate $M(r)/M$, where $M(r)$ is the mass within a sphere of
radius $r$. From top to bottom panels we compare the behaviour of the
Rosseland mean opacity (cm$^2$ gr$^{-1}$), the temperature, and the
density (in g $cm^{-3}$) for $\kappa_{\rm fix}$ (solid line) and
$\kappa_{\rm var}$ (dashed line) cases.}
\label{fig_envelope}
\end{figure}

\subsection{Interpulse period}
\label{ssec_ip}
The interpulse period $\tau_{\rm ip}$ marks the pace of thermal pulses
during the TP-AGB phase, hence affecting both the evolution over the
quiescent regime (e.g., core mass and luminosity increase) and the
events associated to the occurrence of He-flashes (e.g., frequency of
dredge-up episodes).

In the present calculations we adopt the formalism suggested by I04,
which is a modified version of Wagenhuber \& Groenewegen (1988) core
mass-interpulse period relation that accounts for the increase of
$\tau_{\rm ip}$ seen in the more massive models with deep dredge-up.
The I04 (their equation 28) improvement consists in adding a positive
term, depending on the dredge-up efficiency $\lambda$, to the
Wagenhuber \& Groenewegen's relation.

\subsection{Effective temperature}
The effective temperature of our TP-AGB models is derived with the aid
of complete integrations of static envelope models, extending from the
photosphere down to the core. The adopted integration scheme is fully
described in Marigo et al. (1996, 1998, 1999), and Marigo (2002) to
whom the reader should refer for all details. The fundamental output
of the envelope integrations is the effective temperature, which
affects the properties of the TP-AGB evolution, like position in the
HR diagram, mass loss efficiency, pulsation, colours, etc.

One crucial aspect to be remarked regards the radiative opacities at
low temperatures (i.e. $T<10\,000\, {\rm K}$), which are predominantly
due to the molecules (and dust grains for $T<1\,500\, {\rm K}$) that
form in the coolest photospheric layers of AGB stars.  Importantly, we
abandon the unproper choice -- still commonly used in evolutionary
calculations of the AGB phase -- of opacity tables for solar-scaled
chemical composition ($\kappa_{\rm fix}$). The present TP-AGB models
are computed with the adoption of the routine developed by Marigo
(2002), that calculates the low-temperature molecular opacities
consistently coupled with the current chemical composition of the gas
($\kappa_{\rm var}$). In this way we are able to account for the drastic
opacity changes that take place as soon as a model passes from an
oxygen-rich (C/O$\,<1$) to a carbon-rich (C/O$\,>1$) surface composition
as a consequence of the third dredge-up, or when the reverse
transition possibly occurs due to HBB.
 
Figures~\ref{fig_teff_z008} and \ref{fig_teff_z001} display the
evolution of the effective temperature and surface C/O ratio during
the whole TP-AGB evolution of a few selected synthetic models with varying
initial mass and metallicity.  Thermal pulses cause the quasi-periodic
complex behaviour of $T_{\rm eff}$ that essentially mirrors that of
$L$, shown in the corresponding Figs.~\ref{fig_lum_mc_z008} and
\ref{fig_lum_mc_z001}.  As just mentioned, the introduction of
variable molecular opacities is responsable for the sudden change in
the average slope of the $\log T_{\rm eff}(t)$ curve as the
photospheric C/O rises above unity in models where the third dredge-up
take place. In contrast, models without the third dredge-up
(e.g., $M_{\rm i}=1.0\,M_\odot, Z=0.008$) are characterised by a steady
decrease of $T_{\rm eff}$ for most of their evolution (but for the
latest stages of heavy mass loss).  An interesting case is shown by
the $M_{\rm i}=4.0\,M_\odot, Z=0.008$ model experiencing the sequence
C/O$<1\longrightarrow$C/O$>1\longrightarrow$C/O$\,<1$ as a consequence
of the competition between the third dredge-up and HBB. Note the kind
of valley in the $T_{\rm eff}$ curve specularly corresponding to the
part of the C/O curve above unity.

As already demonstrated by Marigo (2002) and Marigo et al. (2003), the
adoption of chemically-variable molecular opacities brings a radical
improvement in the treatment of carbon-star evolutionary models,
leading solve several long-lasting discrepancies between theory
and observations.  For instance, the red tail drawn by field carbon
stars in near-infrared colour-colour diagrams of the Magellanic Clouds
(e.g., DENIS, 2MASS surveys) is reproduced, as well as the low C/O and
\Teff\ values typically found in Galactic AGB C-type stars.

In fact, the reproduction of these basic properties of C stars is 
not possible when using molecular opacities de-coupled from the actual
surface chemical composition of C-rich models, which is still a common
choice in published TP-AGB evolutionary calculations. The sharp
dichotomy between the two cases ($\kappa_{\rm fix}$ and $\kappa_{\rm
var}$) is exemplified in Fig.~\ref{fig_teff_compare}, that compares
the predicted behaviour of $T_{\rm eff}$ in a $M_{\rm i}=1.8,\,
Z=0.008$ TP-AGB model with chemically-variable (model A with
$\kappa_{\rm var}$ in Table~\ref{tab_mod}) and with fixed,
solar-scaled (model D with $\kappa_{\rm fix}$ in Table~\ref{tab_mod})
molecular opacities.  It is well evident that the \Teff\ evolution of
the $\kappa_{\rm fix}$ model is not influenced at all by the behaviour
of the C/O ratio, while in the $\kappa_{\rm var}$ case the rate of
$T_{\rm eff}$ decrease with time becomes significantly faster as the
star enters the C-rich domain. As a consequence, keeping the other
input prescriptions the same, the duration of the TP-AGB phase turns
out shorter in the $\kappa_{\rm var}$ model compared to the the
$\kappa_{\rm fix}$ one.

Indeed, molecular opacities heavily affect the structure of the star's
surface layers, as illustrated in Fig.~\ref{fig_envelope}. The
significant differences between $\kappa_{\rm fix}$ and $\kappa_{\rm
var}$ (top panel) strongly influence the temperature (middle panel)
and density (bottom panel) stratification in the external regions. In
the outermost layers, in particular (with mass coordinate $0.998\la
M(r)/M \le 1$), $\kappa_{\rm var}$ overcomes $\kappa_{\rm fix}$, which
determines a sizeable reduction of the effective temperature, passing
from $T_{\rm eff} \sim 3281\, K$ for $\kappa_{\rm fix}$ to $T_{\rm
eff} \sim 2729\,K$ for $\kappa_{\rm var}$. As further consequence, the
$\kappa_{\rm var}$ model corresponds to a more expanded photosphere
(i.e. $R$ is forced to increase to allow the model radiate the same
$L$ ), hence to lower densities (bottom panel).

\begin{figure}[!tbp]  
\resizebox{\hsize}{!}{\includegraphics{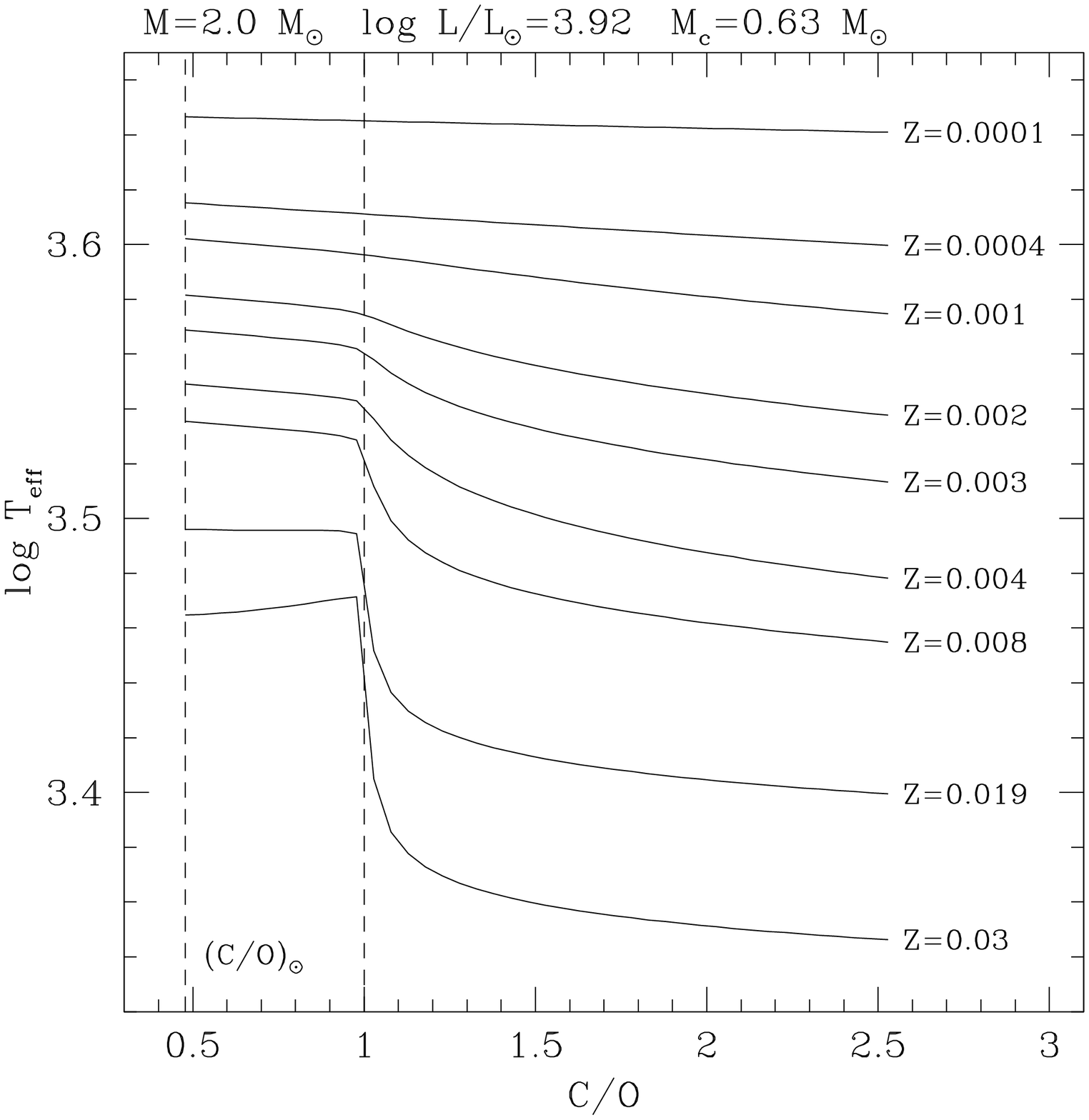}}
\caption{Metallicity dependence of the relationship between the effective 
temperature and the photosphetic C/O ratio.  The curves are obtained
by means of envelope integrations assuming the same set of stellar
input parameters (i.e. total mass, core mass, luminosity), while
varying the initial metallicity, as indicated. Starting from solar
C/O, the carbon abundance is progressively increased to mimic the
effect of the third dredge-up. We notice that the sudden photosperic
cooling caused by molecular opacities as soon as C/O $\,>1$ is quite
pronounced at higher metallicities, while it almost disappears for $Z
<0.0004$.  }
\label{fig_teffz}
\end{figure}

At this point it is important analysing how much such behaviour is
sensitive to the metallicity.  Considering that the formation of
molecules in the atmospheres of AGB stars becomes less efficient at
lower $Z$ -- because of the higher temperatures and the lower
abundances of the involved isotopes -- it follows that the cooling
effect on the atmospheres of carbon stars due to molecular opacities
should be less and less evident at decreasing metallicity.  To give an
example, by comparing Fig.~\ref{fig_teff_z008} and
Fig.~\ref{fig_teff_z001} one already notices that the sudden change in
the mean slope of the effective temperature curve, at the transition
to the C-star domain, is much more pronounced in TP-AGB models with
$Z=0.008$ compared to those with $Z=0.001$.  This point is better
illustrated with the help of Fig.~\ref{fig_teffz} that shows how the
effective temperature of a selected model (with given total mass,
core mass, and luminosity) is expected to behave at increasing C/O
ratio over a wide range of initial metallicities. As already
mentioned, the sudden and extended excursion toward lower effective
temperature as soon as C/O$\ga 1$, characterising models of higher $Z$,
is progressively reduced at decreasing metallicity, eventually
disappearing for $Z < 0.0004$.

\subsection{The third dredge-up}
\label{ssec_3dup}
The treatment of the third dredge-up is essentially the same one as in
Marigo et al. (1999) to which the reader should refer for all details.
In practice we need to specify three quantities, namely: (i) the
minimum temperature at the base of the convective envelope at the stage 
of the post-flash luminosity maximum, $T_{\rm b}^{\rm dred}$, or
equivalently the minimum core mass required for
the occurrence of the mixing event, $M_{\rm c}^{\rm min}$; 
(ii) the efficiency of the third
dredge-up $\lambda$; and (iii) the chemical composition of the
inter-shell.  Compared to our previous models (e.g., Marigo \& Girardi
2001 and references therein), in this work we improve the description
of the third dredge-up, thanks to the recent results of detailed
calculations of the AGB phase, namely K02 and I04. The main points are
summarised as follows.
  
\subsubsection{Occurrence of dredge-up} 
In Marigo et al. (1999) the adopted criterion assumes that $T_{\rm
b}^{\rm dred}$, derived from numerical integrations of static envelope
models, is constant regardless of stellar mass and metallicity,
following the indications from full AGB calculations as reported by
Wood (1981).  The calibration of the parameter from the fits to the
observed CSLFs in the Magellanic Clouds yields $\log T_{\rm b}^{\rm
dred} = 6.4$.

In this study we explore an alternative approach to $\log T_{\rm
b}^{\rm dred}$ by introducing the quantity $M_{\rm c}^{\rm min}$, that
is the minimum core mass required for dredge-up to take place.  This
has been a commonly used parameter in synthetic AGB models over the
years, often assumed to be the same ($\sim 0.58\, M_{\odot}$) for any
stellar mass and metallicity (Groenewegen \& de Jong 1993; Marigo et
al. 1996; Mouhcine \& Lan\c{c}on 2002).
Only very recently extended grids of complete AGB models have become
available (e.g., K02; I04), allowing to relax the crude simplification
of constant $M_{\rm c}^{\rm min}$ in place of a more physically sound
dependence on stellar mass and metallicity.

In the present study we adopt the relations $M_{\rm c}^{\rm
min}(M,Z)$, proposed by K02, based on grids of complete AGB
calculations covering initial masses $M_{\rm i}=1-6\, M_{\odot}$ and
metallicities $Z=0.02, 0.008, 0.004$.  A partial modification of their
analytical fits turns out necessary for the following reasons.
\paragraph{Systematic shift.} 
The reproduction of the observed CSLFs in the Magellanic Clouds
requires that $M_{\rm c}^{\rm min}$ is lower than predicted
(Sect.~\ref{ssec_cslf}; see also K02 and I04).  Our synthetic
calculations indicate that $M_{\rm c}^{\rm min}$ -- as given by
equation 10 in K02 -- must be decreased by an amount $\Delta
M_{c}^{\rm min}$ that depends on metallicity:
\begin{equation}
\Delta M_{\rm c}^{\rm min}/M_{\odot} =   {\rm max}[0.1, 0.1-100\,(Z-0.008)] 
\end{equation}
The negative correction, $- \Delta M_{c}^{\rm min}$, becomes more
important at decreasing
metallicity, being e.g., $(-0.01, -0.05, -0.08)\, M_{\odot}$ for
$Z=(0.008, 0.004, 0.001)$ respectively.  This implies that
for $Z \le 0.001$ the onset of the third dredge-up in stellar models
of any mass would take place since the first thermal pulse.

In practice, we limit this metallicity extrapolation imposing that no 
dredge-up can occur for initial masses $M < 1\, M_{\odot}$, in order
to comply with the observed lack of intrinsic C-stars 
(not related to binary evolution like the CH stars or dwarf carbon stars) 
in Galactic globular clusters.

\paragraph{Correction for overshooting}
As far the evolution prior to the AGB is concerned, K02 do not assume
overshoot from convective cores during H-burning, while a moderate
convective overshoot is included in Girardi et al. (2000) sets of
tracks, from which we extract the initial conditions of our AGB
models.

One important effect due to convective overshoot is that of lowering
the critical mass $M_{\rm HeF}$, that is maximum initial stellar mass
required to ignite core He-burning in degenerate conditions.  At given
metallicity this limit mass is well defined by the minimum 
in the relation between the core mass at the
first thermal pulse, $M_{\rm c}^{\rm 1}$, and the initial stellar mass
$M_{\rm i}$ (see e.g., Marigo et al. 1999; K02). A similar behaviour
is shared by the corresponding $M_{\rm c}^{\rm min}$ vs $M_{\rm i}$
relation (see e.g., K02), so that the minimum of $M_{\rm c}^{\rm min}$
corresponds to $M_{\rm HeF}$.

It follows that the inclusion of convective overshoot in stellar
evolutionary calculations causes, among other effects, a decrease of
$M_{\rm HeF}$, hence a shift of the minimum $M_{\rm c}^{\rm
min}(M_{\rm i})$ towards lower masses.  In other words, the inclusion
of convective overshooting can be approximately described by
considering that a star with initial mass $M_{\rm i}$ behaves like a
star of higher mass $f_{\rm ov}\, M_{\rm i}$ (with $f_{\rm ov} > $1)
without overshooting.
 
\begin{figure}[!tbp]  
\resizebox{\hsize}{!}{\includegraphics{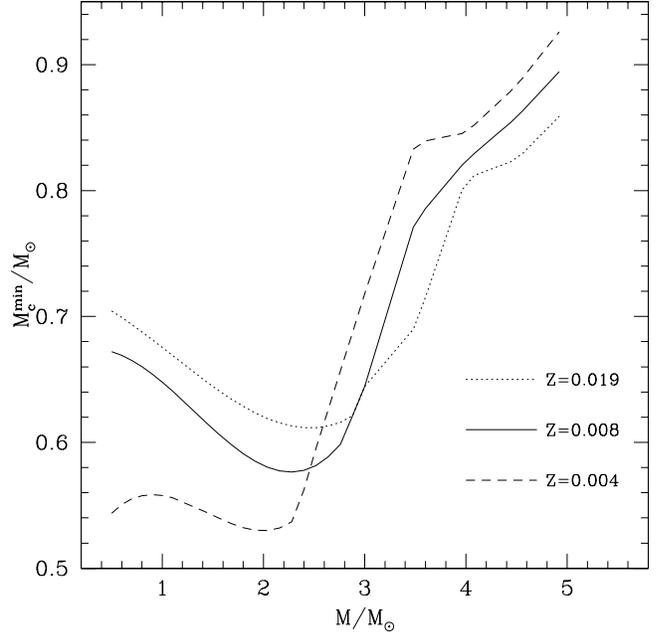}}
\caption{Minimum core mass required for the occurrence of the
third dredge-up as a function of the stellar mass at the first thermal
pulse, for three choice of the initial chemical composition.  Note
that in all curves the minimum corresponds to the critical mass
$M_{\rm He}^{\rm F}$. See text for more explanation. }
\label{fig_mcmin}
\end{figure}

From a comparison between K02 and Girardi et al. (2000) models the
enhancement factor is estimated $f_{\rm ov} \sim 1.2$.  Hence, in
order to mimic the effect of convective overshoot on the relation
$M_{\rm c}^{\rm min}(M_{\rm i})$ we simply adopt the K02 formalism as
a function of the adjusted mass variable $\hat{M}=f_{\rm ov}\, M$:
\begin{equation}
M_{\rm c}^{\rm min}(M) = \left[M_{\rm c}^{\rm min} 
(\hat{M})\right]_{\rm K02} \,\,\,.
\label{eq_mcmin}
\end{equation}
%
%
The resulting $M_{\rm c}^{\rm min}$ at varying initial stellar mass
are shown in Fig.~\ref{fig_mcmin} for $Z=0.019, 0.008, 0.004$. For
metallicities outside the range covered by K02 models, we linearly
extrapolate in $Z$ the coefficients of the fitting polynomial
functions of K02. It should be recalled that for models with initial
mass $M_{\rm i}\ga 3 M_{\odot}$ the core mass predicted by
Eq.~(\ref{eq_mcmin}) is lower than the core mass at the first thermal
pulse $M_{\rm c, 1 TP}$, so that in these cases we set $M_{\rm c}^{\rm
min}=M_{\rm c, 1 TP}$.

\subsubsection{Dredge-up efficiency}
We briefly recall that the efficiency of each dredge-up event is
usually expressed via the quantity $\lambda=\Delta M_{\rm dred}/\Delta
M_{\rm c}$, i.e. the mass brought up to the surface at a thermal
pulse, $\Delta M_{\rm dred}$, normalised to the core mass increment over the
preceeding inter-pulse period, $\Delta M_{\rm c}$.

The usual choice in previous studies has been adopting the same
$\lambda$ for any value of the stellar mass.  The past calibration of
this parameter through the reproduction of the observed CSLFs in the
Magellanic Clouds (see e.g., Marigo et al. 1999) yields $\lambda=0.5$
for $Z=0.008$ and $\lambda=0.65$ for $Z=0.004$.

In the present study we relax the assumption of constant $\lambda$,
accounting for both mass and metallicity dependences found on the
base of K02 and I04 complete AGB models. In summary, once $M_{\rm c}$
has grown above $M_{\rm c}^{\rm min}$, the efficiency $\lambda$
progressively increases towards an asymptotic value $\lambda_{\rm
max}$, with an e-folding interval of $N_{\rm r}$ thermal pulses. This
behaviour is conveniently described by the relation
\begin{equation}
\lambda(N) = \lambda_{\rm max}\left[1-\exp(N/N_{\rm r})\right]  
\end{equation}
where $N$ is the progressive pulse number ($N=0$ as long as $M_{\rm c}
< M_{\rm c}^{\rm min}$).

As to $\lambda_{\rm max}$ we use the analytical fits by K02.  We
notice that at given metallicity the asymptotic quantity increases
with the stellar mass, typically reaching values 0.8--1.0 for $M_{\rm
i}>3\, M_{\odot}$. The metallicity dependence is less strong, such
that at given stellar mass $\lambda_{\rm max}$ tends to be somewhat
higher at decreasing $Z$.

In order to have significantly larger efficiencies at metallicities $Z
< 0.004$ -- that is below the low end of the metallicity range covered
by K02 models -- we modify the original $\lambda_{\rm max}(M, Z)$ 
relation in K02 by applying a positive shift to the mass variable
\begin{equation}
\lambda_{\rm max}(M) = \lambda_{\rm max}(M+\Delta M_{\lambda})\,\,
\end{equation}
where
\begin{equation}
\Delta M_{\lambda}=200\, (0.004-Z)\, M_{\odot} \,\,\,.
\end{equation}

This correction is consistent with the theoretical expectation 
that the third  dredge-up is more efficient at lower metallicities
(see section 3.4 in I04 for a similar adjustment), 
and it is also required to 
reproduce the faint wing of the CSLF in the SMC, which is populated mainly 
by lower-mass stars with metallicities down to $\simeq 0.001$ 
(see Sect.~\ref{sec_calibr}).  

As discussed by K02 there is no simple law able to fit $N_{\rm r}$ as
a function of $M$ and $Z$. A rather satisfactory reproduction of the
results for $N_{\rm r}$ presented by K02 (in their table 5) is
obtained with the function
\begin{eqnarray}
\label{eq_nr}
N_{\rm r} & = & a_1+a_2\,\left[a_3-\exp(a_4 M_{\rm 1TP})\right] \\
\nonumber 
& & \times \left[a_5+a_6 \, \exp(a_7-M_{\rm 1TP})^2\right]  \,\,\,.
\end{eqnarray}
The coefficients ${a_i\,, i=1,\ldots,7}$ are given in
Table~\ref{tab_nr} for $Z=0.019, 0.008,\, {\rm and}\, 0.004$.
Interpolation and extrapolation in $\log Z$ are adopted 
for other metallicities.

\begin{table}[!tbp]  
\caption{Coefficients for Equation~\protect{(\ref{eq_nr})}}
\begin{tabular}{cccc}
\hline
      &        &   Z   &      \\
\cline{2-4}
      &  0.019 & 0.008 & 0.004 \\
\hline
$a_1$ &  4.110 & 2.785 & 2.555 \\
$a_2$ & 42.612 & 10.625 & 809.426 \\ 
$a_3$ & 5.834E-03 & 4.806E-02  & 3.348E-05 \\
$a_4$ & -2.113  & -0.908 & -4.417 \\
$a_5$ &  2.014  & 0.806 &  1.751 \\
$a_6$ & -9.116 & -6.708  & -70.624 \\
$a_7$ & 3.830 & 3.462 & 3.516 \\
\hline
\end{tabular}
\label{tab_nr}
\end{table}

\subsubsection{Chemical composition of the inter-shell}
The abundances of $^{4}$He, $^{12}$C, and $^{16}$O in the intershell
region are derived as a function of the core mass growth $\Delta
M_{\rm c}=M_{\rm c}-M_{\rm c, 1TP}$ on the base of nucleosynthesis
calculations by Boothroyd \& Sackmann (1988). A fit to their figure 9
gives:
\begin{figure}[!tbp]  
\resizebox{\hsize}{!}{\includegraphics{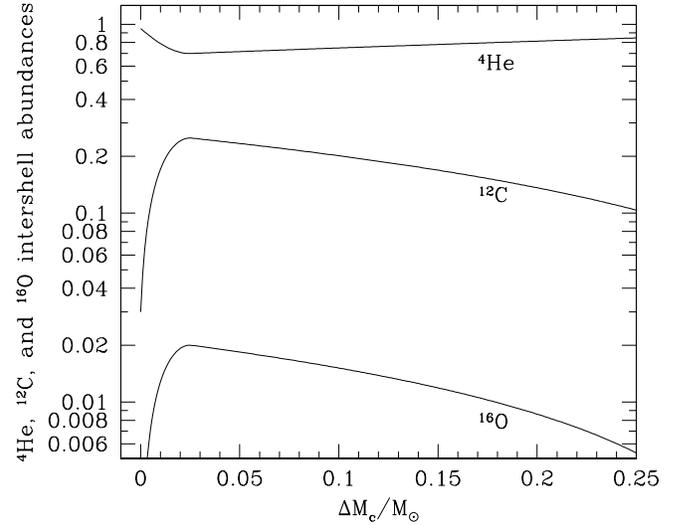}}
\caption{Chemical composition ($^{4}$He, $^{12}$C, and $^{16}$O 
abundances in mass fraction) of the dredged-up material as a function
of the core mass increment during the TP-AGB phase, as predicted by
Eqs.~(\ref{eq_comp1}) and (\ref{eq_comp2}).}
\label{fig_ishell}
\end{figure}
\begin{eqnarray}
\nonumber
X(^{4}{\rm He}) & = &   0.95+400\,(\Delta M_{\rm c})^2-20.0\,\Delta
M_{\rm c}\\
\label{eq_comp1}
X(^{12}{\rm C}) & = &  0.03-352\,(\Delta M_{\rm c})^2+17.6\,\Delta M_{\rm c}\\
\nonumber
X(^{16}{\rm C}) & = &  -32\,(\Delta M_{\rm c})^2+1.6\,\Delta M_{\rm c}
\end{eqnarray}
for $\Delta M_{\rm c} \le 0.025\, M_{\odot}$, and 
\begin{eqnarray}
\nonumber
X(^{4}{\rm He}) & = &   0.70+0.65\,(\Delta M_{\rm c}-0.025) \\
\label{eq_comp2}
X(^{12}{\rm C}) & = &  0.25-0.65\,(\Delta M_{\rm c}-0.025)\\
\nonumber
X(^{16}{\rm O}) & = &  0.02-0.065\,(\Delta M_{\rm c}-0.025)
\end{eqnarray}
for $\Delta M_{\rm c} > 0.025\, M_{\odot}$. These relations are
displayed in Fig.~\ref{fig_ishell}.

Looking at Fig.~\ref{fig_ishell},
we notice that   
the fractional mass of primary $^{12}$C quickly increases 
during the first pulses up to a maximum value of 0.25  
for $\Delta M_{\rm c}=0.025\, M_{\odot}$, and then it starts to decrease
slowly. The same behaviour characterises the evolution
of the  $^{16}$O abundance (with a maximum
value of 0.02), while a mirror-like trend is followed by $^{4}{\rm He}$
that reaches a minimum value of $0.70$.

\subsection{Hot-bottom burning nucleosynthesis}
\label{ssec_hbb}
HBB takes place in the most massive TP-AGB models, with ($M_{\rm i}
\ge 3.5 - 4.5 M_{\odot}$) depending on metallicity, and it corresponds
to H-burning, mainly via the CNO cycle, in the depeest layers of their
convective envelopes.  For the present work we adopt the same
treatment of this process as in Marigo et al. (1999) to which we refer
for more details.

\begin{figure}[!tbp]  
\resizebox{\hsize}{!}{\includegraphics{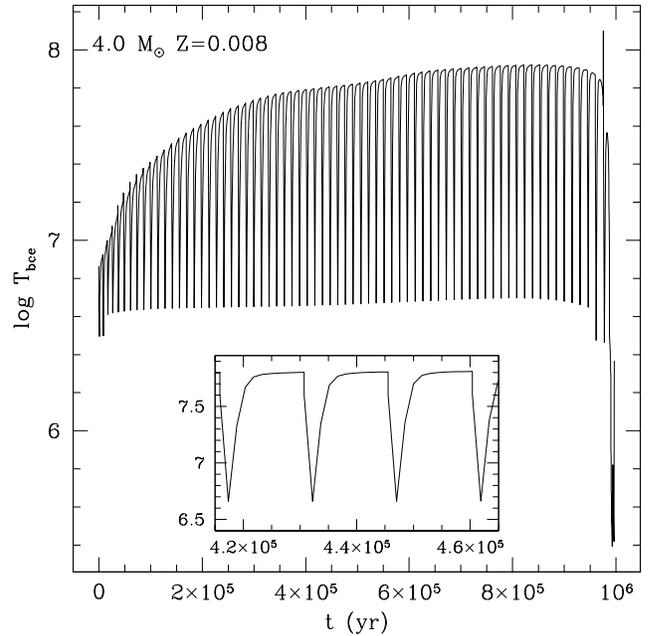}}
\caption{Evolution of the temperature at the base of the convective
envelope during the whole TP-AGB phase of the ($4 M_{\odot},\,
Z=0.008$) model that experiences both the third dredge-up and HBB.
The inset zooms the temporal behaviour of $T_{\rm bce}$ during few
selected pulse cycles. }
\label{fig_hbb}
\end{figure}
   
From an observational point of view, HBB does not only make models
more luminous than expected by the core mass-luminosity relation (see
Sect.~\ref{ssec_lum}) but, more importantly, it produces also notable
chemical changes in the surface chemical composition.  One of the most
significant and well-known effects played by HBB nucleosynthesis is
that of converting C into N, hence delaying or even preventing the
formation of C stars.

However, it should be noticed that, under particular circumstances,
HBB might instead favour the formation of C stars thanks to the
destruction of O in favour of N.  This applies to low-metallicity
massive models in which the temperature at the base of the envelope is
high enough to activate the ON cycle. A few exemplifying cases of
models experiencing both the third dredge-up and HBB are discussed in
Sect.~\ref{ssec_cno} and illustrated in Fig.~\ref{fig_cno}.

It is worth emphasizing that, at each time step during the interpulse
periods, the elemental abundances in the envelope of TP-AGB models
with HBB are determined by solving a network of nuclear reactions (the
p-p chains and the CNO tri-cycle), according to the current
temperature and density stratifications given by numerical
integrations of a complete envelope model (see Marigo et al. 1998 for
a complete description of the method).  This allows a consistent
coupling of the HBB nucleosynthesis with the structural evolution of
the envelope, a technical characteristic that places our synthetic
TP-AGB model closer to full TP-AGB models.  In fact, as already shown
by I04 a parameterised treatment of HBB nucleosynthesis in synthetic
TP-AGB calculations turns out quite complex and sometimes troublesome
for the relatively large number of the parameters involved ($\approx
7$), their intrisic degeneracy, and the non-univocity of the
calibration.

As an example let us consider one of these quantities that is directly
related to the efficiency of HBB, i.e. the temperature at the base
of the convective, $T_{\rm bce}$.  I04 (see their section 3.3.4)
propose a parametric formalism for $T_{\rm bce}$, which in turn
requires additional parameters to be calibrated on the base of full
TP-AGB calculations.

On the other side, by means of envelope integrations carried out with
our synthetic TP-AGB code, we are able to naturally predict the
temporal evolution of $T_{\rm bce}$ during the whole TP-AGB phase, as
displayed in Fig.~\ref{fig_hbb} for the ($M_{\rm i}=4.0 M_{\odot},\,
Z=0.008$) model.  Note the complex behaviour and wide excursion in
$T_{\rm bce}$ due to thermal pulses, which are zoomed in the inset.
Considering only the quiescent pre-flash stages, that ideally describe
the upper envelope of the $T_{\rm bce}$ curve, we see that the
temperature steeply rises during the first pulse cycles, then it
flattens out toward a maximum value, and eventually it drastically
drops as soon as the envelope mass is significantly reduced by mass
loss.

\subsection{Pulsation modes and periods}
\label{ssec_pulsation}
Models for pulsating AGB stars (see e.g., Fox \& Wood 1982) indicate
that the dominant growth rate (GR) among different pulsation modes is
correlated with luminosity, i.e. modes of higher order have the
highest GR at lower luminosities.  This means that if we consider, for
instance, the fundamental (FM) and first overtone (FOM) modes with
growth rates GR$_0$ and GR$_1$, respectively, we expect GR$_1 >$GR$_0$
at lower luminosities while the situation is reversed 
at higher luminosities.

Hence making the reasonable assumption that the active pulsation mode
among the possible modes is that one corresponding to the highest GR,
we derive a picture of pulsation on the AGB: as a star climbs in
luminosity it becomes unstable through a sequence of pulsation modes,
first exciting those of higher order (i.e. third $\rightarrow$ second
$\rightarrow$ first overtone) and eventually switching to the
fundamental mode (see Lattanzio \& Wood 2003).

Due to the lack of sufficient theoretical data on higher order
pulsation modes, in this study we limit to consider the first overtone
mode (FOM) and the fundamental mode (FM). We assume that all stellar
models start their TP-AGB evolution as FOM pulsators.  The transition
from the first overtone to the FM pulsation may take place later,
provided a minimum luminosity, $L_{1-0}$, is reached. This critical
luminosity is defined by the stage at which the growth rates per unit
time, GR, for both modes become equal (GR$_1=$GR$_0$)\footnote{This
latter criterion is the same one as proposed by Bedijn (1988) in his
synthetic AGB model.}.

\begin{figure}[!tbp]  
\resizebox{\hsize}{!}{\includegraphics{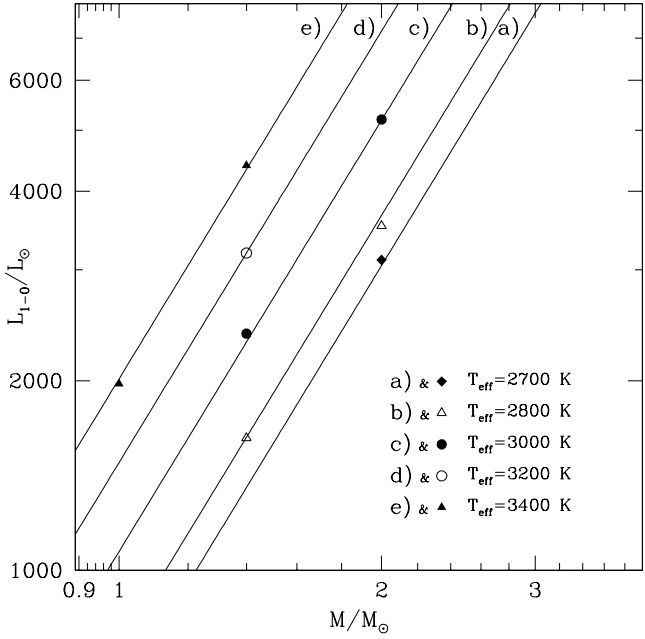}}
\caption{Critical luminosity $L_{1-0}$ defined by the equality of the growth
rates for FO and FM pulsation. Symbols show the predictions 
of nonadiabatic pulsation models by Ostlie \& Cox (1986) for different
values of stellar mass and effective temperature. Lines (a-e) are plotted 
according to the fitting Eq.~(\protect\ref{eq_p1p0}), 
each of them corresponding to 
a given $T_{\rm eff}$ at varying stellar mass. See text for more details. }
\label{fig_p1p0}
\end{figure}

To obtain $L_{1-0}$ we use the results of the linear, nonadiabatic 
pulsation models for Mira variables calculated by Ostlie \& Cox (1986).
A good fitting relation (based on their table 2) providing the critical
luminosity as a function of stellar mass and effective temperature is

\begin{eqnarray}
\label{eq_p1p0n}
\log L_{1-0}/L_{\odot} &=& -14.516+2.277\log M/M_{\odot} \\
	\nonumber & & +5.046\log T_{\rm eff}\,\,. 
\end{eqnarray}
As we notice from Fig.~\ref{fig_p1p0} the switch from the FOM to the
FM is expected to occur at higher luminosity for larger stellar mass
and effective temperature.  Since Ostlie \& Cox (1986) present models
for solar composition only -- with $(Y,Z)=(0.28,0.02)$ --, we include a
metallicity dependence to Eq.~(\ref{eq_p1p0n}) as quantified by Bedijn
(1988) on the base of pulsation models for long-period variables
calculated by Fox \& Wood (1982) for metallicities $Z=0.001, 0.01,
0.02$.  Following Bedijn (1988, see his equation 22) the transition
radius, $R_{1-0}$, from the FOM to the FM scales as
$(Z/0.02)^{-0.042}$, hence we get
\begin{eqnarray}
\label{eq_p1p0}
\log L_{1-0}/L_{\odot} & = & 
-14.516+2.277\log M/M_{\odot} \\
\nonumber
& & +5.046\log T_{\rm eff}-0.084\log(Z/0.02)\, , 
\end{eqnarray}
considering that $L_{1-0} \propto R^2_{1-0} T_{\rm eff}^4$ for given 
effective temperature.

The pulsation periods are derived from theoretical $(P,M,R)$-relations
for long-period variables calculated by Fox \& Wood (1982).
Specifically, FOM periods ($P_1$) are derived through  
the analytical fit of the pulsation constant (Wood et al. 1983):
\begin{eqnarray}
\label{eq_p1}
Q/{\rm days} & = &  (P_1/{\rm days})(M/M_{\odot})^{1/2}(R/R_{\odot})^{-3/2}\\ 
\nonumber
             & = & 0.038 \\
\nonumber
             & & \!\!\!\!\!\!\!\!\!({\rm if}\,\, M\ge 3 M_{\odot};\,\,
            {\rm or}\,\,M=2 M_{\odot}\,\,{\rm and}\,\,P_1<300\,\,{\rm days};\\
\nonumber
             & & \!\!\!\!\!\!\!\!\!{\rm or}\,\,M=1 M_{\odot}\,\,{\rm and}
             \,\,P_1<200\,\,{\rm days})\\
\nonumber
             & = & 0.038+2.5\times10^{-5}(P_1-300) \\
\nonumber
             & & \!\!\!\!\!\!\!\!\!({\rm if}\,\,M=2 M_{\odot}\,\,{\rm and}
             \,\,P_1>300\,\,{\rm days})\\
\nonumber
             & = & 0.038+4.5\times10^{-5}(P_1-150) \\
\nonumber
             & & \!\!\!\!\!\!\!\!\!({\rm if}\,\,M=1 M_{\odot}\,\,{\rm and}
             \,\,P_1>150\,\,{\rm days})\\
\nonumber
             & = & 0.038+5.5\times10^{-5}(P_1-100) \\
\nonumber
             & & \!\!\!\!\!\!\!\!\!({\rm if}\,\,M=0.7 M_{\odot}\,\,{\rm and}
             \,\,P_1>100\,\,{\rm days})\,
\end{eqnarray}
whereas FM periods, $P_0$, are calculated with
\begin{eqnarray}
\label{eq_p0}
\log(P_0/{\rm days}) & = & -2.07+1.94\log(R/R_{\odot}) \\
\nonumber
                   & & -0.9\log(M/M_{\odot})\,\,\,\,({\rm if}\,\,M<1.5 M_{\odot}) \\
\nonumber
                   & = & -2.59+2.2\log(R/R_{\odot}) \\
\nonumber
                   &   & -0.83\log(M/M_{\odot})-0.08\log(Z/10^{-3}) \\
\nonumber
                   &   & +0.25 (Y-0.3)\,\,\,\,({\rm if}\,\,M>2.5 M_{\odot})
\end{eqnarray}
where $Z$ and $Y$ denote metallicity and helium content (in mass
fraction), respectively. These latter formulae are taken from
Groenewegen \& de Jong (1994; for $M\le1.5 M_{\odot}$) and Fox \& Wood
(1982; for $M\ge2.5 M_{\odot}$). For $1.5<(M/M_{\odot})<2.5$, $\log P_0$
is linearly interpolated using $\log M$ as the independent variable.

\begin{figure}[!tbp]  
\resizebox{\hsize}{!}{\includegraphics{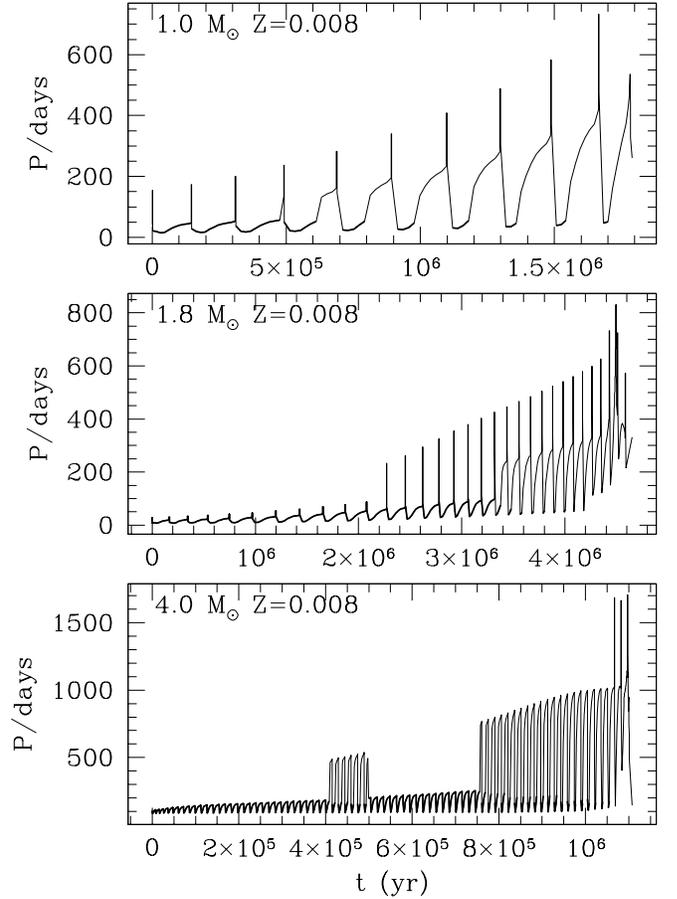}}
\caption{Evolution of pulsation period along the TP-AGB phase
of a few selected models with initial metallicity $Z=0.008$. 
First-overtone and fundamental periods, $P=P_1$ and $P=P_0$, 
are drawn with thick and thin solid lines, respectively. }
\label{fig_period_z008}
\end{figure}
\begin{figure}[!tbp]  
\resizebox{\hsize}{!}{\includegraphics{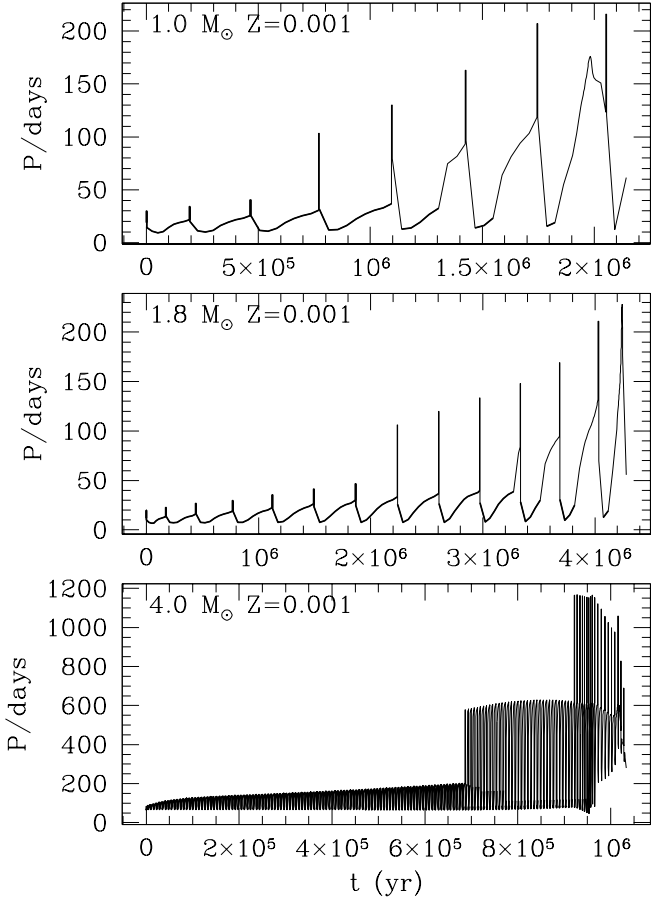}}
\caption{The same as in Fig.~\ref{fig_period_z008}, but for metallicity
$Z=0.001$.}
\label{fig_period_z001}
\end{figure}

\subsubsection{A cautionary remark: how to treat pulsation in 
C-rich models}
\label{sssect_pcrich}

At present all published pulsation models for long-period variables
(e.g., Fox \& Wood 1982; Wood 1990; Ya'Ari \& Tuchman 1996) are
strictly valid only for oxygen-rich chemical compositions, with
photospheric C/O$\,<1$, while models suitably constructed for pulsating
C stars -- and using proper C-rich opacities -- are still
missing. This means that the extension of the $(P,M,R)$-relations
available in the literature, is not straightforward, and possibly
wrong.  Since one of the main goals of the present study is to provide
a more reasonable description of the AGB evolution as a
function of the stellar chemical-type, we face the problem to assign
periods to C-rich models. Given the lack of theory, the only remaining
approach comes then from the empirical ground.

For instance, assuming that the inertial radius $R$, which determines
the pulsation period, coincides with the photospheric radius --
defined by the black-body relation $L = R^2 T_{\rm eff}^4$ (in solar
units) --, and applying Eqs.~(\ref{eq_p1}-\ref{eq_p0}) to compute
periods of C-rich models, then one would expect that, at a given
luminosity, C stars should have larger $R$ owing to their lower
$T_{\rm eff}$, hence longer periods compared to M stars.

As already noted long ago by Wood et al. (1983) this prediction is
contradicted by observations since both groups of stars appear to
populate the same sequences in the $(M_{\rm bol}-P)$ diagram, as shown
for instance in figure 6 of Wood et al. (1983; see also Hughes \& Wood
1990).  This feature has been confirmed by more recent and extended
surveys of long-period variables (e.g., Groenewegen 2004; Fraser et
al. 2005).

As suggested by Wood et al. (1983) the immediate implication is that,
differently from the photospheric radius, the inertial radius of
pulsating AGB stars should be largely insensitive to the photospheric
C/O ratio. In other words, while the photospheric radius of C stars is
expected to be much more extended than that of M stars of the same
luminosity due to an opacity effect, the inertial radii of both
classes of stars should be quite similar.

For all these reasons, in this work we simply decide to calculate
periods as follows (all quantities are in solar units)
\begin{equation}
R = L^{0.5} T_{\rm eff}^2\,\,\,\,\,\,\,\,\,\,\mbox{\rm for M-type models with C/O}<1 
\end{equation}
that is the inertial radius coincides with the photospheric radius, 
\begin{equation}  
R = L^{0.5} T_{\rm eff, M}^2\,\,\,\,\,\mbox{\rm for C-type models with C/O}>1
\end{equation}
that is the inertial radius is calculated by assigning the effective
temperature $T_{\rm eff, M}$ that would compete to an M-type model
with the same luminosity. In these case $T_{\rm eff, M}$ is derived
from envelope integrations always assuming a fictious oxygen-rich
composition (i.e. assuming the solar ratio, C/O$=0.48$).

The predicted period evolution during the whole TP-AGB phase of a few
selected models is shown in Figs.~\ref{fig_period_z008} and
\ref{fig_period_z001}.  The variations of luminosity and photospheric 
radius produced by thermal pulses reflect in the quasi-periodic
behaviour of $P$, in substantial agreement with the predictions of
detailed AGB models and their pulsation properties (e.g., Vassiliadis
\& Wood 1993; Wagenhuber \& Tuchman 1996).  Moreoever, in all models
it is well evident the switch from FOM to FM pulsation regime and
viceversa, corresponding to a mean change in period by a factor of
$\sim 2$.  According to the luminosity criterion given by
Eq.~(\ref{eq_p1p0}) the mode switching would not correspond to a
single event during the TP-AGB evolution, but it could actually take
place several times, driven by the luminosity variations produced by
thermal pulses. As displayed in Figs.~\ref{fig_period_z008} and
\ref{fig_period_z001}, there might be cases in which the 
mode transition may occur during the pulse cycle, with the FOM
pulsation charactering the low-luminosity dips, and the FM pulsation
being recovered at increasing luminosities during the quiescent
stages.  It should be also noticed that the $T_{\rm eff}$-dependence
of $L_{1-0}$ in Eq.~(\ref{eq_p1p0}) favours the switching to the FM
pulsation at decreasing effective temperature.  This explains the
occurrence of a short FM stage experienced by the $(M_{\rm i}=4.0
M_{\odot},\, Z=0.008)$ model at evolutionary ages $t \approx 5\times
10^{5}$ yr, which just corresponds to the transitory fulfillment of
the condition C/O $> 1$ (see Fig.~\ref{fig_teff_z008}), thanks to the
effect of the third dredge-up prevailing over that of HBB.

\subsection{Mass loss rates}
\label{ssec_mlr}
Mass loss rates, $\dot M$, during the TP-AGB evolution are calculated
in a different fashion depending on whether a stellar model is
oxygen-rich or carbon-rich.

\begin{figure}[!tbp]  
\resizebox{\hsize}{!}{\includegraphics{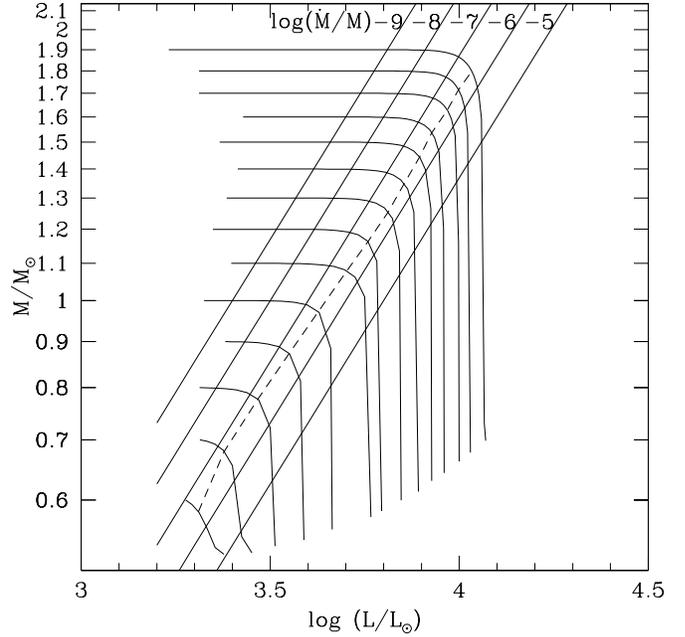}}
\caption{Mass evolution of O-rich TP-AGB models with $Z=0.019$ as a 
function of luminosity, with mass-loss rates calculated on the base of
Bowen \& Willson (1991) dynamical atmospheres for Miras including
dust. The five parallel solid lines define approximate loci of
constant $\dot M/M$ (in units of yr$^{-1}$),
corresponding to evolutionary slopes $\displaystyle d\log M/d\log L=
-0.0018, -0.018, -0.18, -1.8, -18$ (from left to right). The dashed
line is the ``cliff'' line, along which $\displaystyle d\log M/d\log
L=1$ and $\dot M/M = 5.67\times10^{-7}$ yr$^{-1}$.}
\label{fig_mlorich}
\end{figure}
  
\subsubsection{O-rich models}
\label{sssect_omlr}
In the C/O$\,<1$ case we use the results of dynamical atmospheres
including dust for long-period variables calculated by Bowen \&
Willson (1991).  The grid of model atmospheres is suitably constructed
for oxygen-rich Mira-type stars with solar metallicity and it covers
the ranges of initial masses $M_{\rm i}=0.7-2.4 \, M_{\odot}$ and
periods $P=150-800\, {\rm days}$. Pulsation is assumed to occur in the
FM.

By suitably combining their dynamical calculations and with a few
basic formulas\footnote{These refer to the period-mass-radius relation
for FM pulsation, the radius-luminosity-mass relation, and the core
mass-luminosity relation.} to describe the evolution of stellar
parameters on the AGB, Bowen \& Willson (1991) derive a relationship
between the mass loss rate $\dot M$ and the slope of a stellar AGB
evolutionary track in the $\log M$--$\log L$ diagram:
\begin{equation}
\frac{\dot M}{M} = 5.65\times10^{-7} \, \frac{d\log M}{d\log
  L}\,\,\,\,\,\,\,\,\,\,\,\,\,\,\,\,\,\,\,\,\,\,\,\,\,\,\,[{\rm yr}^{-1}]
\label{eq_sl}
\end{equation}     
where $M$ and $L$ are in solar units, and $\dot M$ is given in
$M_{\odot}$ yr$^{-1}$.

To summarise, the adopted procedure to specify $\dot M$ for an
oxygen-rich AGB model is as follows. Given the current values of
stellar mass $M$, photospheric radius $R$, and pulsation period $P$
(obtained according to Sect.~\ref{ssec_pulsation}), we derive the
corresponding luminosity used by Bowen \& Willson (1991, their
equation 3) in their computation of model atmospheres. Then, for any
specified combination of $(L, M)$, the evolutionary slope
$\displaystyle{d\log M}/{d\log L}$ is obtained by interpolation
between grid points defined by the lines of constant $\dot M/M$ shown
in figure 2 of Bowen \& Willson (1991). Finally the mass-loss rate
follows from Eq.~(\ref{eq_sl}).

Figure~\ref{fig_mlorich} shows a grid of TP-AGB evolutionary tracks
with C/O$\,<1$ in the $\log M$--$\log L$ diagram. 
We plot only models at the quiescent stages of pre-flash
luminosity maximum. Along
each track the stellar mass keeps constant until an abrupt change of
slope takes place. This marks the onset of the superwind mass loss,
which determines the quick ejection of the whole envelope and the
termination of the AGB phase.  The sequence of these sharp corners at
varying $M$ and $L$ form the so called ``cliff'' line, as designated
by Willson (2000). It is defined by the condition $\displaystyle d
\log M/dt \sim d \log L/dt$ and it corresponds to a critical mass loss
rate $\dot M_{\rm crit} \sim 5.67\times 10^{-7} \times M $
in units of $M_{\odot}\,{\rm yr}^{-1}$.
Once reached the ``cliff'' point the star enters the superwind phase.

As anticipated by Bowen \& Willson (1991) and thoroughly discussed by
Willson (2000), at decreasing metallicity the mass-loss rate of an oxygen-rich
AGB star is affected by two factors, namely: i) less efficient dust production
due to the lower abundances of involved elemental species, and ii) more compact
structure, i.e. smaller photospheric radius at given luminosity.
It turns out that the latter effect is the largest, producing a shift of the
cliff line to higher $L$ for lower $Z$ (see figure 9 in Willson 2000).
Following Bowen \& Willson (1991) for $Z < 0.1 Z_{\odot}$  the onset of the 
supewind takes place at a critical luminosity, approximately given by
\begin{equation}
 L_{\rm SW}(Z)  =   L_{\rm SW}(Z_{\odot})+\Delta L_{\rm SW}, 
\end{equation}
where
\begin{equation}
 \Delta L_{\rm SW}  =  0.12-0.13\, \log(Z/Z_{\odot}) 
\end{equation} 

A few comments should be made at this point.  First, we should keep in
mind that the results obtained by Bowen \& Willson (1991) and
illutrated in the key ($\log M$ vs $\log L$) diagram (their figure
2) are not only dependent on the details of their model for dynamical
atmospheres, but they also rely on the adopted relations 
to describe the underlying AGB evolution for
various choices of the stellar mass. This fact introduces some
inconsistency in our calculations 
as we are dealing with different synthetic TP-AGB models
(e.g., different core mass-luminosity relation). Nevertheless the
adopted approach is the best possible with the available theoretical
information. Moreover, it produces results that considerably improve
the comparison between predictions and observations (see
Sect.~\ref{ssec_mlrp}).

\begin{figure}[!tbp]  
\resizebox{\hsize}{!}{\includegraphics{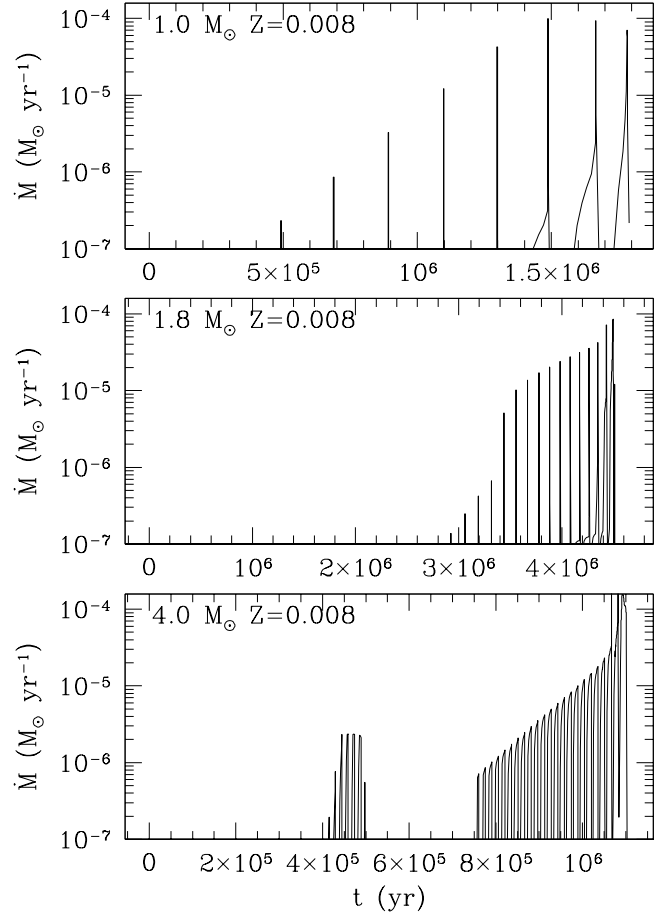}}
\caption{Evolution of the mass-loss rate along the TP-AGB phase
of a few selected models.}
\label{fig_mlr_z008}
\end{figure}
\begin{figure}[!tbp]  
\resizebox{\hsize}{!}{\includegraphics{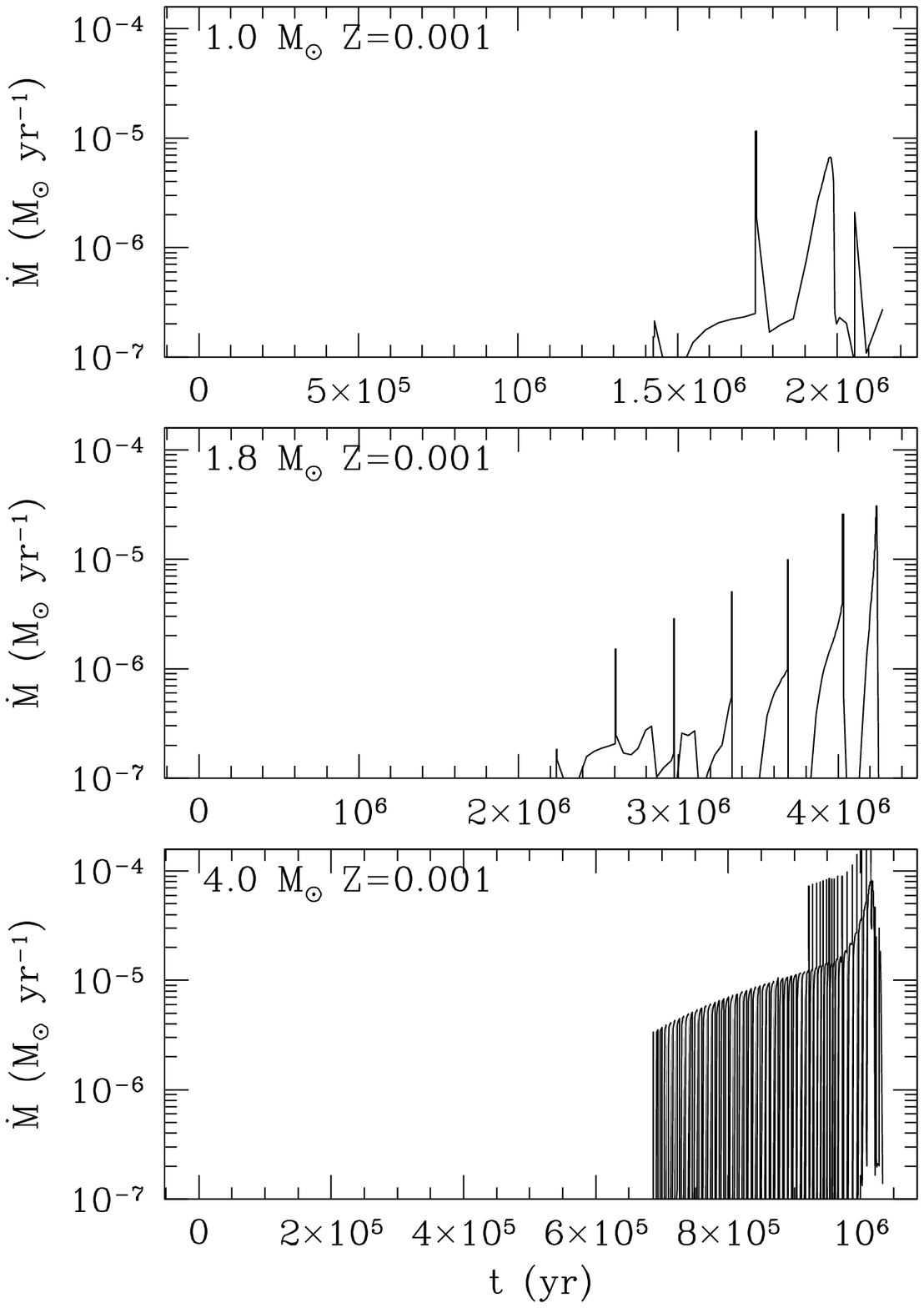}}
\caption{The same as in Fig.~\ref{fig_mlr_z008}, but for metallicity
$Z=0.001$.}
\label{fig_mlr_z001}
\end{figure}

\begin{figure}[!tbp]  
\resizebox{\hsize}{!}{\includegraphics{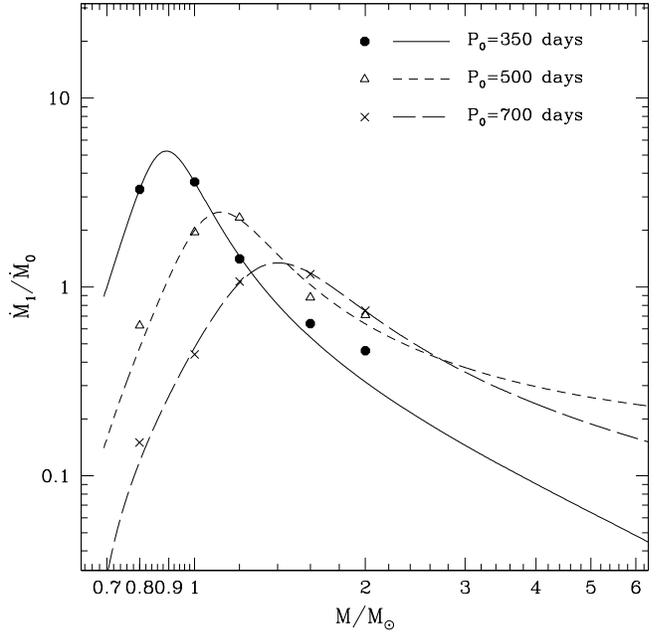}}
\caption{Predicted ratio between 
the mass loss rates $\dot M_{1}$ and $\dot M_{0}$, under the
assumption of FOM and FM pulsation respectively, as a function of the
stellar mass according to the dynamical atmosphere models calculated
by Bowen (1988).  Three groups of models, characterised by the same
$P_0$, are shown together with the corresponding polynomial fits.  }
\label{fig_FOM}
\end{figure}
  
Second, we recall that Bowen \& Willson (1991) calculations are
strictly valid for FM pulsation. There is no extended analysis of the
same kind devoted to pulsation in modes of higher order. A limited set
of dynamical models for FOM pulsation including dust are presented by
Bowen (1988).  By comparing the mass-loss rates calculated by Bowen (1988) 
for FM (table 6) and FOM models (table 8) with the same given
combination 
of stellar mass, radius and
effective temperature\footnote{For assigned fundamental-mode periods 
$P_0 = 350, 500, 700$ days, and stellar masses, 
$M=0.8, 1.0, 1.2, 1.6, 2.0\, M_{\odot}$, the corresponding radii and 
first-overtone mode periods  are
derived from the period-mass-radius relationships by Ostlie \& Cox (1986), 
while it is assumed $T_{\rm eff}=3\:000$ K in all cases.}
we obtain the behaviour shown in Fig.~\ref{fig_FOM}.

We notice that the trend is not monothonic, as $\dot{M}$ for $P=P_1$
(hereafter $\dot M_{1}$) may be lower or larger than $\dot{M}$ for
$P=P_0$ (hereafter $\dot M_{0}$) depending on stellar mass.  Moreover,
the peak in the $\dot{M_1}/\dot{M_0}$ ratio appears to become higher
and to shift towards lower masses at decreasing $P_{0}$.

Polynomial fits of $\dot{M_1}/\dot{M_0}$ for three values
of $P_{0}$ are derived assuming the rational form
\begin{equation}
\frac{\dot{M_1}}{\dot{M_0}} = \frac{b_1+b_2 M+b_3 M^2}{c_1+c_2 M+c_3 M^2}\,\,\,.
\label{eq_fom}
\end{equation} 
The coefficients $a_i$ and $b_i$ are given in Table~\ref{tab_fom}.

\begin{table}[!tbp]  
\caption{Coefficients for Equation~\protect{(\ref{eq_fom})}}
\begin{tabular}{cccc}
\hline
      &        &   $P_0$/days   &      \\
\cline{2-4}
      &  350 & 500 & 700 \\
\hline
$b_1$ &  -0.2946 & -0.2581 & 0.4167 \\
$b_2$ & 0.5488  & 0.1089 & 2.1701 \\ 
$b_3$ & 2.1750E-02 &  0.5608 & -0.4167 \\
$c_1$ & 1.3309  & 3.9290 & 0.8738 \\
$c_2$ & -3.0054  & -7.2055 &  -6.0280 \\
$c_3$ & 1.7392 & 3.4831 & 25.6437 \\
\hline
\end{tabular}
\label{tab_fom}
\end{table}

In practice, in order to estimate the mass-loss rate of a TP-AGB star
pulsating in the FOM mode we proceed as follows.  For given stellar
mass, radius and luminosity we compute the period $P_{0}$ via
eq.~(\ref{eq_p0}) and the corresponding mass-loss rate $\dot M_0$
according to Sect.~\ref{ssec_mlr}.  Then we apply Eq.~(\ref{eq_fom})
to derive $\dot M_1$ (linear interpolation or extrapolation in $\log
P$ are adopted between grid points at given $M$).

Being aware that this recipe based on the few FOM models calculated by
Bowen (1989) does not allow to derive a complete picture of mass loss
under FOM pulsation, nevertheless it represents a first attempt to
address the question.  Moreover, this is likely a more realistic
approach than simply scaling the mass-loss rate for FM pulsators for
$P=P_1\approx P_0/2.2$ (cf. Fig.~\ref{fig_mlro}), which would always
lead to very low mass loss-rates. On the contrary, observations
indicate that low-period AGB variables, like the semiregulars,  
are characterised by mass-loss rates comparable with those of
Miras, up to few $10^{-7}\, M_{\odot}$ yr$^{-1}$ (e.g., Olofsson et al. 2002).
If these stars are assigned a pulsation mode different from the 
FM, likely the FOM, as currently 
suggested (see Lattanzio \& Wood 2003), then this would imply that i) 
either the $\dot M(P)$ relation depends on the pulsation mode, or ii)
$\dot M$ is independent of $P$ and the observed correlation between
$\dot M$ and $P$  just hides the true dependence on other basic
stellar parameters, like mass, luminosity and effective temperature.
It is clear that detailed analyses of dynamical FOM pulsating
atmospheres should be of paramount importance to investigate this
point.

\subsubsection{Carbon-rich models}
In the C/O$\,>1$ case we exploit the results of pulsating dust-driven
wind models for carbon-rich chemistry, developed by the Berlin group
(Fleisher et al. 1992; Sedlmayr \& Winters 1997; Winters et al. 2000;
Wachter et al. 2002; Winters et al. 2003 and references therein).
Extended grids of models are suitably constructed for carbon-rich AGB
stars, yielding the mass-loss rate as a function of main stellar
parameters, such as: effective temperature (2200-3000 K), luminosity
(3000--24000 $L_{\odot}$), mass (0.8--1.8 $M_{\odot}$), photospheric
C/O ratio (1.20--1.80), and pulsation period (100--1000 days).
According to the detailed investigation by Winters et al. (2000, 2003)
one can schematically distinguish two dynamical regimes for the
stellar outflows.

The first regime refers to the so-called B-type models, and it should
apply to low-luminosity, short-period, high-effective temperature red
giants, for which the dominant mass-loss driving mechanism is stellar
pulsation, while radiation pressure on dust grains plays a reduced
role. In fact for this class of models radiation pressure alone is not
able to produce an outflow, such condition being expressed by 
$\alpha<1$\footnote{The standard notation of dynamical wind
models defines
$\alpha \equiv {a^{\rm rad}}/{a^{\rm grav}}$, the ratio of the
radiative acceleration over the gravitational one.}.  During this
regime tenuous winds are generated, characterised by low terminal
velocities ($v_{\infty} < 5$ km s$^{-1}$), and low mass-loss rates,
never exceeding a critical value of $\dot M_{\rm crit} \approx 3\times
10^{-7}\, M_{\odot}$ yr$^{-1}$.

The subsequent development of a stable dust-driven wind from a carbon
star (with $\dot M > \dot M_{\rm crit}$) requires a minimum luminosity
$L_{\rm SW}$, such that the outward acceleration provided by radiation
pressure on dust exceeds the inward gravitational pull ($\alpha \ge
1$).  According to Winters et al. (2000) models this critical
luminosity depends mainly on stellar mass and effective
temperature. As soon as $L \ge L_{\rm SW}$, a carbon star is expected
to enter the dust-driven-wind regime, described by the so-called
A-type models, with mass-loss rates in the typical range $3\times
10^{-7} \la \dot M \la 10^{-4}\, M_{\odot}$ yr$^{-1}$, and terminal
velocities in the range $5 \la v_{\infty} \la 40$ km s$^{-1}$.
 
The picture just depicted leads us to adopt the following recipe to
calculate the mass-loss rate for carbon-rich models evolving on the
TP-AGB.  First we need to specify the critical luminosity $L_{\rm SW}$
that separates the two wind regimes.  Basing on wind models calculated
by Schr\"oder et al. (1999) -- with transition properties between the
B and A classes -- Schr\"oder et al. (2003) propose an approximate
relation, giving $L_{\rm SW}$ as a function of stellar mass $M$ and
effective temperature $T_{\rm eff}$:
\begin{equation}
\label{eq_lsw0}
\log(L_{\rm SW}/L_{\odot})= 3.8 + 4  (\log T_{\rm eff}-3.45) + 
\log(M/M_{\odot})
\end{equation} 
The above relation is derived from a small subset of wind models for
various choices of stellar mass and effective temperature, while
adopting the same period, $P=400$ days, and carbon-to oxygen ratio,
C/O$=1.3$.

In the present work an attempt is made to refine this relation by
including also a dependence on the pulsation period $P$. To this aim
we use a sample of pulsating wind models computed by Winters et
al. (2000; their tables 6--9) at varying input parameters, namely:
stellar mass $M$, effective temperature $T_{\rm eff}$, pulsation
periods $P$ (in days), carbon-to-oxygen ratio C/O, and piston
velocity\footnote{In the piston-approximation, commonly used in
pulsating wind models, the piston amplitude $\Delta v_{\rm p}$ denotes
the velocity amplitude of the innermost grid point, assumed to vary
sinusoidally over the pulsation period $P$.} $\Delta v_{\rm p}$.

From these data we derive the stellar luminosity at which the
time-averaged (over a few pulsation periods) $\alpha \approx 1$, as a
function of $M$, $T_{\rm eff}$, and $P$ (in days), while we consider
fixed both the carbon-to-oxygen ratio (C/O$=1.2$), and the piston
velocity ($\Delta v_{\rm p}=5$ km s$^{-1}$).  The resulting fitting
relation reads
\begin{eqnarray}
\label{eq_lsw}
\log(L_{\rm SW}/L_{\odot}) & = & 3.796+6.614 \log(T_{\rm eff}/2800) \\
\nonumber 
      & &\!\!\!\!\!\!\! +1.096\log(M/M_{\odot})-1.728\log(P/400)
\,\,\,\,\,. 
\end{eqnarray}

\begin{figure}  
\resizebox{\hsize}{!}{\includegraphics{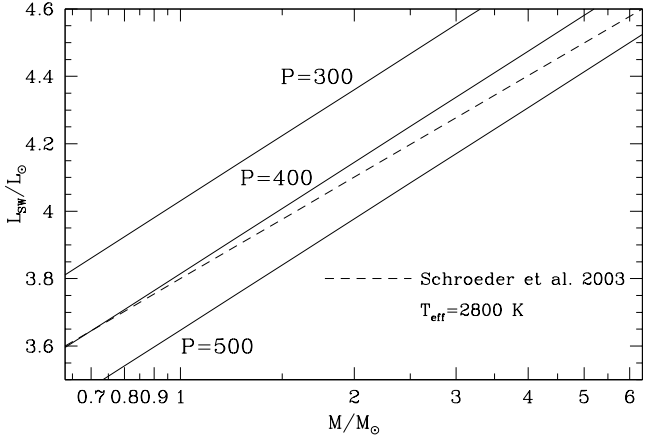}}
\caption{Critical luminosity  for the onset of the super-wind
in C-rich models as a function of the initial
stellar mass, according to Eq.~(\ref{eq_lsw0}) (dashed line) for
$P=400$ days, and  Eq.~(\ref{eq_lsw}) (solid line) for three choices
of $P$, as indicated. In all cases the effective temperature is set
equal to $2\,800$ K.}
\label{fig_lsw}
\end{figure}

For the sake of comparison Fig.~\ref{fig_lsw} displays $L_{\rm SW}$ as
a function of stellar mass, as predicted for $T_{\rm eff} = 2800$ K by
our proposed fit (Eq.~\ref{eq_lsw}) for three choices of the pulsation
period, and by Eq.~(\ref{eq_lsw0}) without period-dependence suggested
by Schr\"oder et al. (2003).  We note that our proposed formula
recovers the Schr\"oder et al. (2003) relation well for $P=400$ days,
while predicting a significant sensitiveness of $L_{\rm SW}$ to period
changes, i.e. the onset of the dust-superwind should occur at lower
luminosities for longer periods.

As long as $L < L_{\rm SW}$ the mass-loss rate is calculated  
as a function of the near infrared $J-K$ color 
\begin{equation}
\label{eq_mlrb}
\log[\dot M/(M_{\odot}\, {\rm yr}^{-1})] = \frac{-6.0}{J-K +0.1} -4.0 
\end{equation}
according to the semi-empirical calibration by Schr\"oder et
al. (2003).  The above equation is the best match to the mass-loss
rates of Galactic disk C stars belonging to the Le Bertre (1997) sample
with moderate mass loss (with colours in the range $2.5 \la J-K \la
4$), that is before the full development of the dust-driven superwind.

In turn, the $J-K$ colour for these models is computed with the aid of
the relation 
\begin{equation}
\label{eq_jkb}
(J-K) = 17.32-4.56\log T_{\rm eff}+0.052\,{\rm C/O}
\end{equation}
proposed by Marigo et al. (2003). The above equation extends the
original colour-$T_{\rm eff}$ transformation for C stars by Bergeat et
al. (2001), by introducing a C/O-dependence to account for the effect
molecular blanketing on the spectra of these stars.

Then, once the luminosity of a carbon-rich model overcomes $L_{\rm SW}$ the
dust-driven superwind is assumed to start. This phase is described 
on the base of pulsating wind models presented by Wachter et al. (2002), 
in which the driving mechanism of mass loss is the radiation pressure on 
dust grains. From those models we derive a fitting relation, expressing 
the mass-loss rate as a function of stellar luminosity $L$, effective 
temperature $T_{\rm eff}$, mass $M$,  pulsation period $P$, and photospheric 
C/O ratio (the piston amplitude is assumed $\Delta v_{\rm p}=5$ km s$^{-1}$
for all models)
\begin{eqnarray}
\label{eq_mlrcsw}
\log[\dot{M}/(M_{\odot}\,{\rm yr}^{-1})] & = &-4.529\\
\nonumber 
& & \!\!\!\!\!\!\!\!\!\!\!\!\!\!\!\!\!\!\!\!\!\!\!\!\!\!\!\!\!\!\!\!-6.849\,\log(T_{\rm eff}/2600\,{\rm K})+1.527\,\log(L/10^4\, L_{\odot})\\
\nonumber
& & \!\!\!\!\!\!\!\!\!\!\!\!\!\!\!\!\!\!\!\!\!\!\!\!\!\!\!\!\!\!\!\!-1.997\,\log(M/M_{\odot})-0.995\,\log(P/650\,{\rm days})\\
\nonumber
& & \!\!\!\!\!\!\!\!\!\!\!\!\!\!\!\!\!\!\!\!\!\!\!\!\!\!\!\!\!\!\!\!+ 0.672 
\log\displaystyle\left(\frac{\rm C/O}{1.5}\right) 
\,\,\,\,.
\end{eqnarray}

The numerical coefficients  are quite similar to those of the fitting formula
for $\dot M$ already proposed by Wachter et al. (2002; see their equation 1)
obtained from the same set of wind models. 
We note that Eq.~(\ref{eq_mlrcsw}) contains an explicit dependence of $\dot M$
on the C/O ratio that is neglected in the Wachter et al. formula.
Our choice is motivated by the fact that, though the influence of C/O is 
expected to be rather small for relatively low ratios ($<2$ as is the case of
Galactic C stars), it may instead become important for larger C/O ratios, 
e.g., in carbon stars belonging to metal-poor populations.

A few examples of the predicted mass-loss rates are displayed
in Figs.~\ref{fig_mlr_z008} and \ref{fig_mlr_z001}. In all cases, regardless
of the C/O value, we note that i) the mass-loss rate is modulated
by the variations
of luminosity and effective temperature driven by the He-shell flashes, ii)
most of the stellar mass is lost during the quiescent stages of high
luminosity preceeding the occurrence of the thermal pulses, and iii)
the super-wind regime is attained during the FM pulsation ($P=P_0$).

\section{Basic model predictions}
\label{sect_modpre}
The synthetic model just described in Sect.~\ref{sec_syntagb} has been
applied to calculate the TP-AGB evolution of an extended set of stellar models,
with initial masses in the interval $0.5-5.0\,M_{\odot}$, and  for 7 
choices of metallicity, 
$Z=0.0001,\;0.0004,\;0.001,\;0.004,\;0.008,\;0.019,\;0.03$.
The TP-AGB phase is followed from the first thermal pulse
up to the complete ejection of the envelope by stellar winds. 

In the following we describe the basic properties of our TP-AGB
models, emphasizing the new aspects that derive from a number of
updates applied to our synthetic code (described in
Sect.~\ref{sec_syntagb}), the most important of which are: (i) an
improved description of the pulsation periods, so as to account for
the switching from the FOM to the FM of pulsation; (ii) a better
recipe to predict the mass-loss rates, distinguishing between O-rich
and C-rich models; (iii) the use of variable molecular opacities; (iv)
a more detailed treatment of the third dredge-up and HBB.

\subsection{Transition luminosities}
\label{ssect_ltrans}
\begin{figure*} [!tbp]   
\begin{minipage}{0.33\textwidth}
	\resizebox{\hsize}{!}{\includegraphics{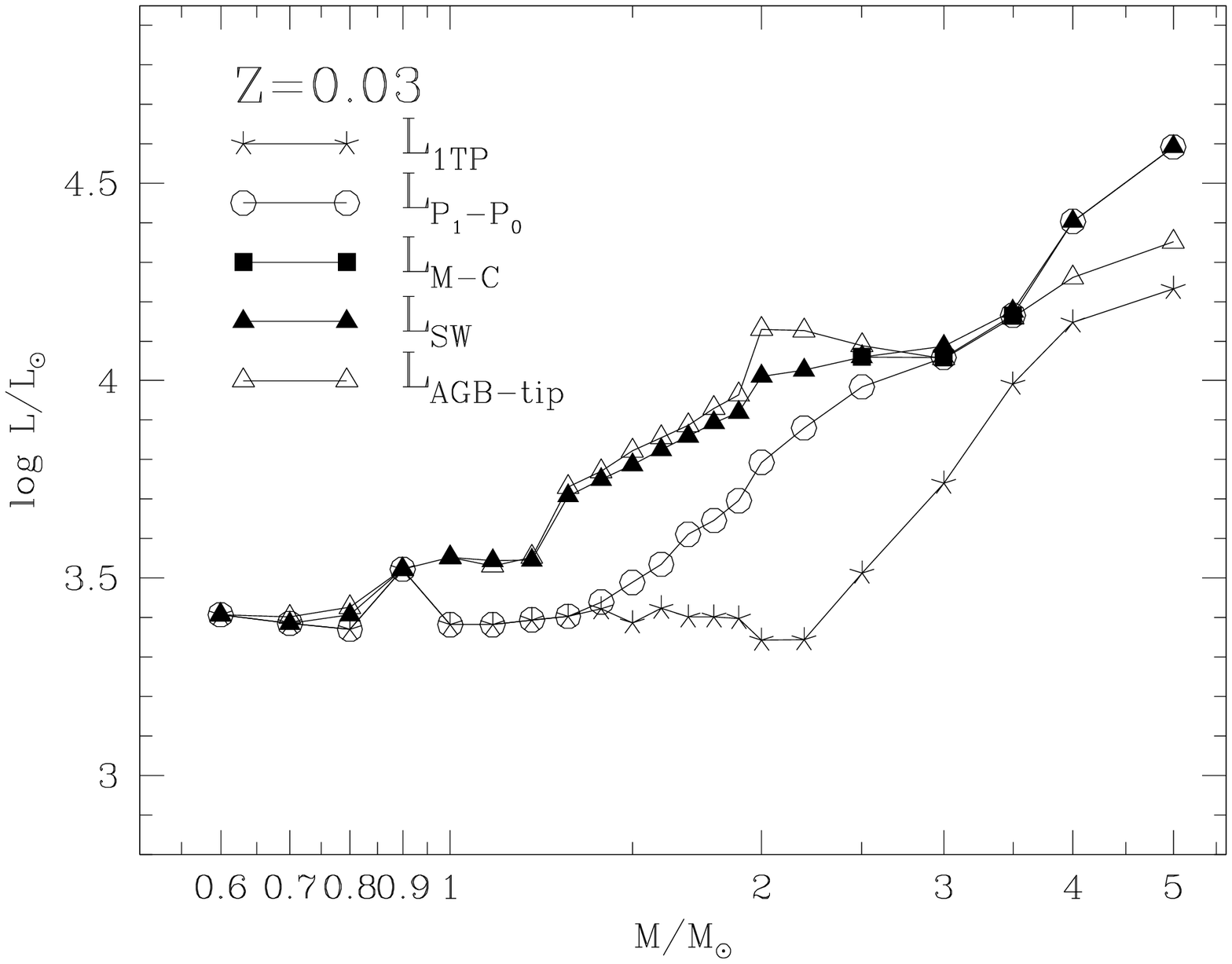}}
\end{minipage} 
\hfill
\begin{minipage}{0.33\textwidth}
	\resizebox{\hsize}{!}{\includegraphics{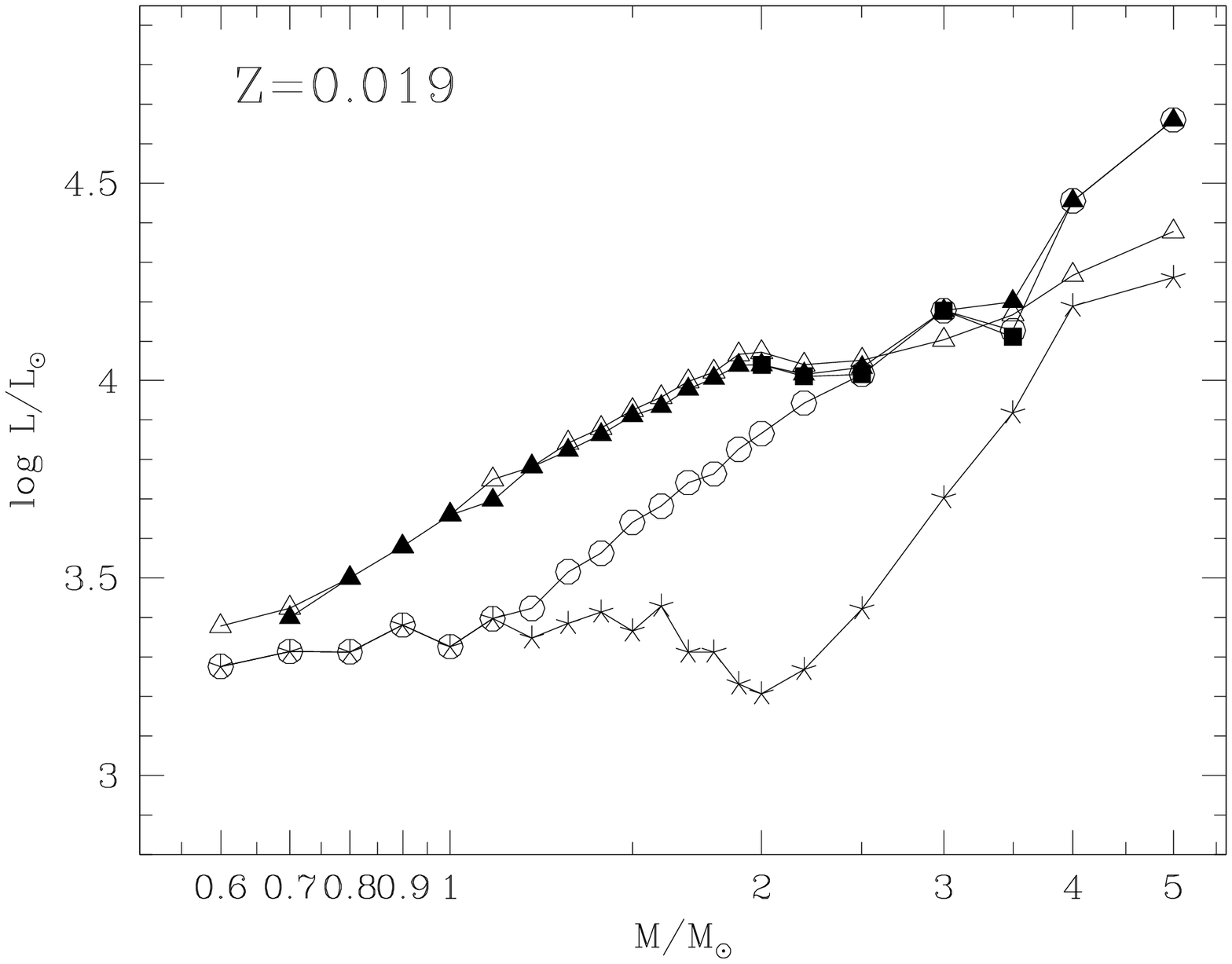}}
\end{minipage} 
\hfill
\begin{minipage}{0.33\textwidth}
	\resizebox{\hsize}{!}{\includegraphics{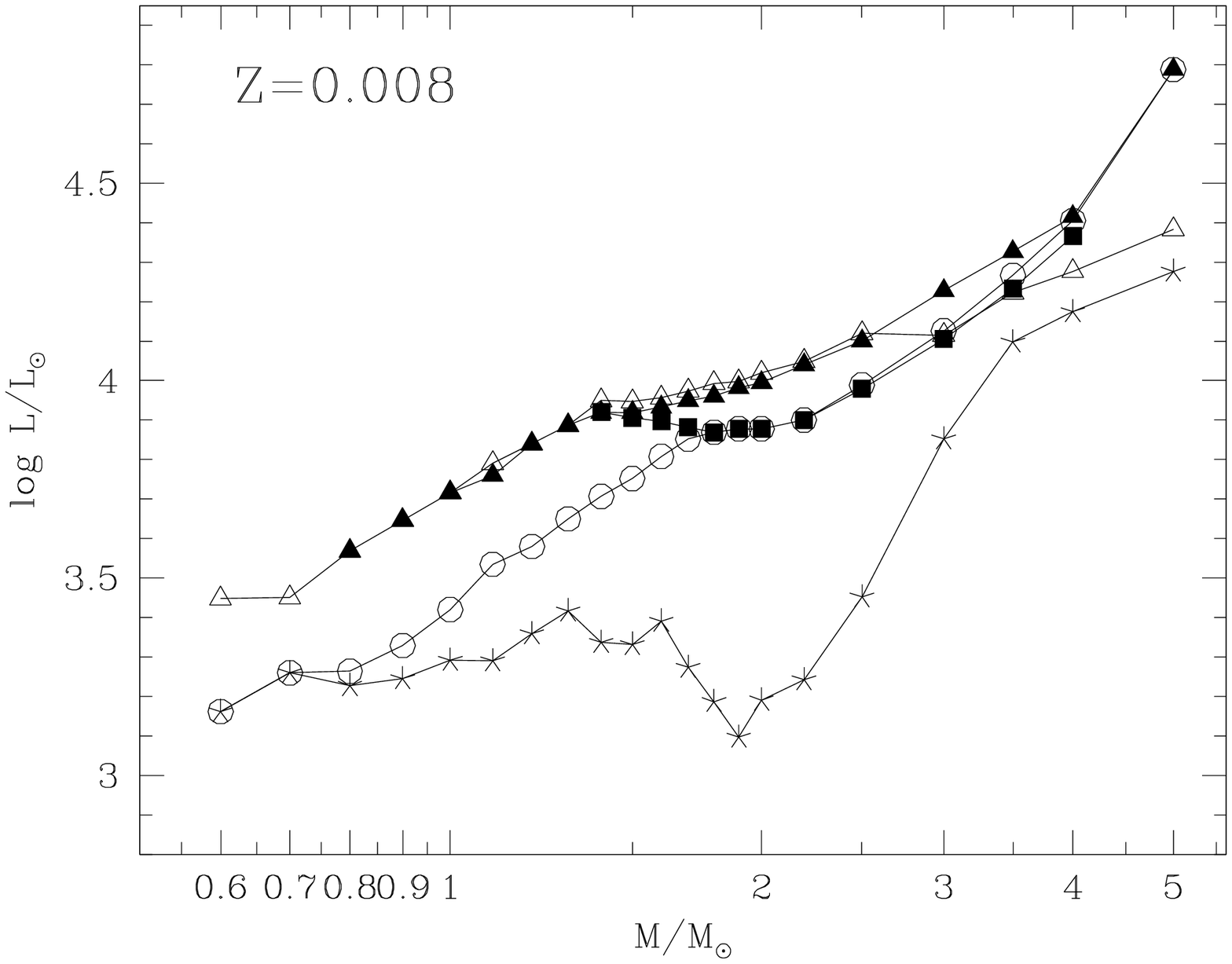}}
\end{minipage} 
\\\
\begin{minipage}{0.33\textwidth}
	\resizebox{\hsize}{!}{\includegraphics{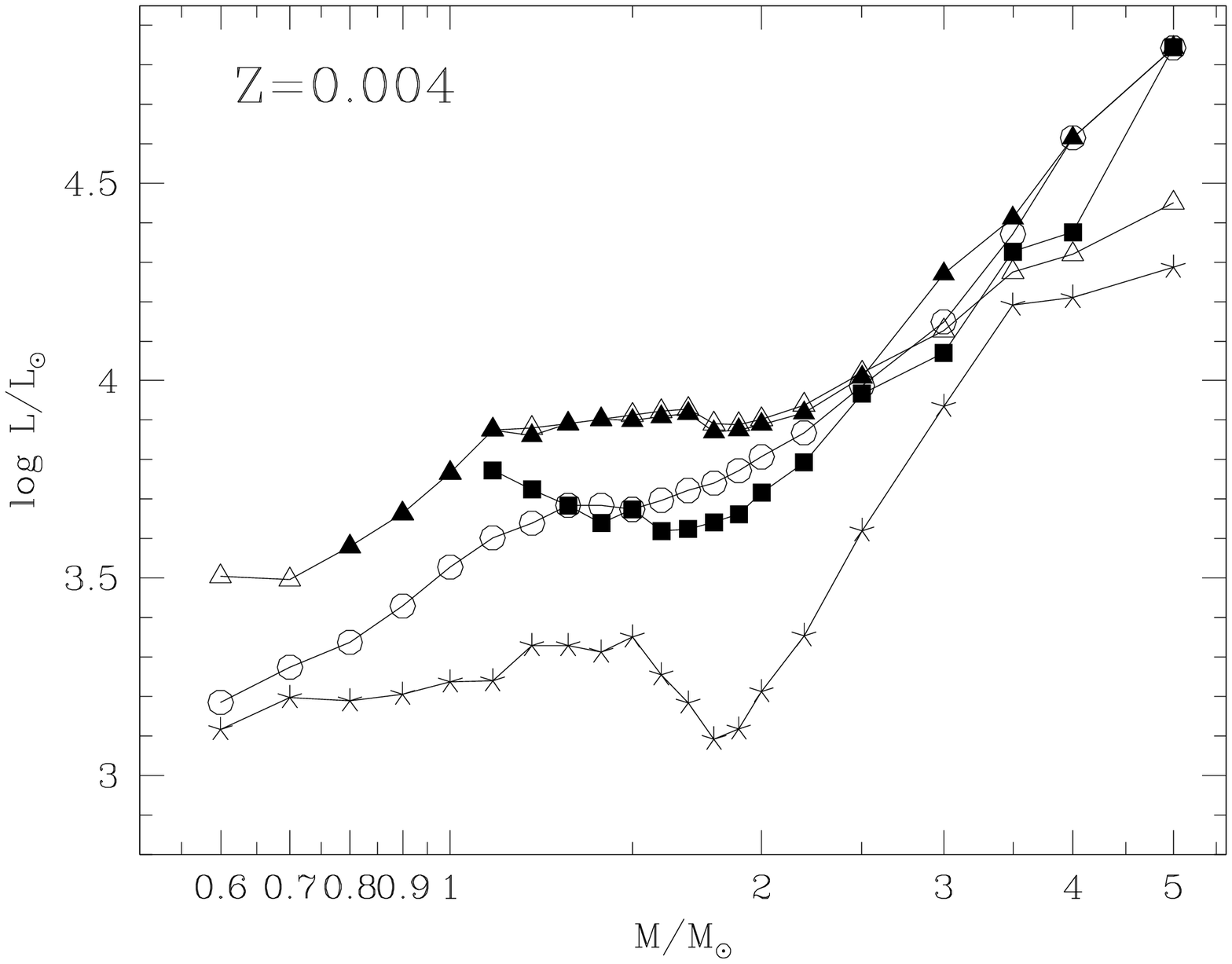}}
\end{minipage} 
\hfill
\begin{minipage}{0.33\textwidth}
	\resizebox{\hsize}{!}{\includegraphics{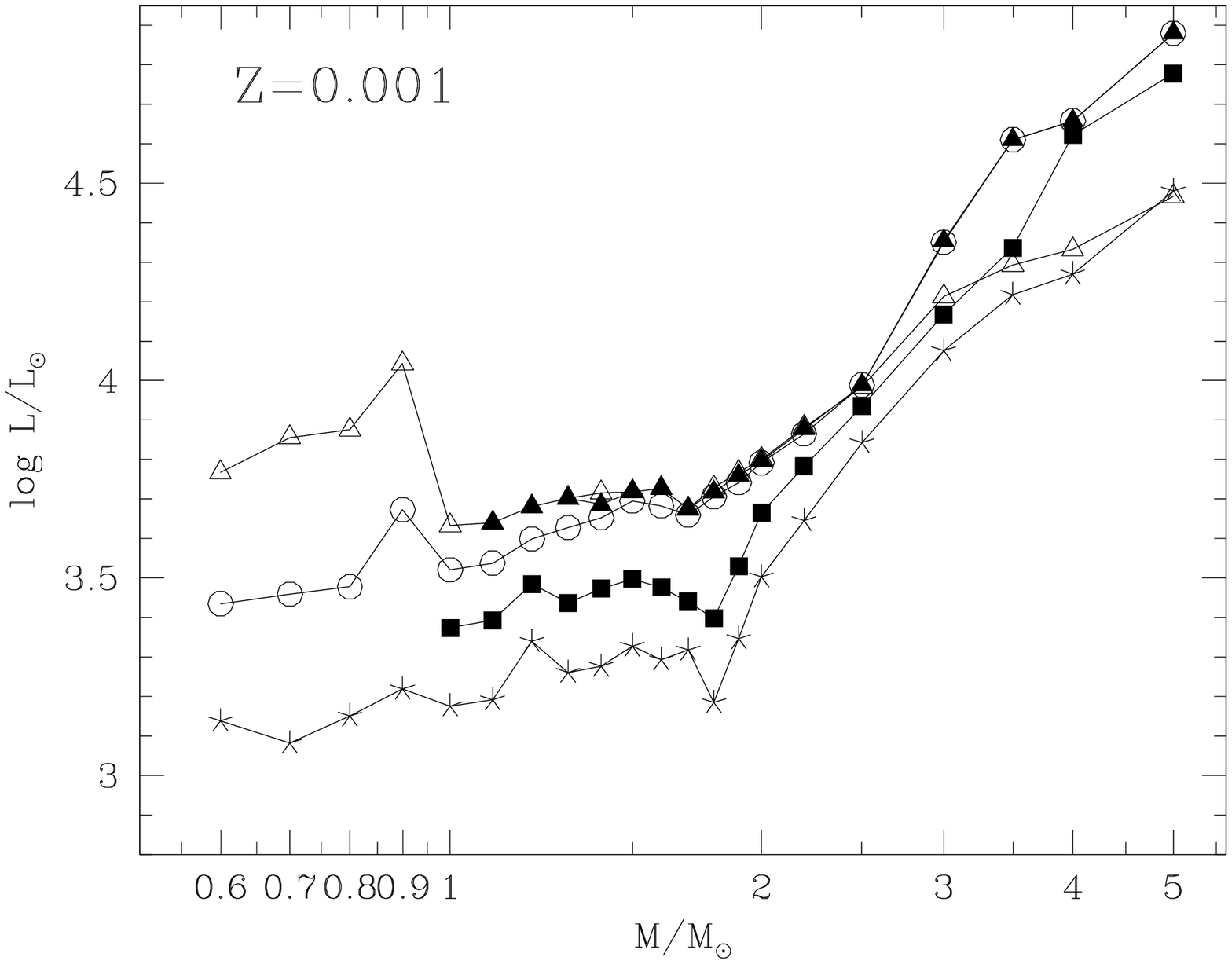}}
\end{minipage} 
\hfill
\begin{minipage}{0.33\textwidth}
	\resizebox{\hsize}{!}{\includegraphics{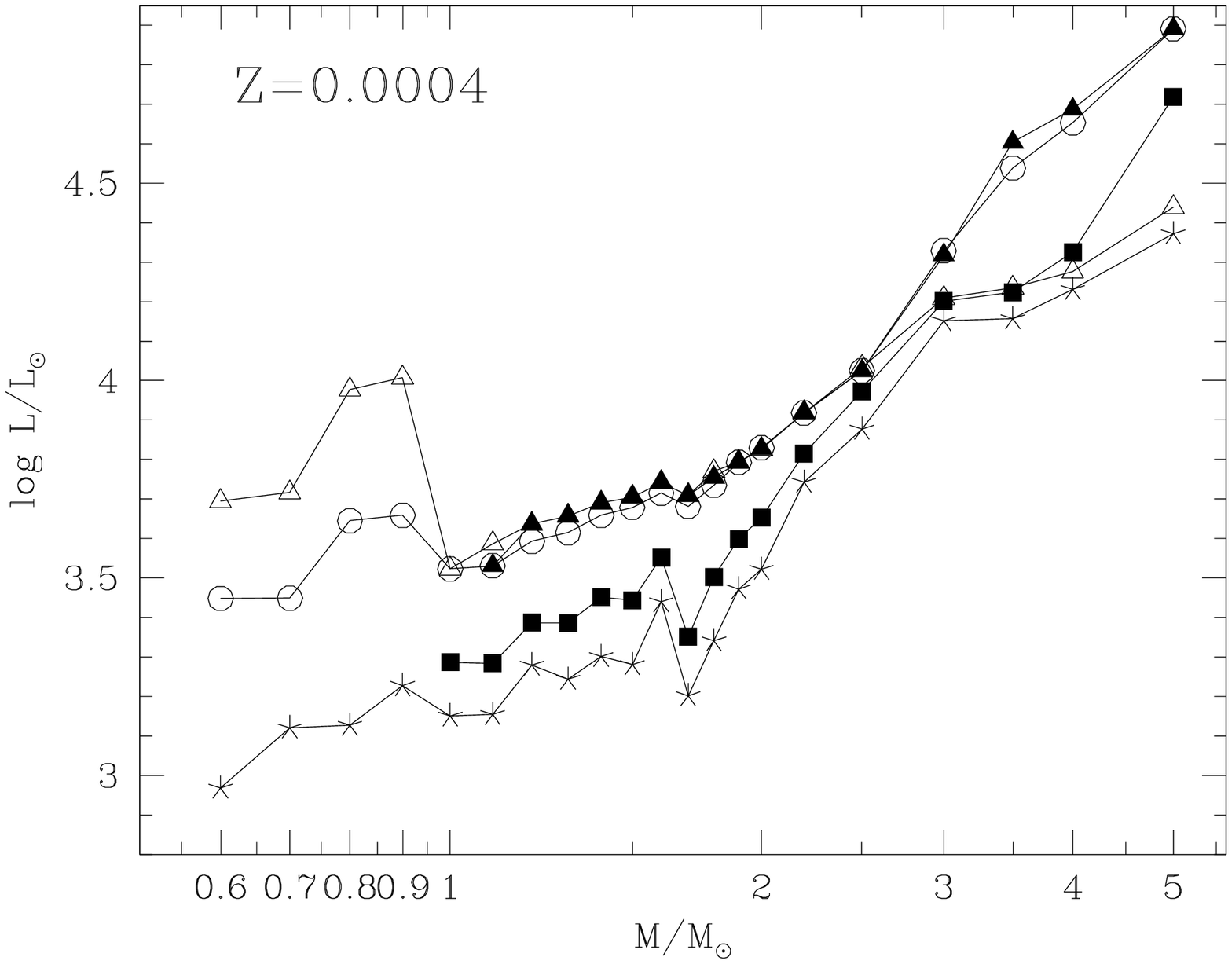}}
\end{minipage}
\\ 
\begin{minipage}{0.33\textwidth}
	\resizebox{\hsize}{!}{\includegraphics{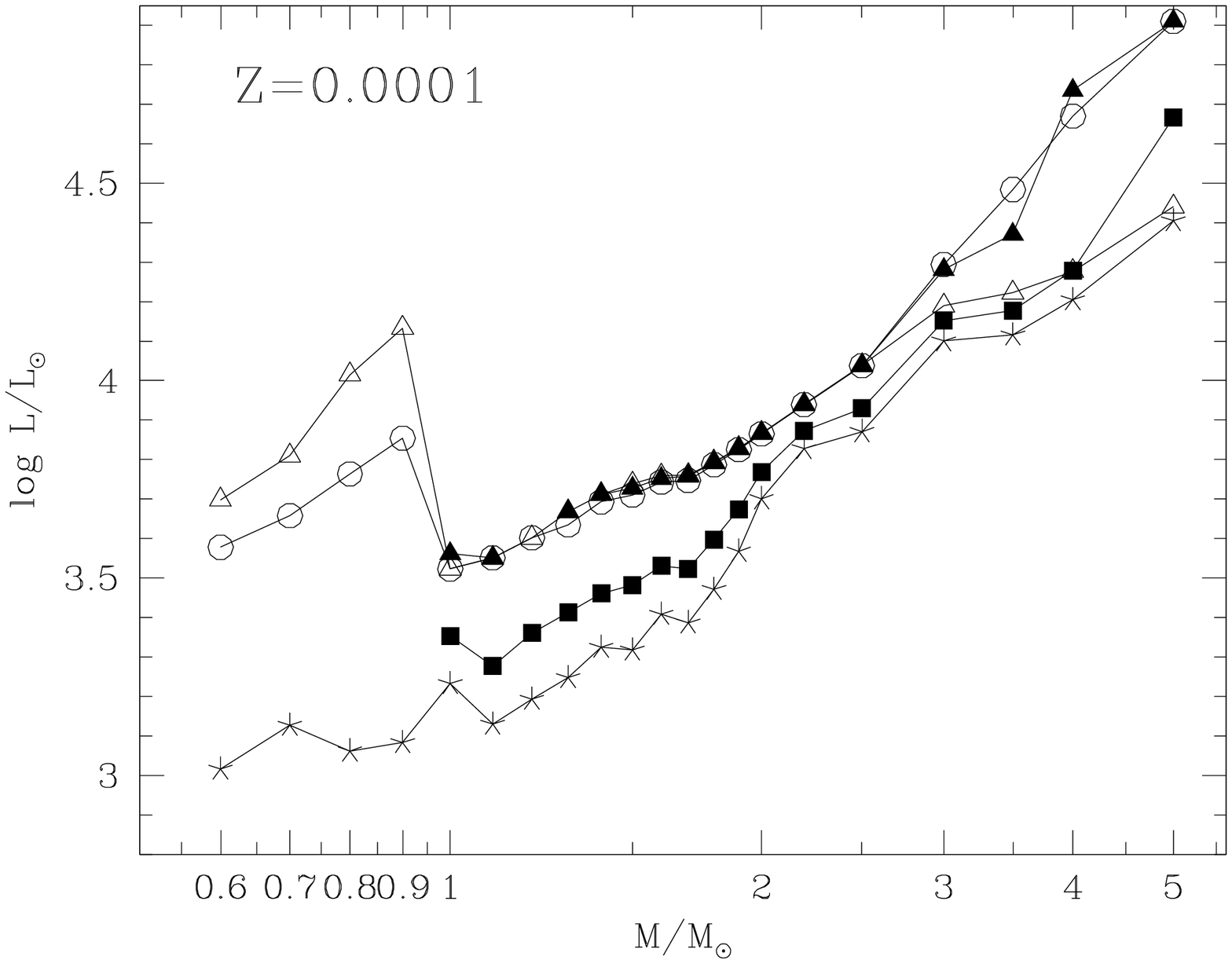}}
\end{minipage} 
\hfill
\begin{minipage}{0.66\textwidth}
\caption{Transition luminosities as a function of initial stellar mass and 
metallicity, namely: $L_{\rm TP,1}$ at the first thermal pulse;
$L_{P_1-P_0}$ at the switch from FOM to FM pulsation; $L_{\rm M-C}$ at
the transition from O-rich to C-rich surface chemical composition;
$L_{\rm SW}$ at the onset of the dust-driven superwind regime; $L_{\rm
max}$ at the tip of the AGB. For the sake of simplicity, the plot 
is constructed considering the maximum quiescent luminosities reached 
just before the occurrence of thermal pulses, even if the transitions
may actually occur at some other phase of the corresponding pulse cycles}
\label{fig_ltrans}
\end{minipage} 
\end{figure*}

The classical picture of TP-AGB evolution is one of luminosity
generally increasing with time as the core mass grows along the
evolution.  This picture has already been demonstrated to fail under a
series of circumstances (e.g., flash-driven luminosity variations,
very deep dredge-up, extinction of HBB, etc.), however it is still
useful to define a few transition luminosities that mark the onset of
particular regimes during the TP-AGB phase. The several panels in
Fig.~\ref{fig_ltrans} show a few characteristic luminosities as a
function of the stellar mass, for the 7 metallicities explored in this
work. These critical luminosities correspond to the quiescent pre-flash
values at the stages of:
\begin{itemize}
\item The first termal pulse, $L_{1 TP}$;
\item The switch from fundamental mode to first-overtone mode pulsation,  
$L_{P_1}-L_{P_0}$;
\item The transition from oxygen-rich to carbon-rich surface chemical
composition,  $L_{\rm M-C}$;
\item The onset of the superwind phase of mass loss, $L_{\rm SW}$, 
with $\dot M > 5\,10^{-7}\, M_{\odot}$ yr$^{-1}$;
\item The luminosity at the end of the AGB phase, $L_{\rm AGB-tip}$.  
\end{itemize}

From comparing the various panels in Fig.~\ref{fig_ltrans} we derive
the following indications:
\begin{itemize}
\item The ``chemical'' transition from M- to C-type takes place at lower
luminosities with decreasing $Z$. Moreover, the minimum mass to become
a carbon star decreases with metallicity, being $\sim 2.0\, M_{\odot}$
for $Z=0.019$ and $\sim 0.8\, M_{\odot}$ for $Z \le 0.001$.
\item The transition from the FOM to the FM pulsation occurs 
at higher luminosities with decreasing $Z$. In particular low-mass
stars (say with $M_{\rm i} \la 1-1.5\, M_{\odot}$) of low $Z$ enter
the TP-AGB phase as FOM pulsators, while they should already appear as
FM pulsators at solar or super-solar metallicity.  At larger masses
the transition to FM pulsation always takes place during the TP-AGB
phase, regardless of $Z$.
\item Combining the above trends for $L_{P_1-P_0}(M_{\rm i},Z)$ and 
$L_{\rm{M}-\rm{C}}(M_{\rm i},Z)$ it follows that at lower
metallicities carbon stars of any mass should first populate the FOM
sequence, and then jump to the FM sequence in the $P-L$ diagram.  At
increasing metallicities the $L_{\rm{M}-\rm{C}}$ transition tends to
occur at higher luminosities at given stellar mass, so that carbon
stars should predominantly appear as FM pulsators.  
\item The $L_{\rm SW}$ for both oxygen-rich and carbon-rich models
is always attained after the transition from FOM to FM pulsation, and
it practically determines the final luminosity, $L_{\rm tip-AGB}$
attained at the termination of the AGB phase.
\item For carbon-rich models  $L_{\rm SW}$ is very close
to the $L_{\rm{M}-\rm{C}}$ transition at higher metallicities, e.g., as
a star with initial $Z \ge 0.019$ becomes a carbon star the
significant decrease in $T_{\rm eff}$ favours the quick development of
the dusty superwind regime.  At decreasing metallicities the surface
cooling effect caused by the transition from M- to C-type domain
is less pronounced, so that the onset of the superwind is delayed.
\item   For oxygen-rich models $L_{\rm SW}$ takes place, on average, 
at larger luminosities with decreasing $Z$ due to the higher $T_{\rm
eff}$.  We note that at very low metallicities, say $Z\le 0.001$,
low-mass models with $M_{\rm i}\le 0.8\, M_{\odot}$ do not reach the
minimum mass-loss rates ($\approx 5\,10^{-7}\, M_{\odot}$ yr$^{-1}$)
required for the development of the dusty superwind. In these cases
the reduction of the envelope mass and the consequent termination of
the AGB phase are mostly determined by the core mass growth, i.e. the
rate of outward displacement of the H-burning shell.
\end{itemize} 
  
\begin{figure*}[!tbp]  
\begin{minipage}{0.33\textwidth}
	\resizebox{\hsize}{!}{\includegraphics{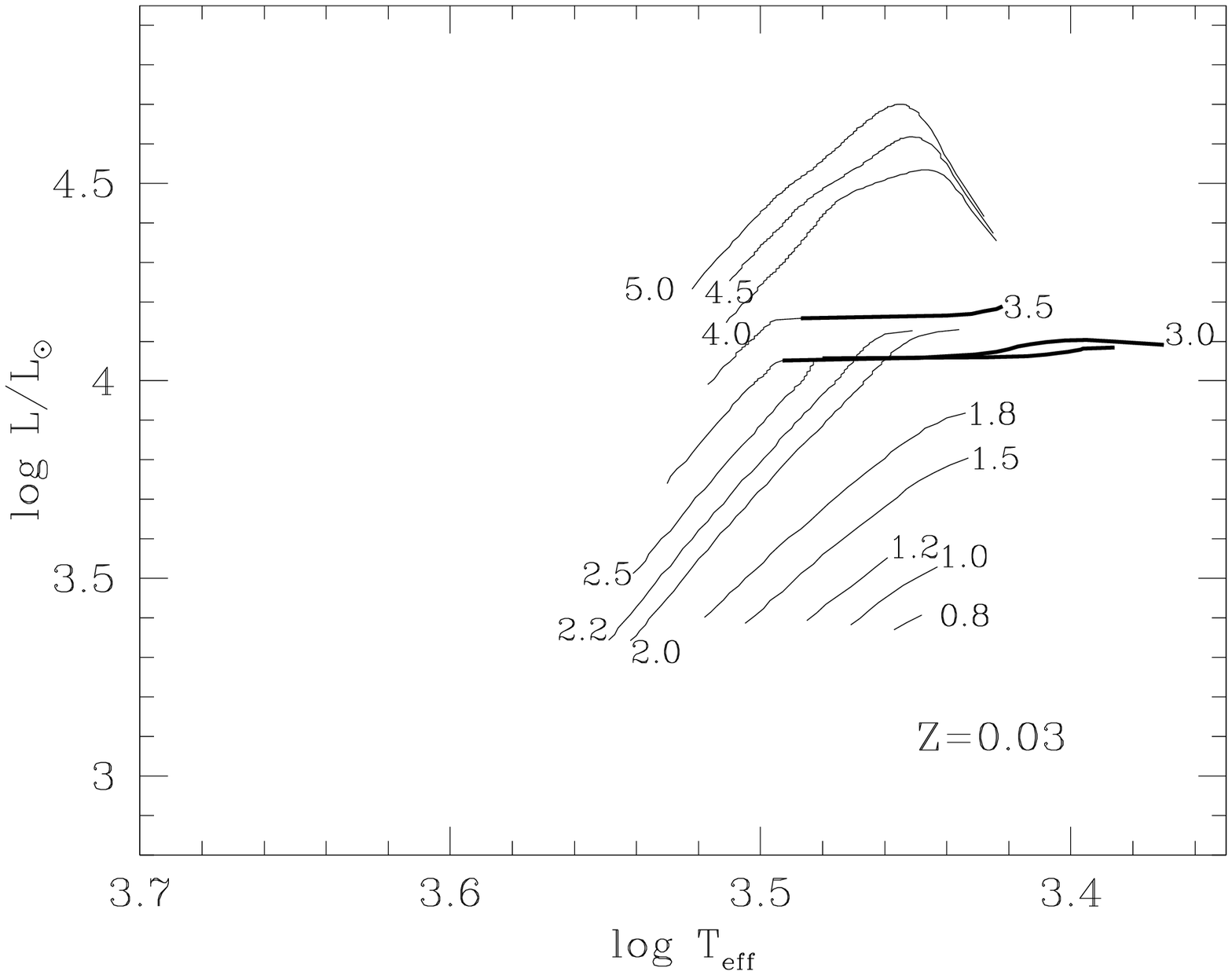}}
\end{minipage} 
\hfill
\begin{minipage}{0.33\textwidth}
	\resizebox{\hsize}{!}{\includegraphics{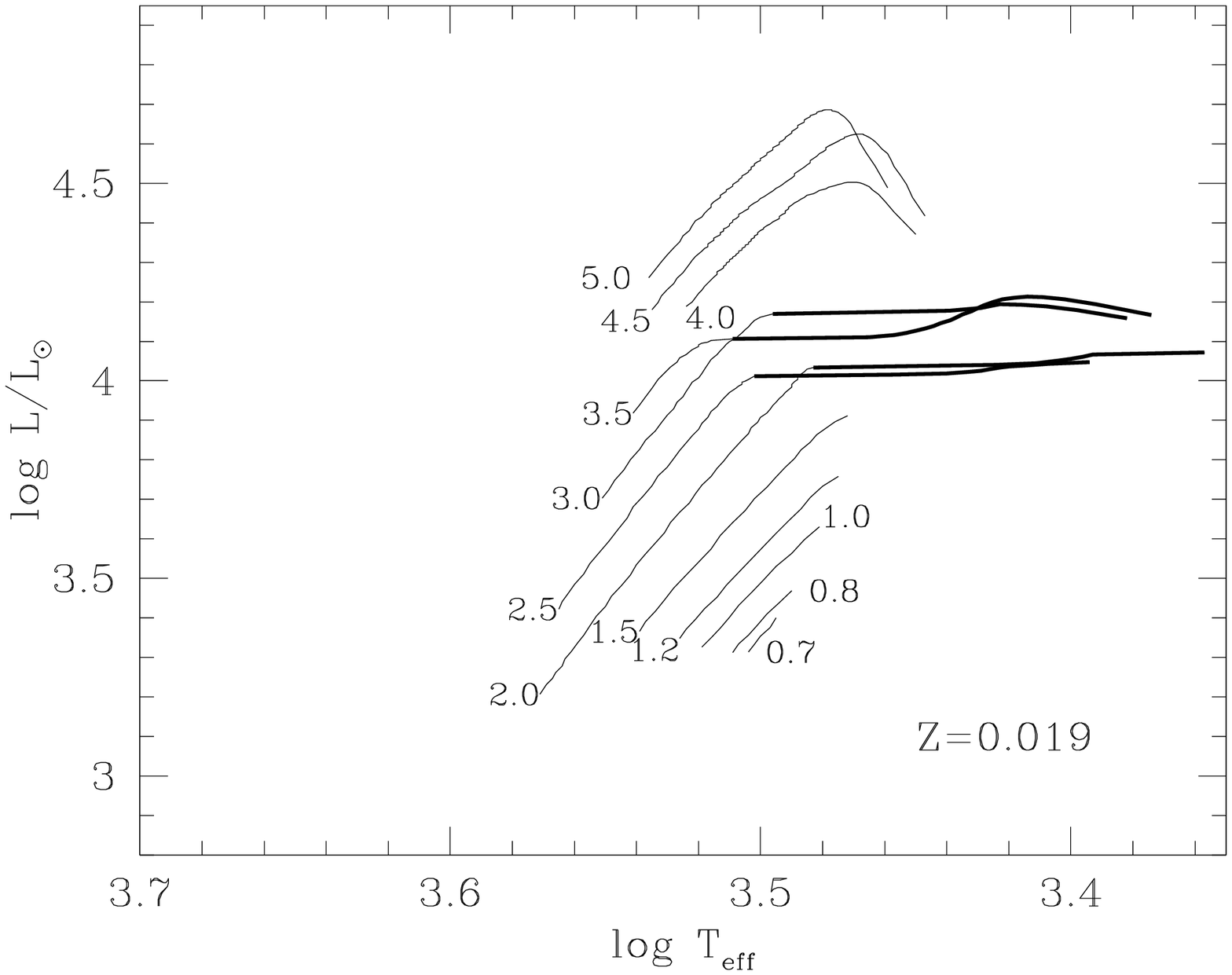}}
\end{minipage} 
\hfill
\begin{minipage}{0.33\textwidth}
	\resizebox{\hsize}{!}{\includegraphics{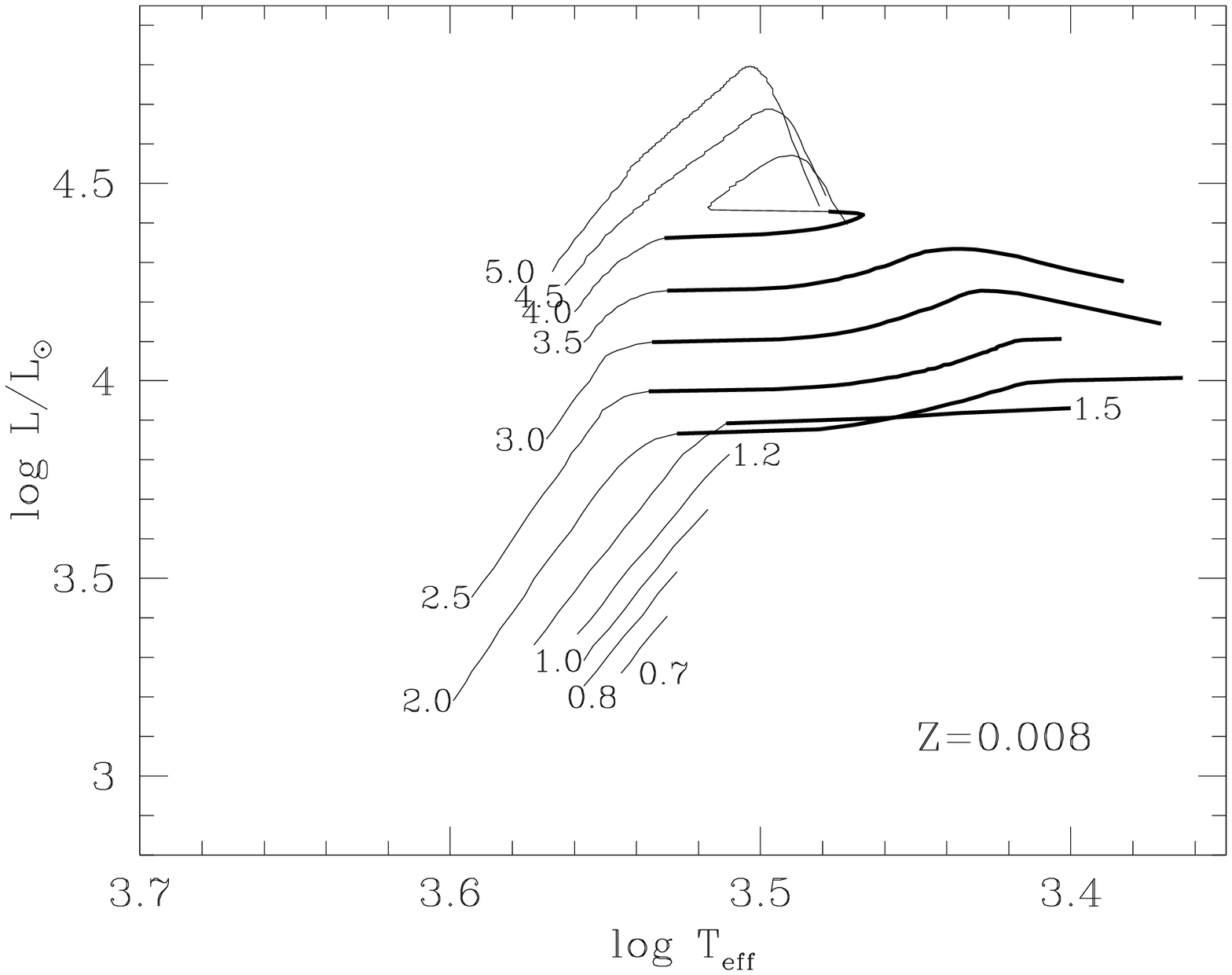}}
\end{minipage} 
\\\
\begin{minipage}{0.33\textwidth}
	\resizebox{\hsize}{!}{\includegraphics{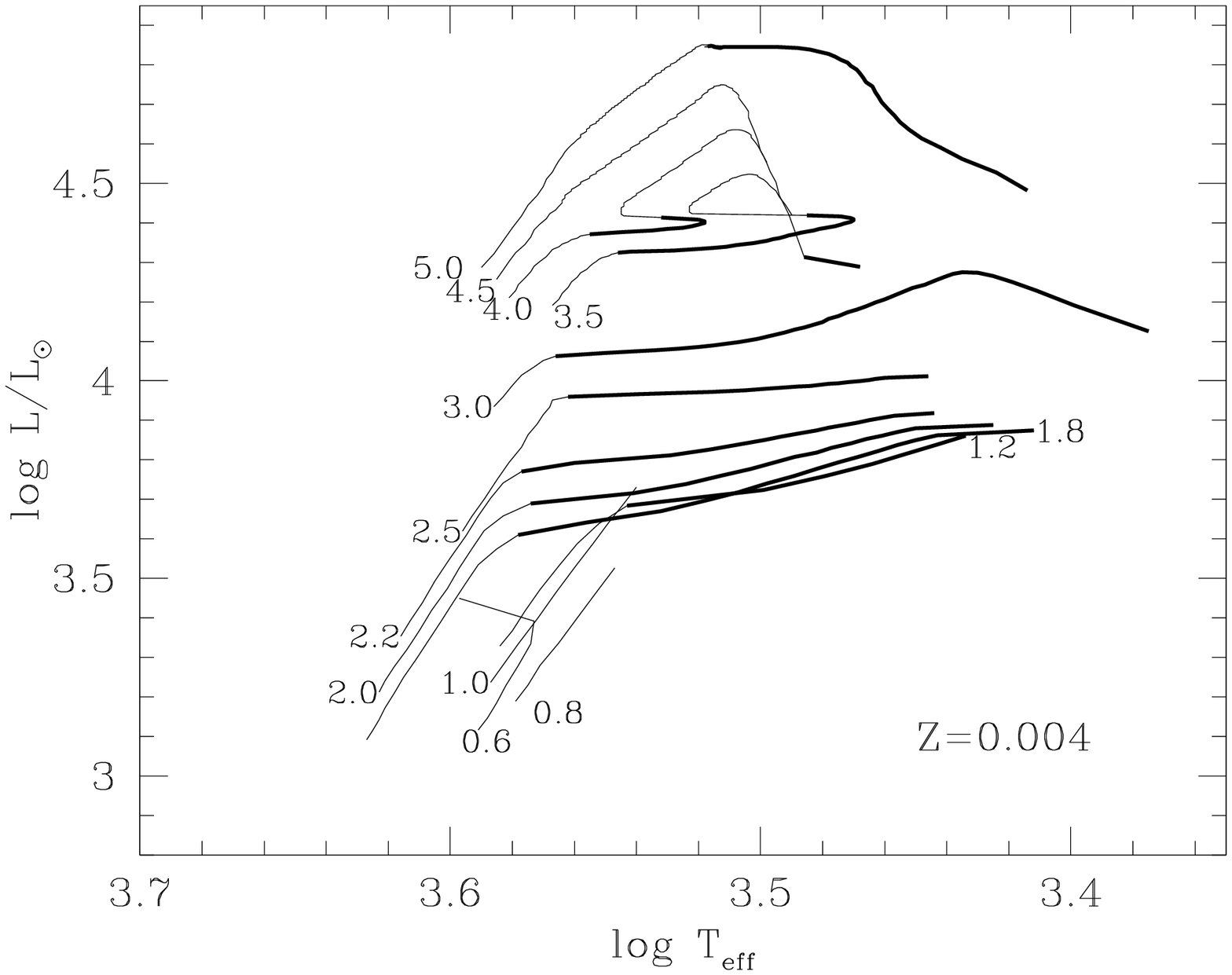}}
\end{minipage} 
\hfill
\begin{minipage}{0.33\textwidth}
	\resizebox{\hsize}{!}{\includegraphics{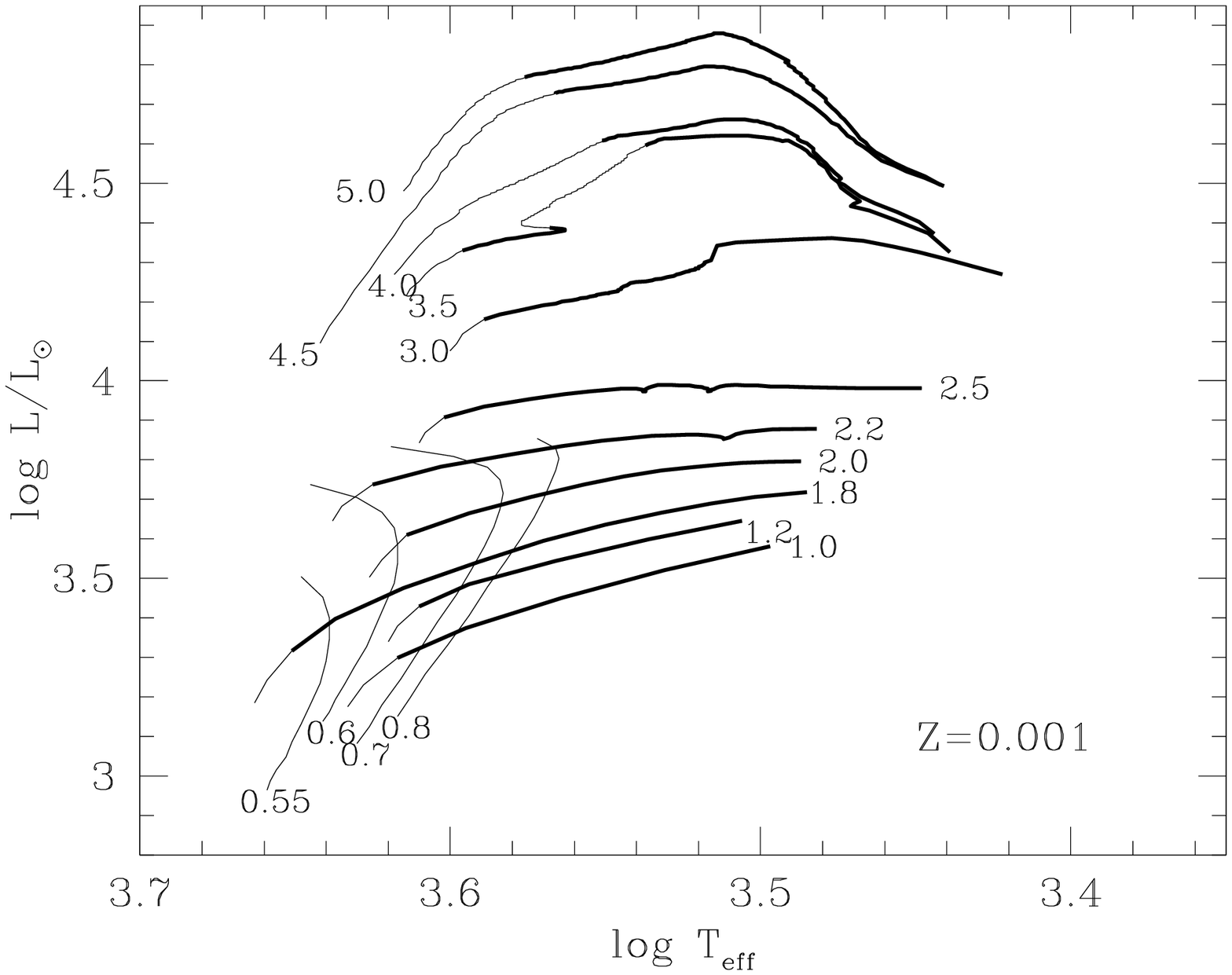}}
\end{minipage} 
\hfill
\begin{minipage}{0.33\textwidth}
	\resizebox{\hsize}{!}{\includegraphics{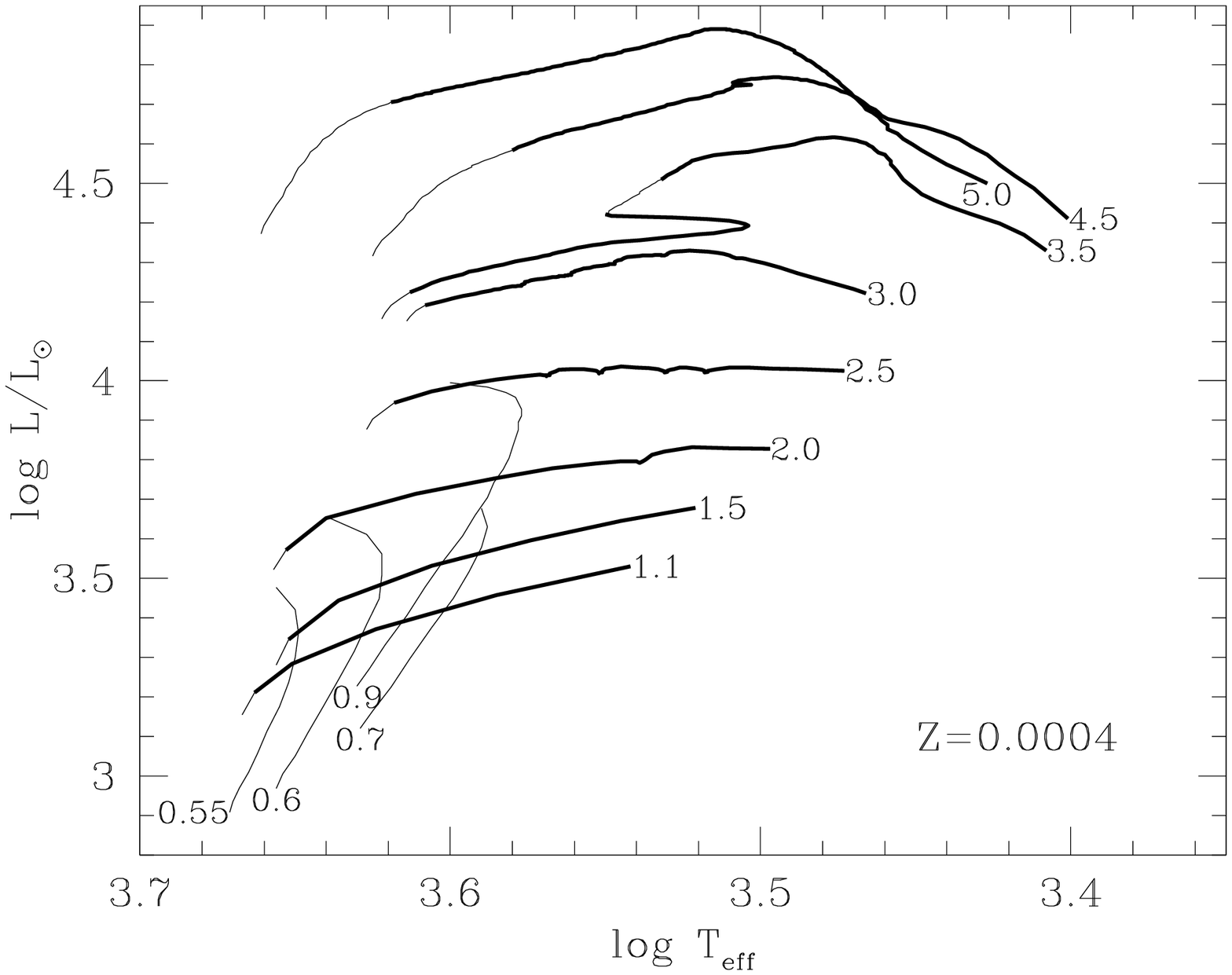}}
\end{minipage}
\\ 
\begin{minipage}{0.33\textwidth}
	\resizebox{\hsize}{!}{\includegraphics{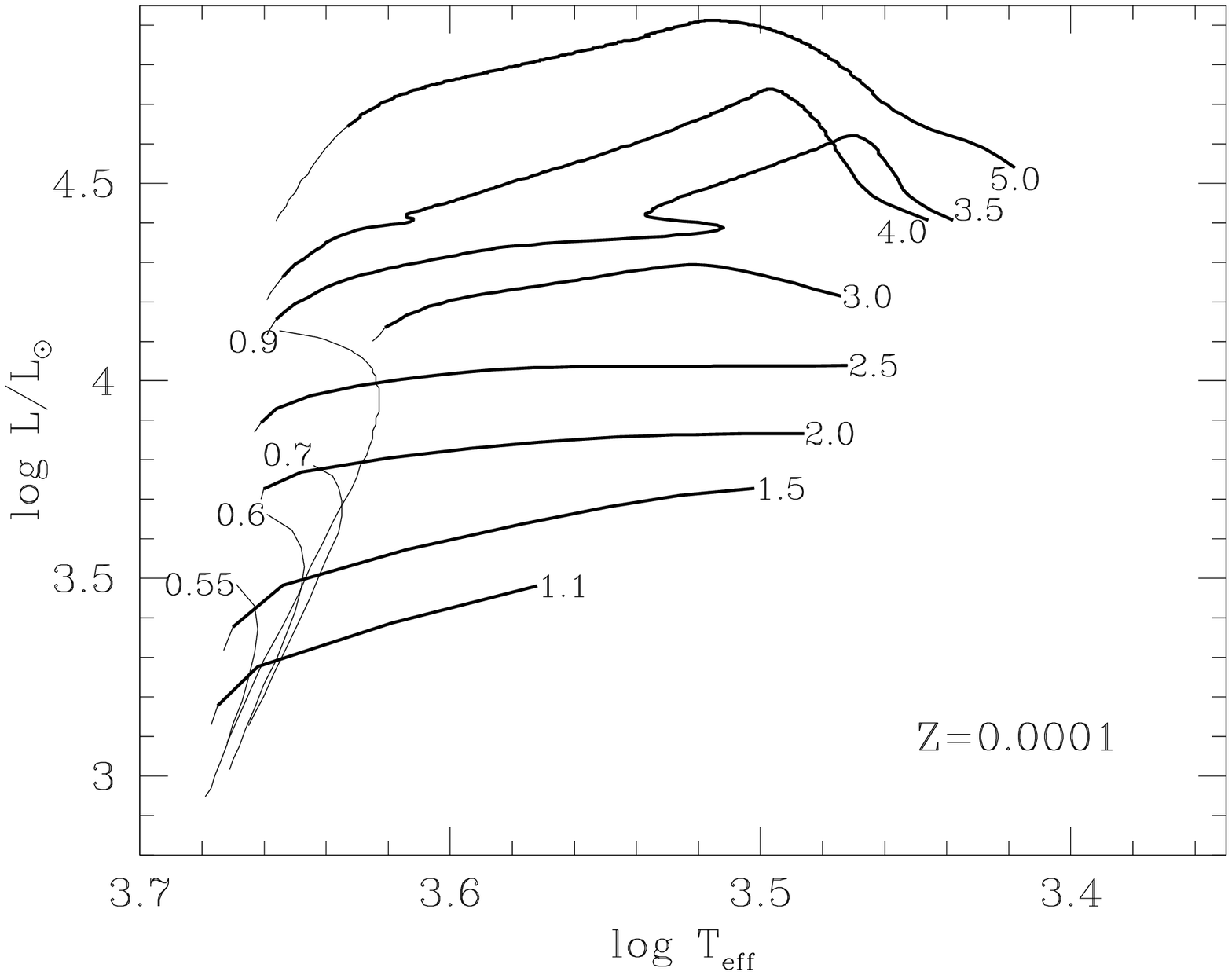}}
\end{minipage} 
\hfill
\begin{minipage}{0.66\textwidth}
\caption{TP-AGB tracks in the H-R diagram as a function of stellar
mass and metallicity, as labelled. Luminosity and effective
temperature correspond to the quiescent values just before the
predicted occurrence of thermal pulses.  Surface O-rich and C-rich
configurations are distinguished with thin and thick lines,
respectively.}
\label{fig_hr}
\end{minipage} 
\end{figure*}

\begin{figure*}[!tbp]    
\begin{minipage}{0.33\textwidth}
	\resizebox{\hsize}{!}{\includegraphics{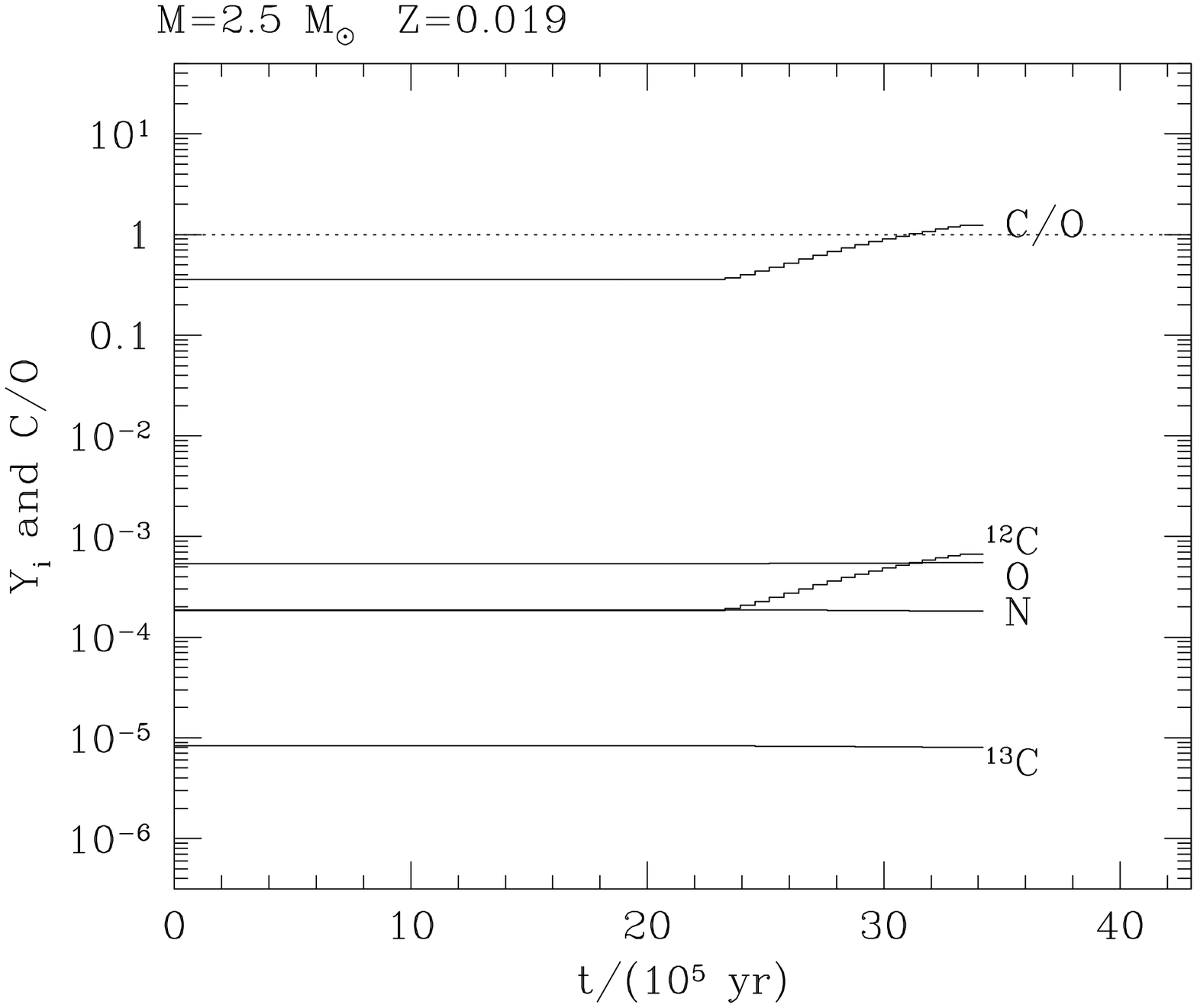}}
\end{minipage} 
\hfill
\begin{minipage}{0.33\textwidth}
	\resizebox{\hsize}{!}{\includegraphics{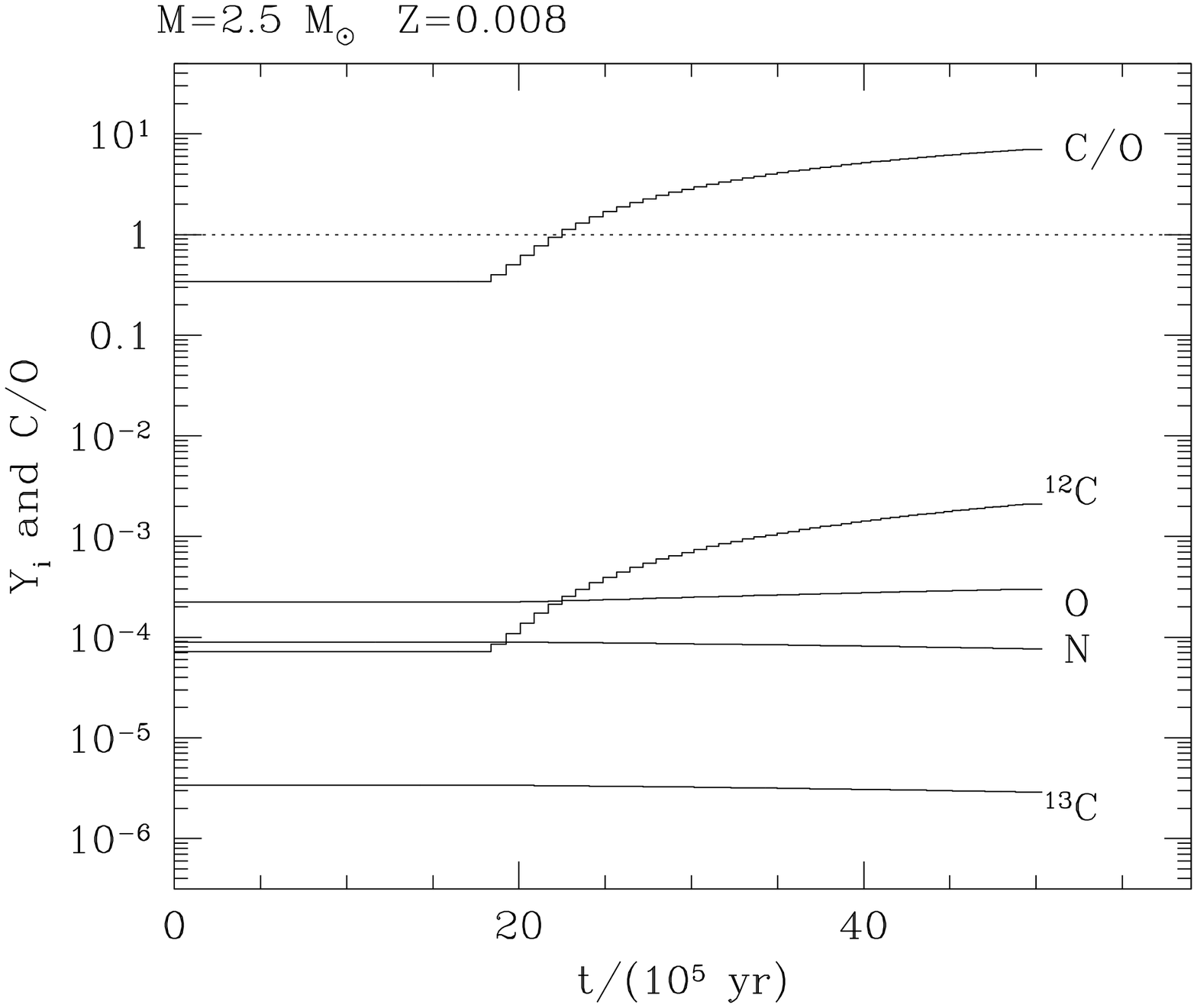}}
\end{minipage} 
\begin{minipage}{0.33\textwidth}
	\resizebox{\hsize}{!}{\includegraphics{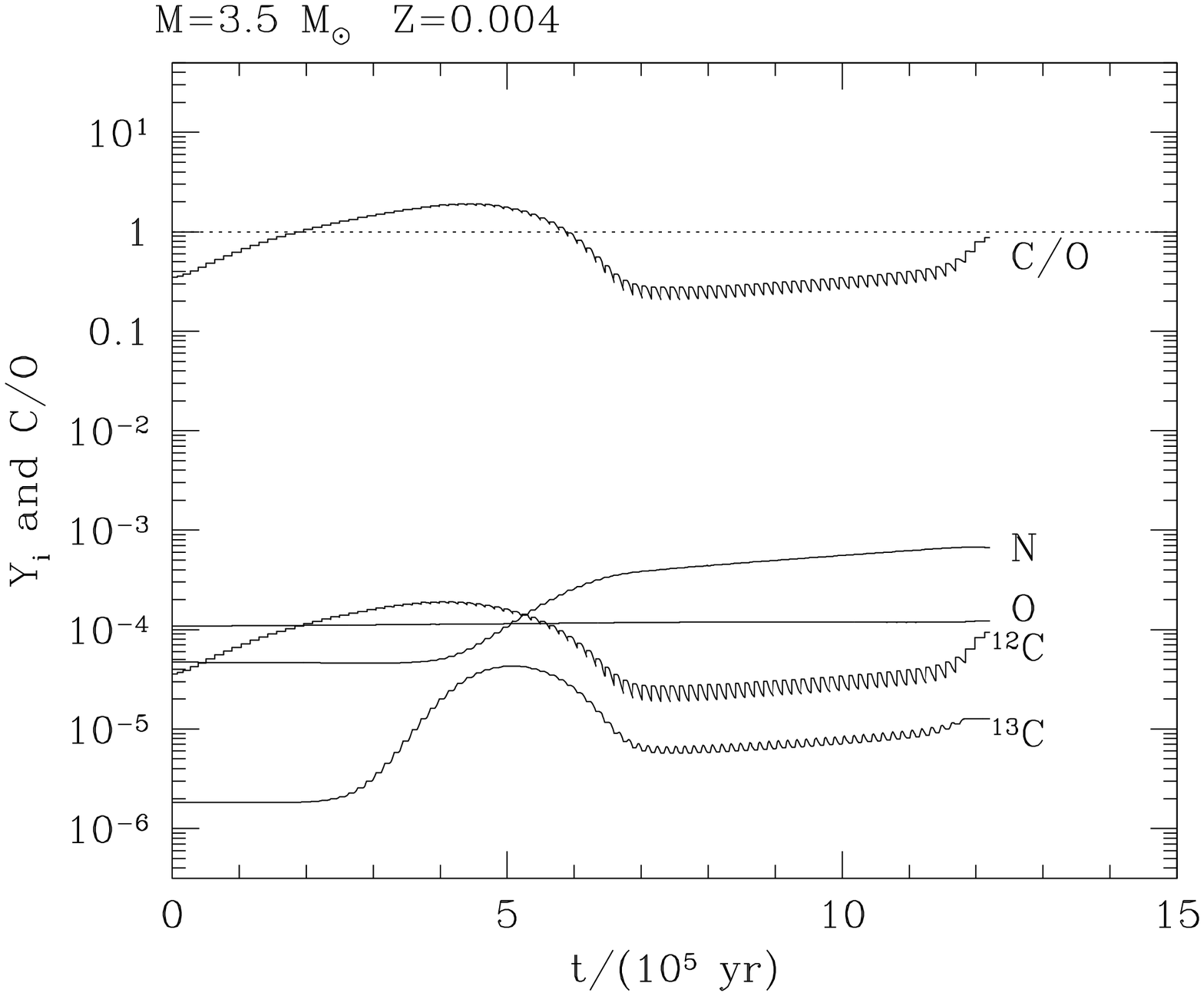}}
\end{minipage} 
\hfill
\begin{minipage}{0.33\textwidth}
	\resizebox{\hsize}{!}{\includegraphics{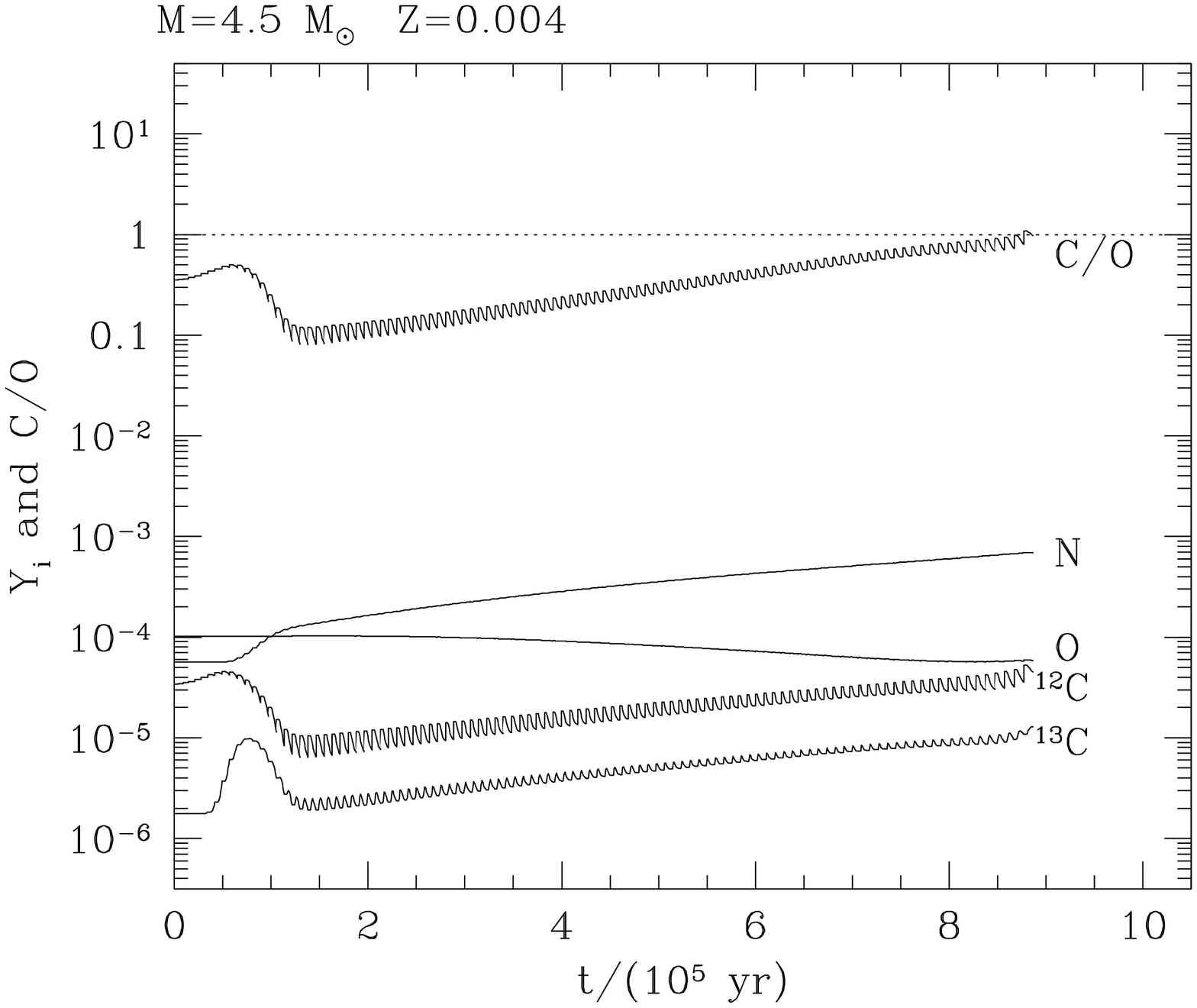}}
\end{minipage} 
\begin{minipage}{0.33\textwidth}
	\resizebox{\hsize}{!}{\includegraphics{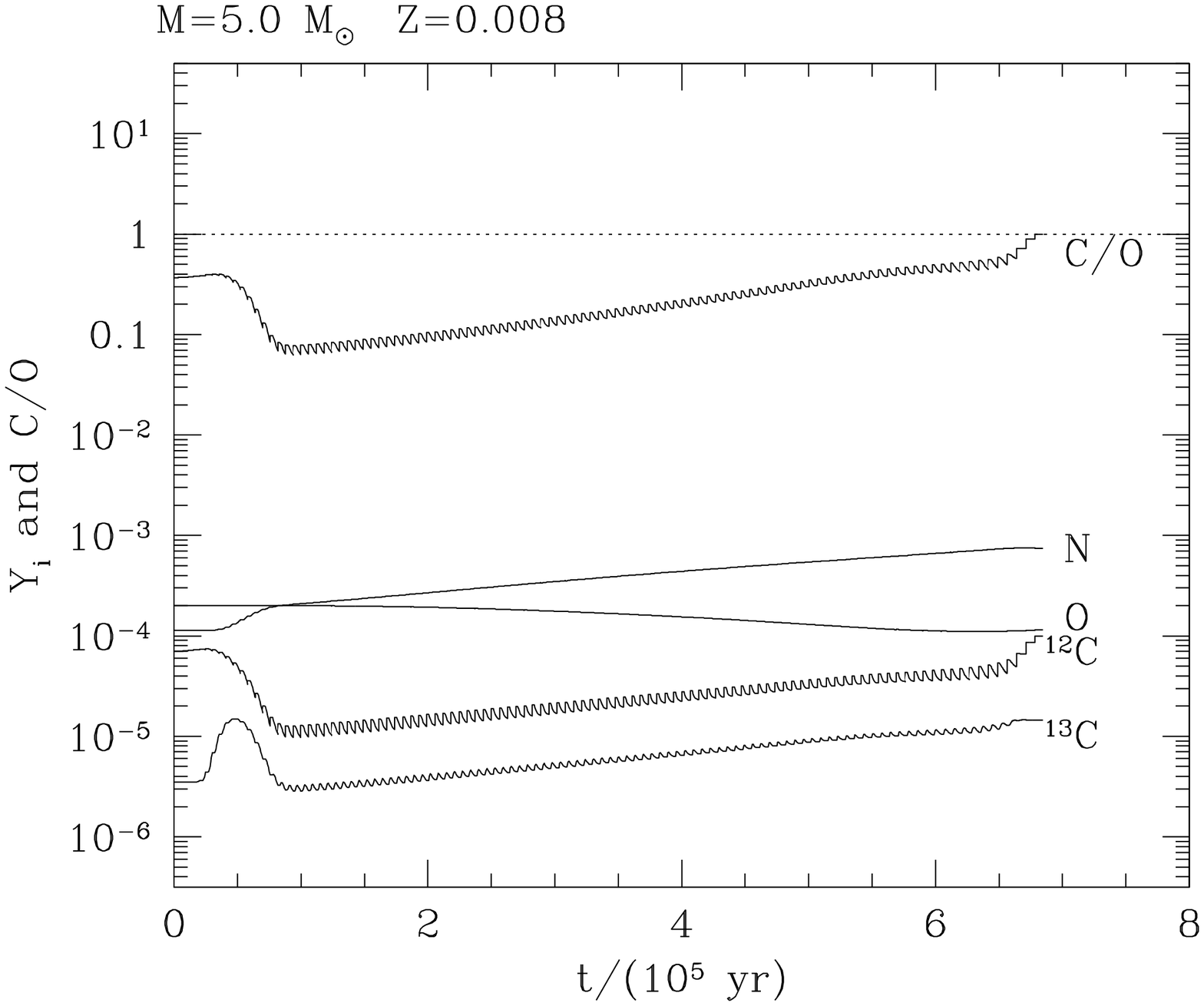}}
\end{minipage} 
\begin{minipage}{0.33\textwidth}
	\resizebox{\hsize}{!}{\includegraphics{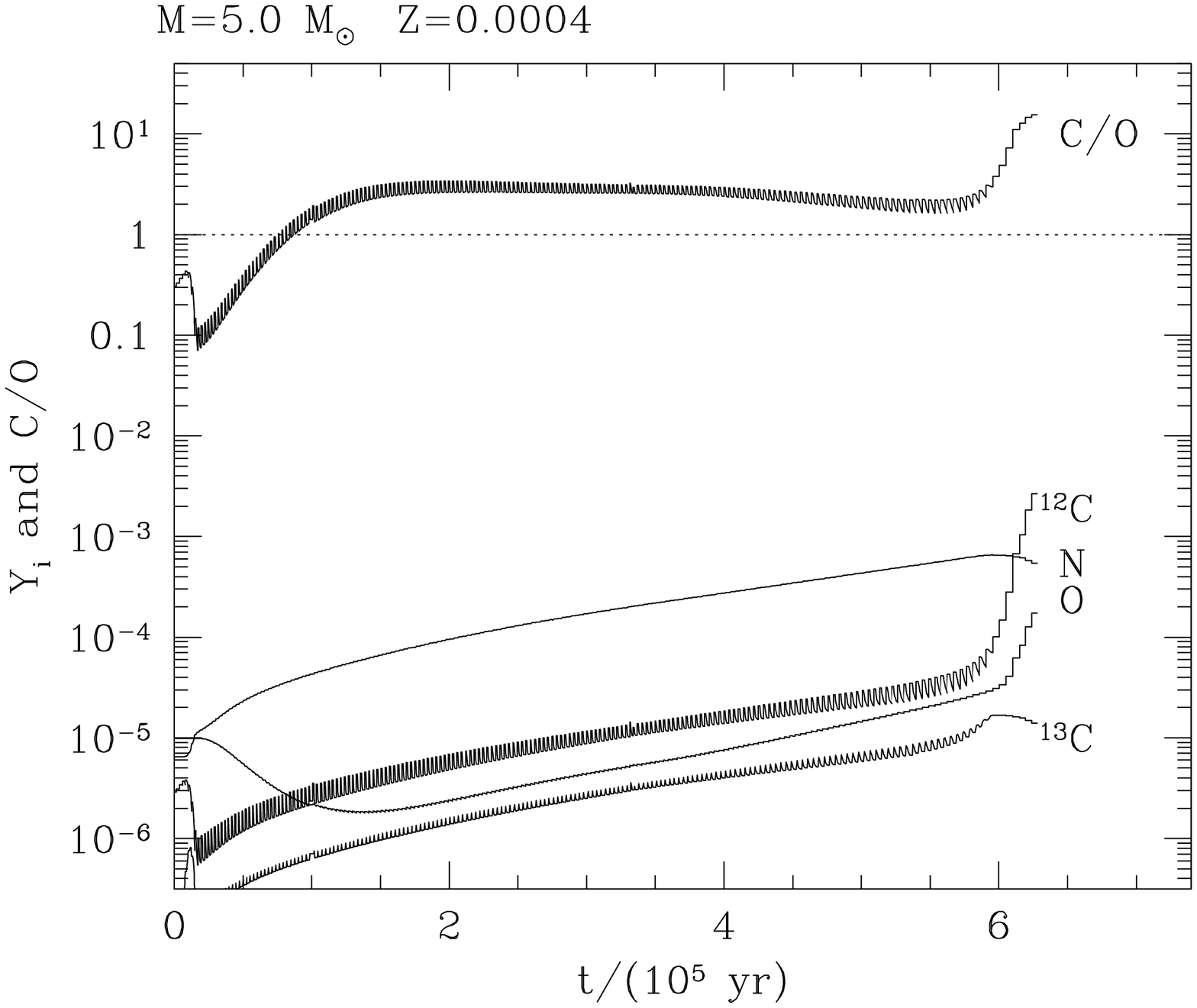}}
\end{minipage}    
\caption{Evolution of surface CNO abundances (in mole gr$^{-1}$) and 
C/O ratio during the TP-AGB phase of a few models with various initial
masses and metallicities.}
\label{fig_cno}
\end{figure*}

\subsection{TP-AGB tracks in the H-R diagram}
\label{ssec_hr}
Figure~\ref{fig_hr} displays the complex morphology of the TP-AGB
tracks in the H-R diagram at varying initial stellar mass and
metallicity. Again, for the sake of simplicity, only quiescent stages 
-- i.e. pre-flash luminosity maxima -- are considered. 

As long as the mass loss remains so modest that the TP-AGB evolution
can be safely considered at constant mass, the main parameter that
affects the evolution in effective temperature is the surface chemical
composition, specifically the C/O ratio (see Sect.\ref{ssec_cno}).
Lower-mass stars that do not experience the third dredge-up, hence
always remaining O-rich, are characterised by a slight decrease in
$T_{\rm eff}$ as they evolve at increasing luminosities along their
tracks, with a slope $d\log L/d\log T_{\rm eff}$ essentially
determined by the stellar mass and initial metallicity.
 
This behaviour changes drastically as a model makes the transition
from the O-rich to the C-rich configuration, which is is accompanied
by a marked displacement toward lower effective temperatures (thick
lines).  As already discussed, this is the direct result of the
enhanced molecular opacities as soon as as the photopsheric ratio
passes from C/O$\,<1$ to C/O$\,>1$.  We also note that at decreasing
metallicity, on average, the C-rich tracks describe 
a wider excursion in $T_{\rm eff}$, compared to
C-rich models with higher $Z$.
This is due to the fact that at lower $Z$
the transition to the C-star domain takes place earlier, so that
models leave the O-rich branch toward lower $T_{\rm eff}$s 
already during the initial stages of the TP-AGB.

One remark should be made at this
point.  There are some intermediate mass models that, after having become
carbon-rich as a consequence of the third dredge-up, are re-converted
to the oxygen-rich configuration due to the prevailing effect of HBB
(see Sect.~\ref{ssec_cno}).  This is the case, for instance, of the
($4 M_{\odot}, Z=0.008$) and ($3.5 M_{\odot}, Z=0.004$) models. As
shown in Fig.~\ref{fig_hr} the consequence of such sequence of
chemical transitions (C/O$<1\rightarrow$C/O$>1\rightarrow$C/O$\,<1$)
produces visible loops in the H-R diagram.

At increasing mass and metallicity, say $M_{\rm i}>4 M_{\odot}$ and
$Z>0.004$, TP-AGB tracks do not describe a significant cooling in
$T_{\rm eff}$ as HBB prevents the C/O ratio to increase above unity
(except in the very last stages).  However, we should remark that at
very low metallicities ($Z\le 0.001$) the role of HBB in keeping the
surface C/O below $1$ may not be effective any longer. Rather,
efficient nuclear burning at the base of the convective envelope may
favour itself the formation of C-rich models via the activation of the
ON cycle, as oxygen begins to be efficiently burnt in favour of
nitrogen (see Fig.~\ref{fig_cno}).  Due to this, metal-poor tracks of
larger masses describe in Fig.~\ref{fig_hr} pronounced excursions
toward lower $T_{\rm eff}$.

Finally, it is worth noticing that the luminosity evolution of more
massive models is significantly affected by the occurrence of
HBB. This process makes these models to be over-luminous compared to
what predicted by the core mass-luminosity relation in quiescent
conditions. As we see in fact in Fig.~\ref{fig_hr}, despite their core
mass does not essentially grow because of the deep dregde-up, the
corresponding tracks are characterised first by a relatively steep
rise in luminosity as a consequence of HBB.  Then, after reaching a
maximum the tracks exhibit a final decline in luminosity, finally
converging to the values predicted by the core mass-luminosity
relation.  This takes place at the onset of the superwind regime of
mass loss, when the drastically reduced envelope mass determines the
extinction of HBB.
    
\begin{figure*}[!tbp]  
\begin{minipage}{0.49\textwidth}
	\resizebox{0.92\hsize}{!}{\includegraphics{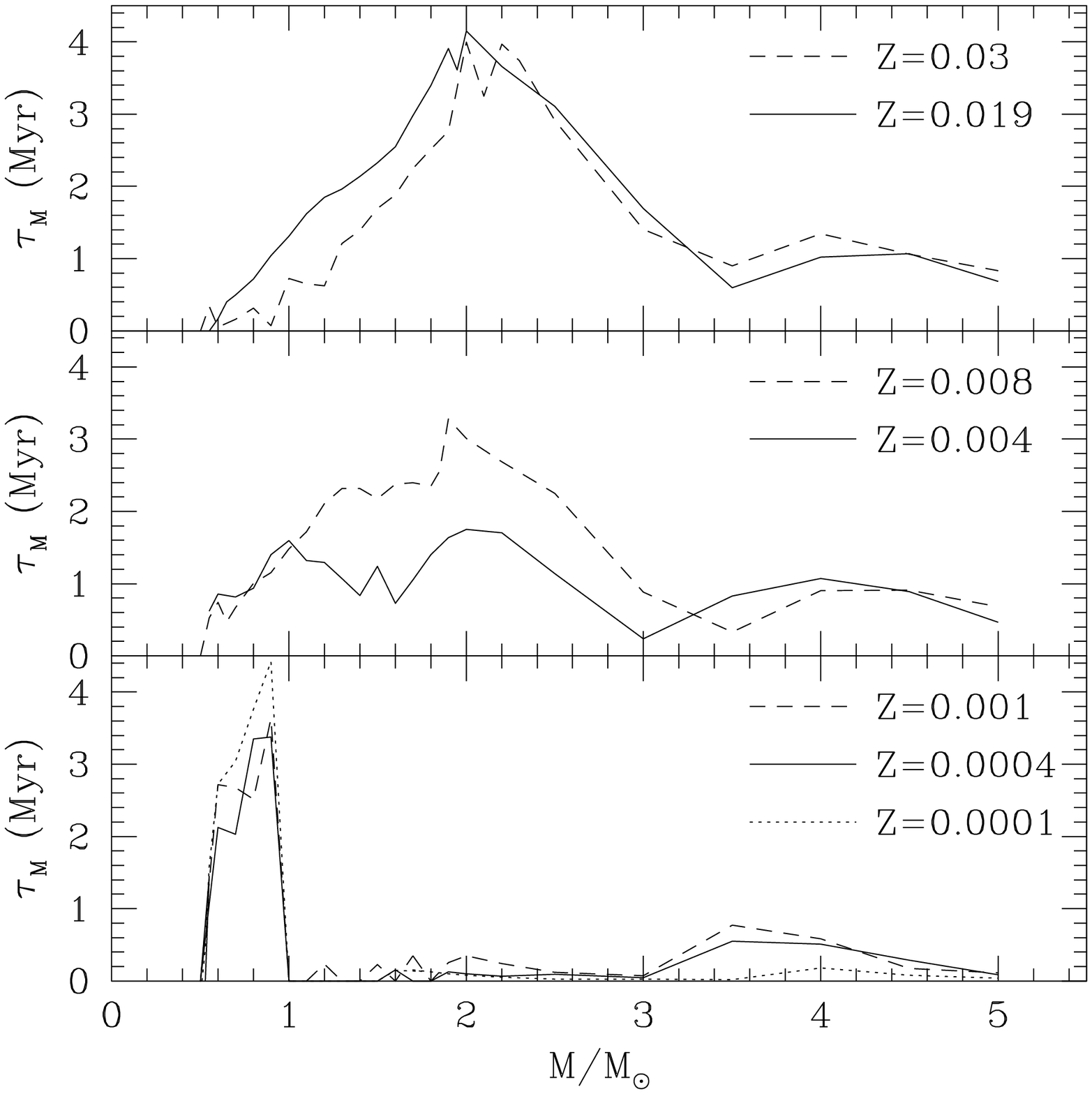}}
\end{minipage} 
\hfill
\begin{minipage}{0.49\textwidth}
	\resizebox{0.92\hsize}{!}{\includegraphics{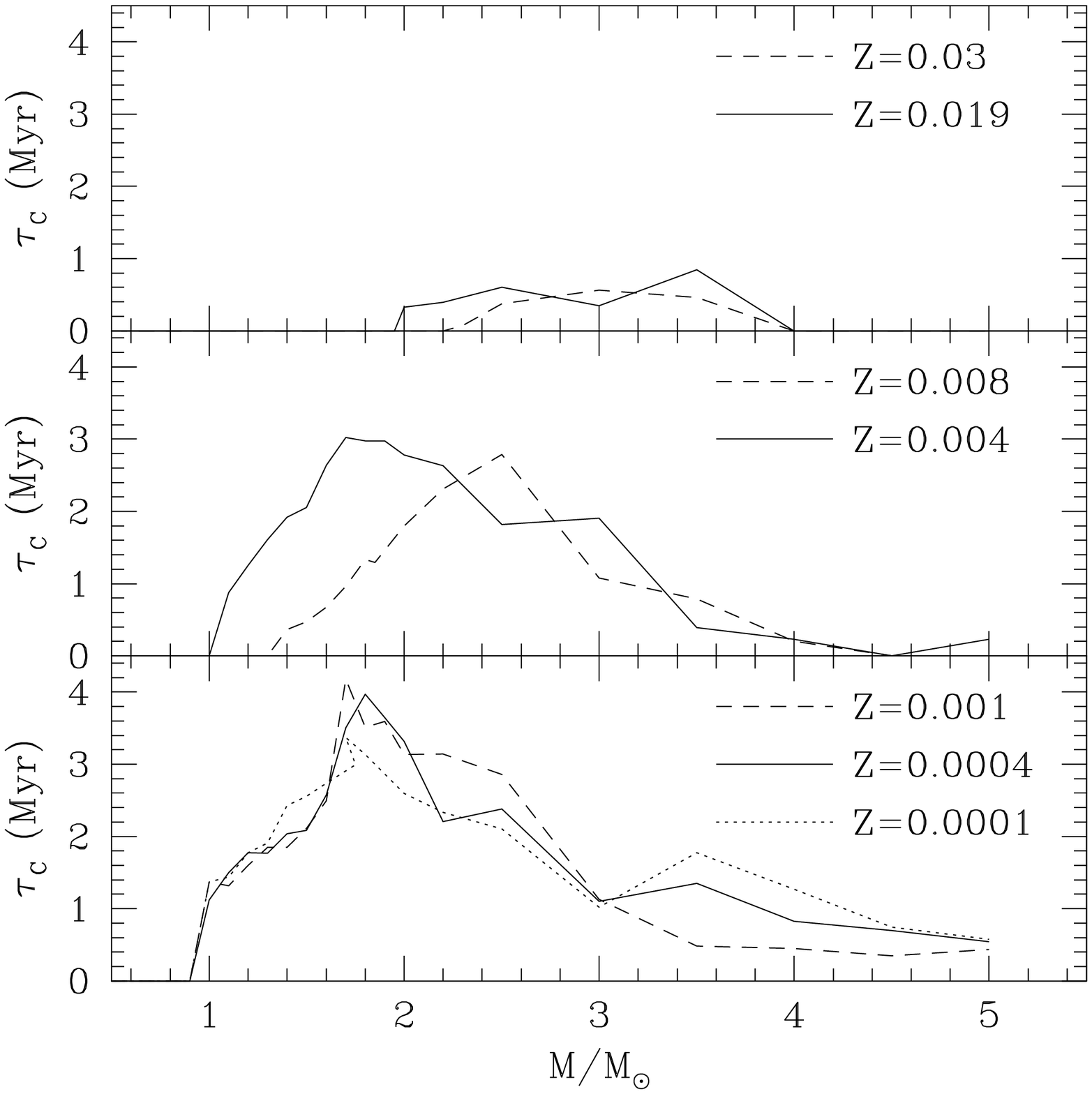}}
\end{minipage} 
\caption{Predicted duration of the phases spent by TP-AGB models 
as M-type stars ($\tau_{\rm M}$: left panels) and C-type stars
($\tau_{\rm C}$: right panels), as a function of both stellar mass and
metallicity.  For the sake of comparison with observations we consider
stellar models with luminosities higher than $\log L/L_{\odot}=3.33$,
i.e., brighter than the RGB-tip at $M_{\rm bol} = -3.6$ }
\label{fig_taumc}
\end{figure*}

\subsection{Surface chemical composition}
\label{ssec_cno}
As already mentioned in Sect.~\ref{ssec_hr}, one key parameter that
controls the morphology of the TP-AGB tracks in the H-R diagram is the
photospheric C/O ratio, mainly affecting the evolution in $T_{\rm
eff}$.  This latter may remain unaltered during the whole TP-AGB phase
-- as is the case of lower mass models --, or it may be affected by
the third dredge-up and HBB -- as in the
most massive models (refer to Sect.~\ref{ssec_hbb}).

It is interesting to compare a few emblematic cases, illustrating the
evolution of the CNO abundances and the C/O ratio in the convective
envelope of models with different masses and metallicities.  The
results are schematically summarised as follows.

Models that experience only the third dredge-up show a progressive
increase of the $^{12}$C abundance, hence of the C/O ratio which may
overcome unity thus leading to the formation of carbon stars. This
case applies, for instance, to the $M_{\rm i}=2.5\,M_{\odot}$ models
shown in the bottom panels of Fig.~\ref{fig_cno}. Note that the
duration of the C-star phase is much longer for $Z=0.008$ than for
$Z=0.019$. No important variations are expected in the abundances of N
and O.

In models that experience both the third dredge-up and HBB the
evolution of the surface CNO abundances depends on the competition
between the two processes. We distinguish three possible channels.

1) If the third dredge-up initially dominates, then the C/O ratio is
allowed to increase and the model enters the C-star domain.  The
situation may be later reversed as HBB becomes more and more efficient
in converting C in favour of N, so that the C/O ratio falls below
unity. This case applies to the models with intermediate mass, like
the $M_{\rm i}=3.5 M_{\odot},\, Z=0.004$ one shown in the middle-left
panel of Fig.~\ref{fig_cno}.

It is worth remarking here that the evolution of these
models, hence their nucleosynthesis, strongly depends on the
adopted molecular opacities {\em and} mass-loss rates.
In fact, as shown by Marigo (2007), the use of
a mass-loss formalism highly sensitive to $T_{\rm eff}$,
e.g. Vassiliadis \& Wood (1993), coupled to
{\em variable} molecular opacities could weaken, or even quench,
HBB shortly after the surface C/O ratio has overcome unity as a
consequence of the third dredge-up. In that case, C/O does not decrease
any longer and the model is expected to remain C-rich until the end of
its evolution (see figure 2 in Marigo 2007). This specific point deserves
more detailed investigation.

2) The early transition to the C-star phase may be even prevented in
models of larger mass when HBB is already strong enough to burn the
newly dredged-up $^{12}$C. These models are then predicted to keep the
surface C/O ratio below unity for the whole TP-AGB phase, possibly
except for the very last stages of intense mass loss when HBB stops
and the third dredge-up makes them appear as obscured carbon stars
(van Loon et al. 1997; Frost et al. 1998).  This picture is
exemplified, for instance, by the $M_{\rm i}=4.5 M_{\odot},\, Z=0.004$
and $M_{\rm i}=5.0 M_{\odot},\, Z=0.008$ models displayed respectively
in middle-right panel and bottom-left panel in Fig.~\ref{fig_cno}.

3) At very low metallicities HBB may even concur -- together with the
third dredge-up -- to favour the formation of massive carbon stars. In
fact, the high temperatures ($\ga 7.9\,10^{7}\, K$) reached at the
base of the convective envelope lead to the activation of the
ON-cycle, with the consequent conversion of oxygen into nitrogen.  In
this case the rise of the surface C/O ratio is mainly due to the
depletion of the initial abundance of oxygen. 
According to our calculations this condition is met by
models with $M_{\rm i} \ga 4.0\, M_{\odot}$ and $Z\la 0.001$
(see Ventura et al. 2002 for similar results).  
An example is given by the $M_{\rm i}=5.0 M_{\odot},\, Z=0.0004$, shown
in the bottom-right panel of Fig.~\ref{fig_cno}.
We see that, after an initial decrease of carbon,  the CN cycle attains
its equilibrium regime -- i.e. both
$^{12}$C$/^{13}$C and C/N reach their equilibrium values -- and on
average carbon starts to increase steadily. Then, the activation of the ON cycle leads
to a significant drop of oxygen which makes the 
C/O ratio increase above unity. Subsequently the ON-cycle  
goes to equilibrium as well so that carbon keeps more abundant than oxygen despite 
the steady increase of this latter.

\subsection{Duration of the TP-AGB phase}
\label{ssec_tpagbtime}
Figure~\ref{fig_taumc} shows the predicted TP-AGB lifetimes for both
the oxygen-rich (with $M_{\rm bol} < -3.6$) and C-rich phases,
$\tau_{\rm M}$ and $\tau_{\rm C}$, as a function of the stellar mass,
for all metallicities here considered.  It should be remarked that
these results are derived from the calibration of our TP-AGB
tracks, which will be  discussed in Sect.~\ref{sec_calibr}. Moreover,
the TP-AGB lifetimes displayed in Fig.~\ref{fig_taumc} account 
also for the last
evolutionary stages characterised by the superwind mass-loss rates,
i.e. the contribution from the dust-enshrouded, optically-obscured
evolution is included in the computation of $\tau$.

The termination of the TP-AGB phase, hence its duration, is determined
by all those factors that concur to make the star lose its envelope.
Clearly the main agent is mass loss by stellar winds, (and its complex
dependence on basic stellar parameters like luminosity, effective
temperature, chemical composition, pulsation, dust formation, etc.),
but a certain role is also played by the rate of outward advancement
by the H-burning shell that reduces the envelope mass from inside (and
its dependence on factors like the metal content, and the occurrence
of the third dredge-up and HBB). Actually it is worth mentioning that
for $Z\le 0.004$ low-mass TP-AGB models, say with $M \le 0.7
M_{\odot}$, never reach superwind mass-loss rates so that they leave
the AGB because their envelope is eaten by the growth of the
H-exhausted core.

The behaviour with metallicity of $\tau_{\rm M}$ at given mass, displayed
in the left panel of Fig.~\ref{fig_taumc}, depends on wether a TP-AGB
model undergoes or not the third dredge-up.  Lower-mass models that
keep C/O$\,<1$ up to the end of the AGB tend to have longer $\tau_{\rm
M}$ at decreasing metallicity because of the lower rates of mass loss
and/or core-mass growth.
 
In models of higher mass, that experience the third dredge-up,
$\tau_{\rm M}$ is crucially affected by this latter process -- which
determines the stage at which the surface C/O$\,>1$ -- while it is
essentially independent of the adopted mass-loss prescription.  
For these models the resulting
trend of $\tau_{\rm M}$ is controlled by the mass and metallicity
dependence of the main two dredge-up parameters, $\lambda$ and $M_{\rm
c}^{\rm min}$.

As we see there is a general tendency to have shorter $\tau_{\rm M}$
at decreasing metallicity because of the earlier onset and larger
efficiency of the dredge-up at lower $Z$.  It is interesting to note
that for $Z\ga 0.008$ the $\tau_{\rm M}(M_{\rm i})$ curve displays a
peak at about $M_{\rm i}\simeq 1.9-2.2\, M_{\odot}$. This finding is to be
at least partly ascribed to the relation $M_{\rm c,1TP}(M)$ 
at the first thermal pulse, which presents a minimum just in correspondence to these
stellar masses, so that a relatively large growth of the core
mass is required before the $M_{\rm c}^{\rm min}$ is attained.  
As we discuss in
Sect.~\ref{ssec_lifetimes}, this feature is in agreement with the 
observed maximum of M-type star counts in the Magellanic
Clouds' clusters.

Concerning $\tau_{\rm C}$, shown in the right panel of
Fig.~\ref{fig_taumc}, we note that the mean trend with metallicity is
reversed compared to $\tau_{\rm M}$, i.e. the C-rich phase is expected
to last longer at decreasing metallicity.  For instance at $M_{\rm
i}=2.5\,M_{\odot}$, $\tau_{\rm C}$ passes from $0.6$~Myr for $Z=0.019$
to $2.8$~Myr for $Z=0.008$.  However, we also see that moving to very
low $Z$, $\tau_{\rm C}$ does not increase drastically, rather it keeps on
average almost invariant, e.g., $\tau_{\rm C}\sim 2.4$~Myr for
$Z=0.0004$. This is the net result of the interplay between mass loss and 
dredge-up through their dependence on  metallicity.

\subsection{Mass loss rates, periods and luminosities of variable AGB stars}
\label{ssec_mlrp}

\begin{figure}[!tbp]  
\resizebox{\hsize}{!}{\includegraphics{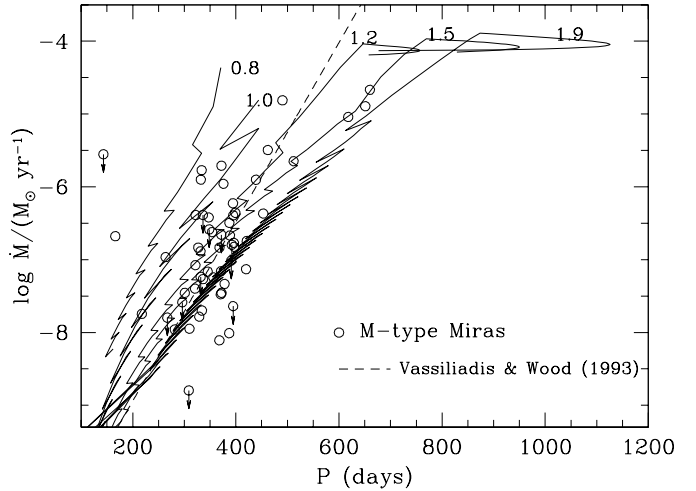}}
\caption{Mass-loss rate versus period for O-rich Mira variables.
Observed data (circles) correspond to Galactic M-type Miras (Le Bertre
\& Winters 1998; Groenewegen et al. 1999).  For the sake of comparison
we show the semi-empirical relation for Miras proposed by Vassiliadis
\& Wood (1993; dashed line), together with the theoretical predictions
for evolving O-rich TP-AGB models with $Z=0.019$ and various initial
mass, based on Bowen \& Willson (1991) wind models. }
\label{fig_mlro}
\end{figure}

\begin{figure}[!tbp]  
\resizebox{\hsize}{!}{\includegraphics{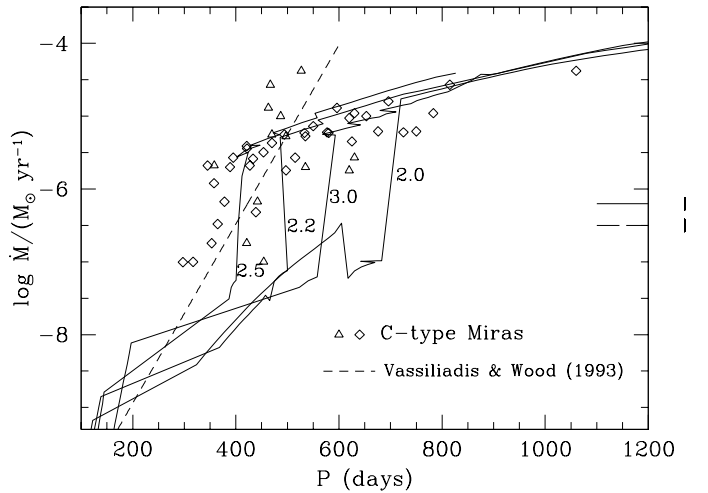}}
\caption{Mass-loss rate versus period for C-rich Mira variables.
Observed data correspond to Galactic C-rich Miras from Groenewegen et
al. (1998; squares), and Sch\"oier \& Olofsson (2001; triangles).  For
the sake of comparison we show the semi-empirical relation for Miras
proposed by Vassiliadis \& Wood (1993; dashed line), together with the
theoretical predictions for evolving C-rich TP-AGB models with
$Z=0.019$ and various initial mass, based on Wachter et al. (2002)
wind models. }
\label{fig_mlrc}
\end{figure}

\begin{figure*}[!tbp]  
\begin{minipage}{0.73\textwidth}
\resizebox{\hsize}{!}{\includegraphics{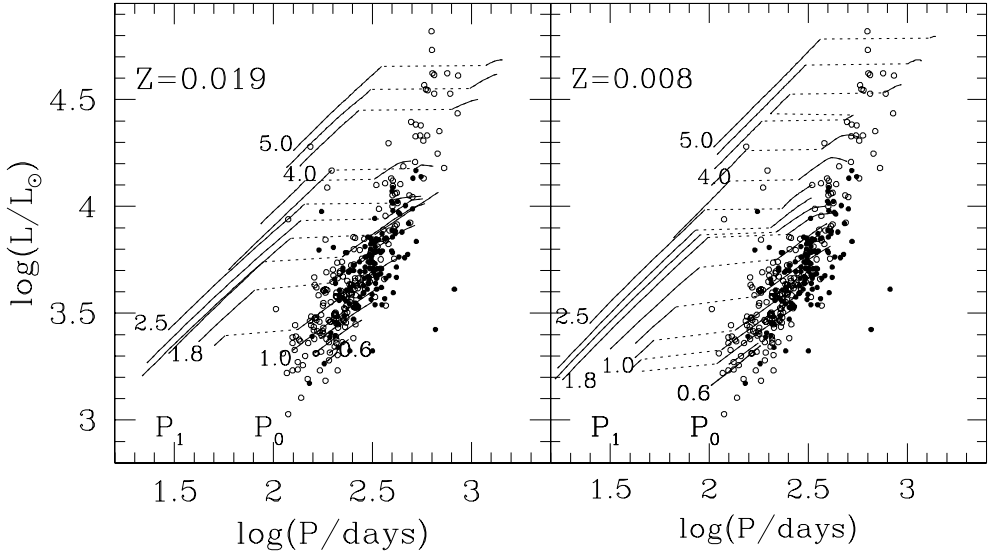}}
\end{minipage}
\hfill
\begin{minipage}{0.25\textwidth}
\caption{Period-luminosity relation of Mira variables in the LMC.
Observed data  correspond to M-type (empty circles) and C-type Miras 
(filled circles)
LMC (Feast et al. 1989; Hughes \& Wood 1990), assuming a distance
modulus $\mu_0 = 18.5$. Lines show the theoretical tracks
described by TP-AGB models for two choices of 
metallicity and with initial stellar masses ranging from 
0.6 to 5.0 $M_{\odot}$ (a few values are indicated).
It is assumed that pulsation occurs initially in 
the FOM ($P_1$ sequence) and then it switches to the FM 
($P_0$ sequence). See text for more esplanation.
}
\label{fig_pl}
\end{minipage}
\end{figure*}

The use of theoretically-derived mass loss rates constitutes a major
change with respect to our previous works. Leaving a detailed
discussion of this subject to a future paper, below we simply
illustrate the implications of our present choice.

Figure~\ref{fig_mlro} compares a few of our O-rich TP-AGB tracks of
solar composition with the data for Galactic M-type Miras in the $\dot
M$--$P$ diagram. It can be noticed that the models reproduce well the
ranges of mass-loss rates and periods covered by the data. In particular,
the significant dispersion in $\dot M$ at given $P$ exhibited by the
Miras is accounted by the models at varying mass. The observed spread
cannot be obtained, instead, by applying, for instance, the
Vassiliadis \& Wood (1993) formalism that gives $\dot M$ as a sole
function of $P$. Moreover, the onset of the superwind phase (with
$\dot M \approx 10^{-5}-10^{-4}\, M_{\odot}$\,yr$^{-1}$) is naturally
predicted with the adopted scheme and allows the reproduction of the
OH/IR data over a wide period range.

Similar comments can be made about mass loss rates on the C-rich
phase, as illustrated in Fig.~\ref{fig_mlrc}. What is more evident
in this case is that models reproduce well the plateau at $\dot M
\approx 10^{-5}\, M_{\odot}$\,yr$^{-1}$in $\dot M$, corresponding to
the superwind phase of C-type stars.

Extensive observations of Miras in the LMC have shown that these
variables populate a well-defined period-luminosity ($P-L$) relation
(e.g., Feast et al. 1989; Hughes \& Wood 1990; MACHO survey), as
displayed in Fig.~\ref{fig_pl} for both oxygen- and carbon-rich objects.

TP-AGB evolutionary tracks at varying initial mass and  with 
$Z=0.019$ and $Z=0.008$ are superimposed for the sake of comparison.
According to the assumed description of pulsation   
(see Sect.~\ref{ssec_pulsation}), tracks split into two bundles
in the  $P-L$ diagram, the one at shorter $P$ corresponding to the 
FOM, and the other at longer  $P$  corresponding to the FM. 
Clearly this latter group turns out to be 
consistent with the observed location of Miras, supporting 
the indication that these variables are pulsating in the fundamental
mode.

A detailed investigation of the several variability sequences, in addition
to that of Miras, 
that populate the  $P-L$ diagram 
(e.g. Wood et al. 1999)  are beyond the scope of the present study.
To this aim it is necessary 
to perform population synthesis simulations that  
couple updated TP-AGB tracks with LPV models for different pulsation
modes. This kind of analysis is underway.

%

\subsection{The initial--final mass relation}
\label{ssec_minifin}

The initial--final mass relation (IFMR) is affected by the competition
between core growth by shell burning and core reduction by third
dredge-up events, the final result depending on the termination of the
AGB evolution by stellar winds.  Since dredge-up is more efficient at
higher masses and lower metallicities, one may naively expect that
this competition will favour flatter IFMR at lower metallicities.
Actually, this may not happen because the IFMR also carries a strong
imprinting of the core mass at the first thermal pulse, which is in
general larger at lower metallicities (see e.g., figure 1 in Girardi
\& Bertelli 1998).  Moreover, the larger efficiency of HBB 
at lower metallicities may reduce the TP-AGB lifetime in such a way
that less time is left for the core mass to significantly change along
the TP-AGB. Finally, the IFMR is modulated by the TP-AGB lifetimes,
the complex dependence of which as a function of mass and metallicity
is shown in Fig.~\ref{fig_taumc}.  Therefore, the actual shape of the
IFMR and its variation with metallicity depends on several parameters
and it cannot be easily predicted via simple reasonings.

\begin{figure}  
\resizebox{\hsize}{!}{\includegraphics{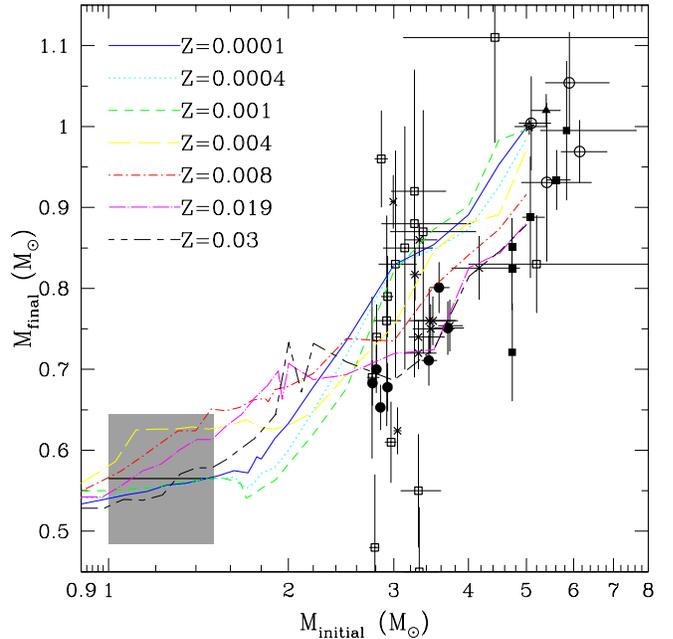}}
\caption{
The initial--final mass relation from our models (lines) for several
metallicities, together with the empirical data for white dwarfs in
open clusters and binary systems: the Hyades (full circles, Claver et
al. 2001), M~37 (open squares, Kalirai et al. 2005), Praesepe
(crosses, Claver et al. 2001, and Dobbie et al. 2006), M~35 (full
squares, Williams et al. 2004), Sirius B (star, Liebert et al. 2005a),
Pleiades (full triangle, Dobbie et al. 2006), and NGC~2516 (open
circles, Ferrario et al. 2005). The gray area indicates the position
(mean and 1$\sigma$) of the primary mass peak of field white dwarfs as
measured by Liebert et al. (2005b), that we indicatively associated to
the 1.0--1.5~\Msun\ interval.  }
\label{fig_minifin}
\end{figure}

Figure \ref{fig_minifin} shows the IFMR as predicted by present models
of various metallicities, compared to {{\bf} recent observational data
for white dwarfs in binary systems and open clusters.  In general,
both the cluster ages and white dwarf progenitor masses have been
estimated (or checked) by using tracks and isochrones from Girardi et
al. (2000), so that the plot should not contain inconsistencies
associated to the use of different prescriptions for convective
overshooting.}  It is worth noticing that:

1) The predicted IFMR have a significant dependence on
metallicity. For a given initial mass the allowed range of final
masses in about 0.1~\Msun\ wide.  The dependence is in the sense the
IFMR is steeper at low metallicities. Although the final mass
generally increases with the initial mass, there are limited mass
ranges in which this trend is reversed (e.g., for the $Z=0.03$
tracks in the $2.2-3$~\Msun\ interval), as well as small
track-to-track variations that arise from the complex interplay
between the evolutionary parameters involved.

2) The data seems to be too dispersed to allow a firm testing of the
model predictions, but this dispersion largely depends on the data for
the M~37 cluster (from Kalirai et al. 2005), which -- apart from a few
low-mass outlyiers -- lie systematically above the mean relation drawn
by the remaining objects. As remarked by Dobbie et al. (2006), the
mass determinations of M~37 white dwarfs may still change considerably
since its initial metallicity is not well determined. Excluding the
M~37 white dwarfs, the IMFR is well defined for the complete range of
$M_{\rm i}>2.5$~\Msun. On the other hand, good-quality data covering the
$1.5-2.5$~\Msun\ mass range are presently not available. Notice that
we have represented the $<1.5$~\Msun\ mass range by field white
dwarfs, whose initial mass is not directly measured.

3) If we disconsider the M~37 white dwarfs, the comparison between the
data and the model IFMRs for close-to-solar metallicities (the
$Z=0.019$ and $0.03$ curves) turns out to be very satisfactory over
the entire $M_{\rm i}<5$~\Msun\ mass range. It seems however advisable
to extend the TP-AGB model computations to slightly larger initial
masses, in order to explain the few points observed at $M_{\rm
i}>5.5$~\Msun. In this regard, more recent evolutionary tracks from
Padova (Bertelli et al., in preparation) suggest that the highest mass
for entering in the TP-AGB phase, using Girardi et al.'s (2000)
prescription for overshooting, is located somewhere between 5.5 and
6~\Msun, rather than equal to the 5~\Msun\ mentioned by Girardi et al.

Future works in this series will improve on this latter point, and
present detailed comparisons with the mass distribution of field white
dwarfs.


\section{Calibration of TP-AGB model parameters}
\label{sec_calibr}


In this section, we first briefly present the set of Magellanic Cloud
data used to constrain the TP-AGB model parameters, together with the
methods employed to simulate the data starting from the TP-AGB
tracks. The calibration process is fully described in the Appendix.

We remark that previous sets of calibrated TP-AGB models (e.g.,
Groenewegen \& de Jong 1993; Marigo et al. 1999) have used the C-star
luminosity functions in the Magellanic Clusters as, basically, the
only observational constraint.  To the CSLFs, we now add the C- and
M-type lifetimes as to be described below. Previous works have also
adopted too simple descriptions for the star formation rate {{\bf} (SFR)} and
age-metallicity relations {{\bf} (AMR)} of both galaxies, like for instance a
constant SFR, and a constant metallicity with varying age. We now
improve upon this approach by adopting SFR and AMR independently
derived by other authors to simulate the data.

\subsection{Carbon star luminosity functions in the Magellanic Clouds}
\label{ssec_cslf}

The CSLF in several Local Group galaxies have been reviewed and
intercompared by Groenewegen (2002; see also Groenewegen 1999), who
kindly provided us with his latest data files for both Magellanic
Clouds. The observed CSLF in the LMC makes use of the Costa \& Frogel
(1996) $m_{\rm bol}$ data, which refers to $895$ carbon stars among
the original sample of $1035$ stars distributed in 52 LMC fields over
an area of $8^{\circ} \times 8^{\circ}$.  This sample of carbon stars
can be considered almost complete and not limited by apparent
magnitude, since the faintest carbon stars are at least 2~mag brighter
($I \sim 15$) than the faint limit of detection ($I \sim 17$) of the
original surveys (Blanco et al. 1980, Blanco \& McCarthy 1983). The
fraction of optically-obscured carbon stars, which would have been
missed in the survey because of dust obscuration, has been estimated
to be lower than $3~\%$ for $M_{\rm bol} < -6$ (Groenewegen \& de Jong
1993). The conversion from the photometry to bolometric magnitudes has
been performed by Costa \& Frogel (1996), who mention a mean error on
the evaluation of $m_{\rm bol}$ of approximately $0.34$~mag. Finally,
the absolute bolometric magnitudes, $M_{\rm bol}$, are then obtained
by adopting a distance modulus of $\mu_0 = 18.5$ for the LMC. The
resulting CSLF extends approximately from $M_{\rm bol}=-3$ up to
$M_{\rm bol}=-6.5$, with the peak located at around $M_{\rm
bol}=-4.875$.

The observed CSLF in the SMC has been derived by Groenewegen (1997)
from a sample of $1636$ stars observed by Rebeirot et al.\ (1993), and 
adopting bolometric correction by Westerlund et al.\ (1986).
The adopted distance modulus for the SMC is $\mu_{0} = 19.0$.  The
first bin of this CSLF contains all stars fainter than $M_{\rm bol} =
-2.5$, which comprise less than 3~\% of the total sample of SMC carbon
stars. Their intrinsic magnitudes can be as low as $M_{\rm bol} =
-1.8$, as found by Westerlund et al.\ (1995). The formation of carbon
stars at such low luminosities can hardly be understood as the result
of the third dredge-up in single AGB stars, and most likely they arise
from the transfer of mass in close binary systems (see e.g.\ de Kool
\& Green 1995; Frantsman 1997). These low-luminosity carbon stars will
not be considered in this work, since we are dealing with single star
evolution. For all practical comparisons, we will consider the CSLF in
the SMC as extending down to $M_{\rm bol} = -2.5$. The resulting CSLF
extends up to $M_{\rm bol}=-6.5$, with the peak located at $M_{\rm
bol}=-4.375$.

{{\bf} Recently Guandalini et al. (2006) have claimed that bolometric
corrections for C stars are systematically underestimed because neglecting
the flux in the mid-IR, and suggested that accounting for
this flux would imply overall much brighter CSLFs.  Although their
arguments are unimpeachable, it is worth noticing that their study is
based on a sample of C stars in the Milky Way selected to have mid-IR
photometry. Such a sample is much more prone to contain IR-bright
objects -- of cool temperatures and undergoing significant mass loss
and dust obscuration --, than samples extracted from optical to near-IR
surveys. This is also indicated by the fact that the majority of their 
stars have $J-K>2.0$ (see their figure 7), which corresponds to the colour
interval of C stars strongly affected by circumstellar dust. The bulk
of C stars observed in the Magellanic Clouds, instead, have $J-K$
comprised between 1.2 and 2.0 (e.g. Nikolaev \& Weimberg 2000), and it
is expected to be overall much hotter and cleaner from dust than
Guandalini et al.'s (2006) sample.  Moreover, we remark that both
observations (van Loon et al. 2005) and population synthesis models
based on the present evolutionary tracks (Marigo et al., in prep.)
agree in estimating as $\approx 10-20\%$ the fraction of
dust-enshrouded C-stars in the Magellanic Clouds, i.e. the bulk of
their C-star population consists of optically-visible objects with
just mild mid-IR emission. Therefore, we consider it unlikely that the
consideration of the mid-IR flux would significantly affect the CSLFs
here discussed.  }

\subsection{TP-AGB lifetimes: M- and C-type phases}
\label{ssec_lifetimes}

Limits to the lifetimes of the M- and C-type TP-AGB phase have been
recently derived by Girardi \& Marigo (2007), using available data for
C and M giants with $\Mbol<-3.6$ in Magellanic Cloud clusters (Frogel
et al. 1990), together with the clusters' integrated $V$-band
luminosities (from Bica et al. 1996 and van den Bergh 1981). They
essentially find the C-star phase to have a duration between $2$ and
$3$ Myr for stars in the mass range from $\sim1.5$ to $2.8\,M_\odot$,
with some indication that the peak of C-star lifetime shifts to lower
masses as we move from LMC to SMC metallicities. The M-giant lifetimes
is found to peak at $\sim2$~\Msun\ in the LMC, with a maximum value of
about 4~Myr, whereas in the SMC their lifetimes are poorly constrained
by the data. Therefore, in the following we will only deal with the
C-star lifetimes for both LMC and SMC, and with the M-star lifetime
above $\Mbol<-3.6$ for the LMC. 

Our aim is to obtain theoretical TP-AGB lifetimes that fit the Girardi
\& Marigo (2007) values, within the 67~\% confidence level of the
observations (1$\sigma$ for the most populated age bins), over the
complete $0.8-5$~\Msun\ interval for the formation of TP-AGB
stars. Although this 67~\% confidence level interval may look quite
ample, they constitute very useful constraints. In fact, Girardi \&
Marigo (2007) show that many TP-AGB models in the literature do
present too short C-star lifetimes, and/or fail to reproduce the
observed trends of the lifetimes with the initial mass.

\subsection{Modelling the CSLF and lifetimes}
\label{ssec_modelling}

In order to simulate the CSLF starting from TP-AGB tracks, we follow a
complex but well-tested procedure, which is somewhat more elaborated
than the one described in Marigo et al. (1999). First, the tracks on
their quiescent phases of evolution, are assembled together with the
previous evolution given by Girardi et al. (2000) tracks. Theoretical
isochrones are then constructed via interpolation for any intermediate
value of age and metallicity (Marigo \& Girardi, in preparation). The
isochrone section corresponding to the C-star phase -- with their
quiescent luminosities -- is then isolated; for each point of it, the
pulse-cycle luminosity variations are reconstructed using Wagenhuber
\& Groenewegen (1998) formula together with basic stellar parameters
like the metallicity and core mass. This allows us to attribute to
each quiescent point in the isochrone the complete luminosity
probability distribution function, and hence to construct the CSLF for
each single-burst stellar population via integration along the
corresponding isochrone. Then, we integrate over population age,
weighting each age by its relative SFR and using the AMR to select the
right metallicity value at each age.

As for the LMC, we have adopted the SFR determined by Holtzman et
al. (1999) in an outer field close to Hodge~4, using deep HST
data. For the SMC, we have used the global SFR as determined by Harris
\& Zaritsky (2004) from the Magellanic Clouds Photometric Survey. In 
both cases, the SFR have been reconstructed using objective algorithms
of CMD fitting. Their results are mostly determined by the star counts
in the main phases of pre-AGB evolution (main sequence, subgiant and
red giant branches, red clump) and, importantly to us, are unlikely to
be sensitive to the (relatively very few) TP-AGB stars present in
their data.


We adopt the AMRs from the ``bursting models'' of chemical evolution
for the LMC and SMC by Pagel \& Tautvaisiene (1998). These relations,
although of theoretical nature, are designed to reproduce the
age-metallicity data for star clusters in both Clouds.
 
The TP-AGB lifetimes in the C- and M-type phases are derived directly
from the tracks, first identifying the turn-off mass and population
age of each TP-AGB track, then interpolating the turn-off mass and
TP-AGB lifetimes for each metallicity--age value given in the AMR.
As a result, we derive the expected TP-AGB lifetimes 
(above $M_{\rm bol}=-3.6$) as a function of
turn-off mass, that can be directly compared to the data tabulated by 
Girardi \& Marigo (2007).

\begin{figure*}[!tbp]  
\begin{minipage}{0.47\hsize}
	\resizebox{\hsize}{!}{\includegraphics{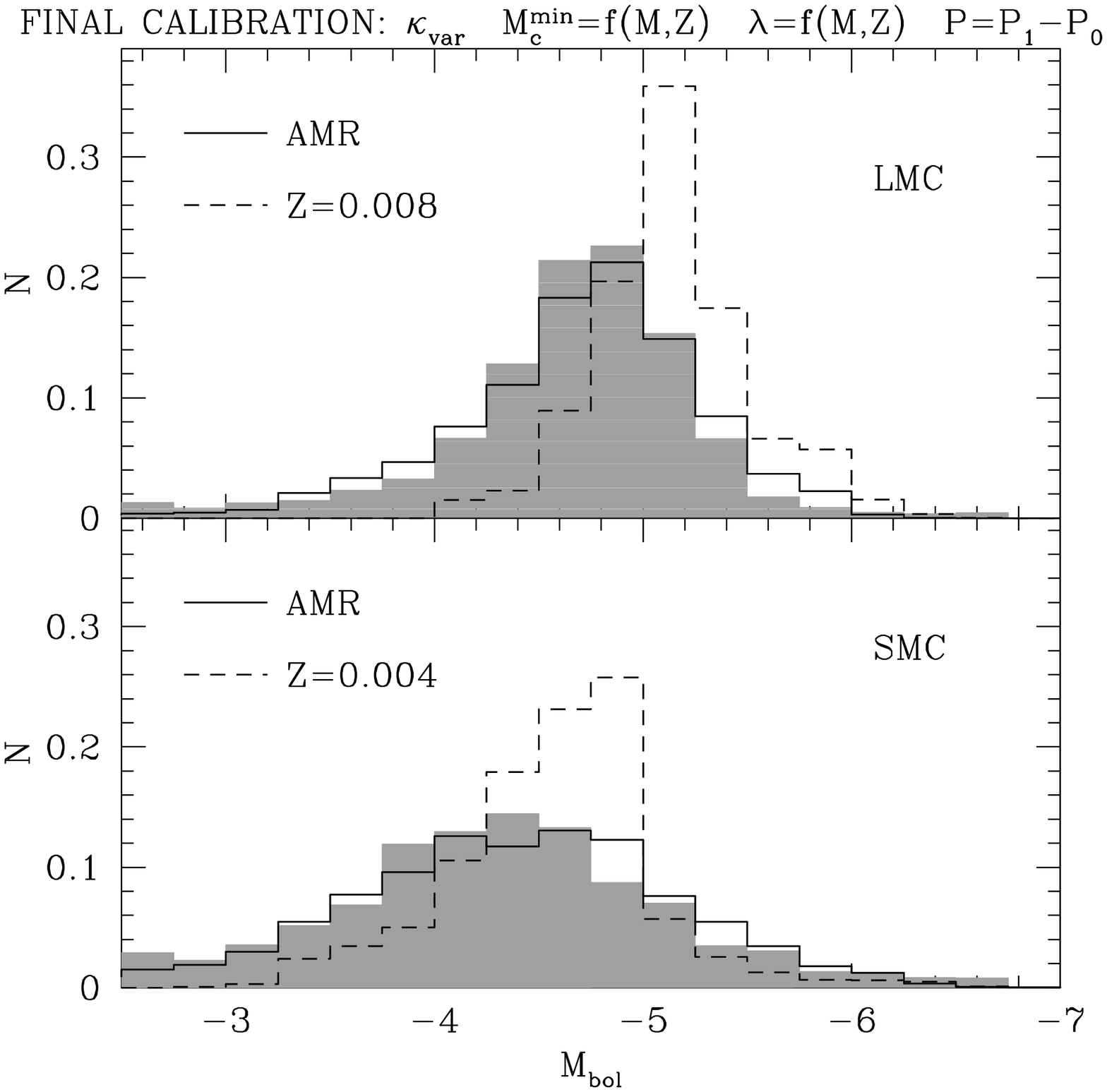}}
\end{minipage} 
\hfill
\begin{minipage}{0.47\hsize}
	\resizebox{\hsize}{!}{\includegraphics{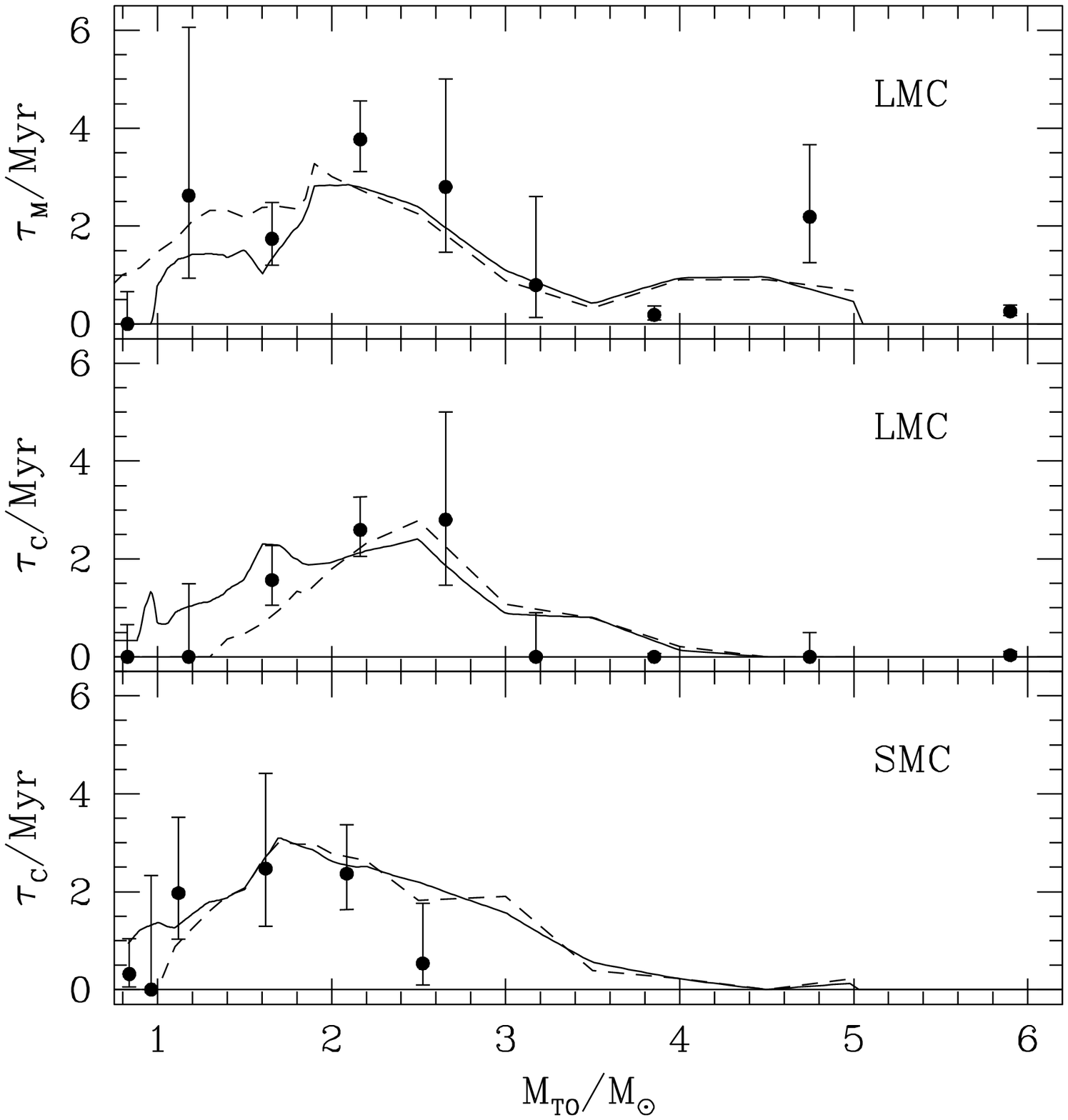}}
\end{minipage} 
\caption{
The final model calibration.
Left panels: The observed C star luminosity functions (CSLF) in the
Magellanic Clouds (Groenewegen 2002; shaded areas), as compared to the
synthetic ones derived from present models.  Right panels: The observed C-
and M-type lifetimes in the Magellanic Clouds (Girardi \& Marigo
2007), as compared to the same models. 
}
\label{fig_cslf_calib2006}
\end{figure*}

We notice that, with the present prescriptions for the mass loss
rates, the superwind phase (with
$\dot{M}\ga5.6\times10^{-7}\,M_\odot\,{\rm yr}^{-1}$) for which the
stars are likely to be optically obscured, lasts less than 20 percent
of the TP-AGB lifetime for all tracks with $Z\la0.008$. For models
with $1.5\la (M/M_\odot) \la2.5$ of such metallicities, which enter
the superwind phase as C stars and are the most relevant for the
comparison with MC data, the superwind phase lasts typically less than
15 percent of the C-star lifetime. These numbers are in rough agreement
with those derived from observations in the MCs (van Loon et
al. 2005). A detailed analysis of this point is left to subsequent
papers in this series. We advance however that the neglection of
objects that are likely optically-obscured in the theoretical CSLFs
and lifetimes, as we verified, would not significantly impact on our
calibration iter.
The Appendix summarizes the calibration iter, starting from the
past calibration of Marigo et al. (1999) and arriving at the final result 
depicted in Fig.~\ref{fig_cslf_calib2006}.

\subsection{Implications from the calibration}
The calibration sequence described in the Appendix 
yields a number of interesting and useful
indications, namely: 

1) The combined use of both CSLFs and stellar lifetimes in the
Magellanic Clouds as calibrating observables strengthens the
reliability of the underlying TP-AGB description. In fact,  we find 
that sets of models, consistent with the data for the C- and
M-star lifetimes, may instead not reproduce the corresponding
luminosity functions and viceversa.

2) Due to the pronounced metallicity dependence of the third
dredge-up, any theoretical investigation of the C-stars' populations
should account for the history of chemical enrichment in the host
galaxies, then relaxing the too simple assumption of constant
metallicity.  Indeed, the simultaneous reproduction of the CSLFs in
both Magellanic Clouds provides a strong indication in this respect,
as we see in Fig.~\ref{fig_cslf_calib2006} when comparing the results
for $Z={\rm const.}$ to those obtained including proper AMRs in the
simulated galaxies.  In particular, we see that the excess of bright C
stars and the deficiency of fainter objects, predicted while assuming
$Z={\rm const.}$, are removed when accounting for the
metallicity-dependence of the third dredge-up.  This means that the
extent of the long-standing theoretical difficulty in producing faint
carbon stars, known as `the carbon star mystery'' since Iben (1981),
could be actually less important than so far believed, just because of
a metallicity mismatch in the comparison between observations and
predictions.  Moreover, the fact that  the
populations of C stars in the two galaxies share a common metallicity
range, say around $Z=0.004$, 
should make
our metallicity calibration for the third dredge-up more constrained.

3) The peak of the M-star lifetimes in the LMC clusters (limiting to
stars brighter than the RGB-tip) is recovered only assuming that the
third dredge-up in stars with $M_{\rm i}\sim 1.7-2.5 M_{\odot}$ and
$Z\approx 0.008$ does not take place as soon as they enter the TP-AGB
phase, but somewhat later.  The $M_{\rm c}^{\rm min}(M,Z)$ formalism
derived by K02 turns out to be consistent with this indication.  In
this way, we alleviate the problem of under-estimating the M-star
populations that affects the simulations presented in Cioni et
al. (2006ab).

4) The decreasing lifetimes of C stars with $M_{\rm i} >2.5 M_{\odot}$
seen in both Magellanic Clouds is due to the concomitant effects of
their brighter luminosities -- that favour the earlier onset of the
superwind -- and, in models of larger masses (with $M_{\rm i}\ga 4.0
M_{\odot}$), of the occurrence of HBB -- that may later reconvert a C
star into an M star, or even prevent the transition to the C-star
domain.



\section{Final remarks}
\label{sec_end}

The set of TP-AGB evolutionary tracks presented in this paper
constitutes a very useful database for population studies, due to
several distiguishing features:

1) We have used updated ingredients and improved treatment of some
key physical processes in the synthetic AGB code, as extensively
discussed in Sects.~\ref{sec_syntagb} and \ref{sect_modpre}. The most
important updating is no doubt the use of variable molecular opacities
instead of the scaled-solar tables still used in complete AGB
models. Previous sets of models using these opacities have already
been used (Marigo 2002; Marigo et al. 2003; Cioni et al. 2006ab) and
distributed by us (see Cioni et al. 2006a and {\tt
http://pleiadi.oapd.inaf.it}), but this is the first time {\em
calibrated} TP-AGB tracks using this approach are released.

2) The present model calibration (Sect.~\ref{sec_calibr}) is based not
only on the reproduction of the CSLFs in both Magellanic Clouds, but
also on the data for C- and M-star counts in Magellanic Cloud
clusters. It means that present models have not only the right
luminosities of the TP-AGB phase (as ensured by the CSLF calibration),
but also that they present, as far as possible, the right lifetimes at
Magellanic Clouds' metallicities. The contribution of any post-MS
evolutionary stage to the integrated light of a single-burst stellar
population is proportional to the product of luminosity and lifetime
-- also known as the {\em nuclear fuel} in the context of the fuel
consumption theorem (Renzini \& Buzzoni 1986). Therefore, evolutionary
population synthesis models that use the present TP-AGB tracks will
present the right contribution of TP-AGB stars to the integrated light
at Magellanic Clouds metallicities, and hopefully, also a good
description of the way this contribution varies with metallicity.

3) The present tracks nicely complement the sets of low-and
intermediate-mass models of Girardi et al. (2000) and Girardi (2002, 
unpublished), that deal with the evolution from the zero-age main
sequence up to the first thermal pulse. In fact, when coupling both sets
of tracks together one will find no discontinuities in quantities like
the core mass and chemical surface composition. Discontinuities in the
HR diagram may appear due to the use of slightly different
low-temperature opacities; we checked however that they are tiny and
certainly much less serious than those one have when amalgamating
tracks from different sources. Therefore, one has the possibility of
building consistent tracks all the way from the ZAMS to the end of the
TP-AGB. Of course, this sequence can also be extended with any
suitable set of post-AGB tracks (e.g., Vassiliadis \& Wood 1994),
so as to cover the subsequent evolution up to the WD stages.

4) We make a first attemp to describe the switching of pulsation modes
between first overtone and fundamental one along the TP-AGB
evolution. These changes have some consequence in terms of the history
of mass loss, and open new possibilities to better calibrate TP-AGB
tracks by means of comparison with the extensive variability surveys
in the Magellanic Clouds.

Forthcoming papers will present extensive and ready-to-use theoretical
isochrones and chemical yields from these tracks, as well as several
population synthesis applications.

\begin{acknowledgements}
We thank  the anonymous referee and Martin Groenewegen for their 
useful remarks.
This study was partially supported by the University of Padova
(Progetto di Ricerca di Ateneo CPDA052212).
\end{acknowledgements}


\Online
\begin{appendix}

\section{The model calibration}
\label{ssec_calres}
This section summarises the calibration iter, mainly involving the
parameterised decription of the third dredge-up. 
We also discuss the strong influence
played by other parameters, like the assumed pulsation mode on the 
mass-loss rates, 
and the age-metallicity relation of the host galaxy.
    
\begin{figure*}[!tbp]  
\begin{minipage}{0.47\hsize}
	\resizebox{\hsize}{!}{\includegraphics{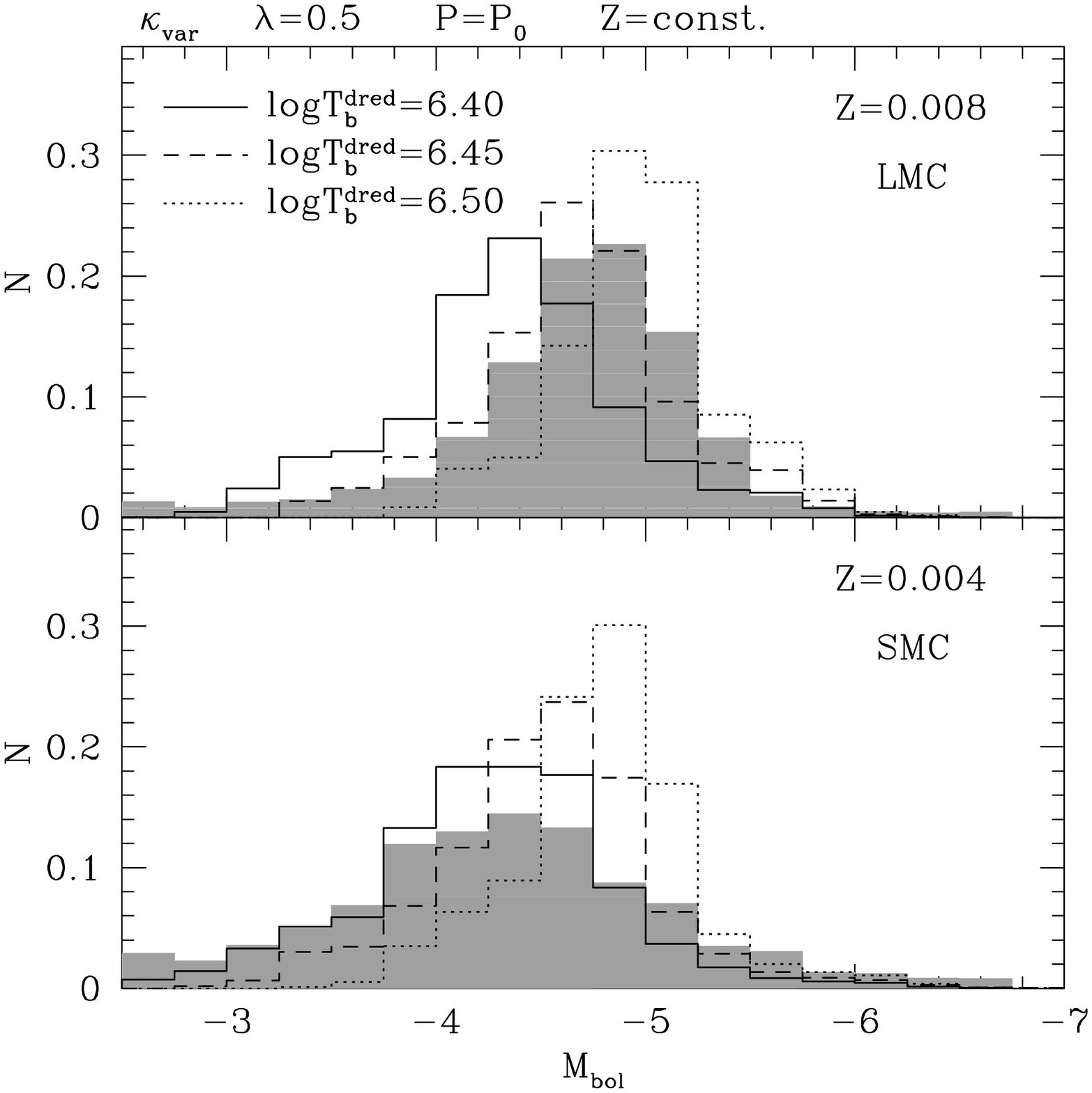}}
\end{minipage} 
\hfill
\begin{minipage}{0.47\hsize}
	\resizebox{\hsize}{!}{\includegraphics{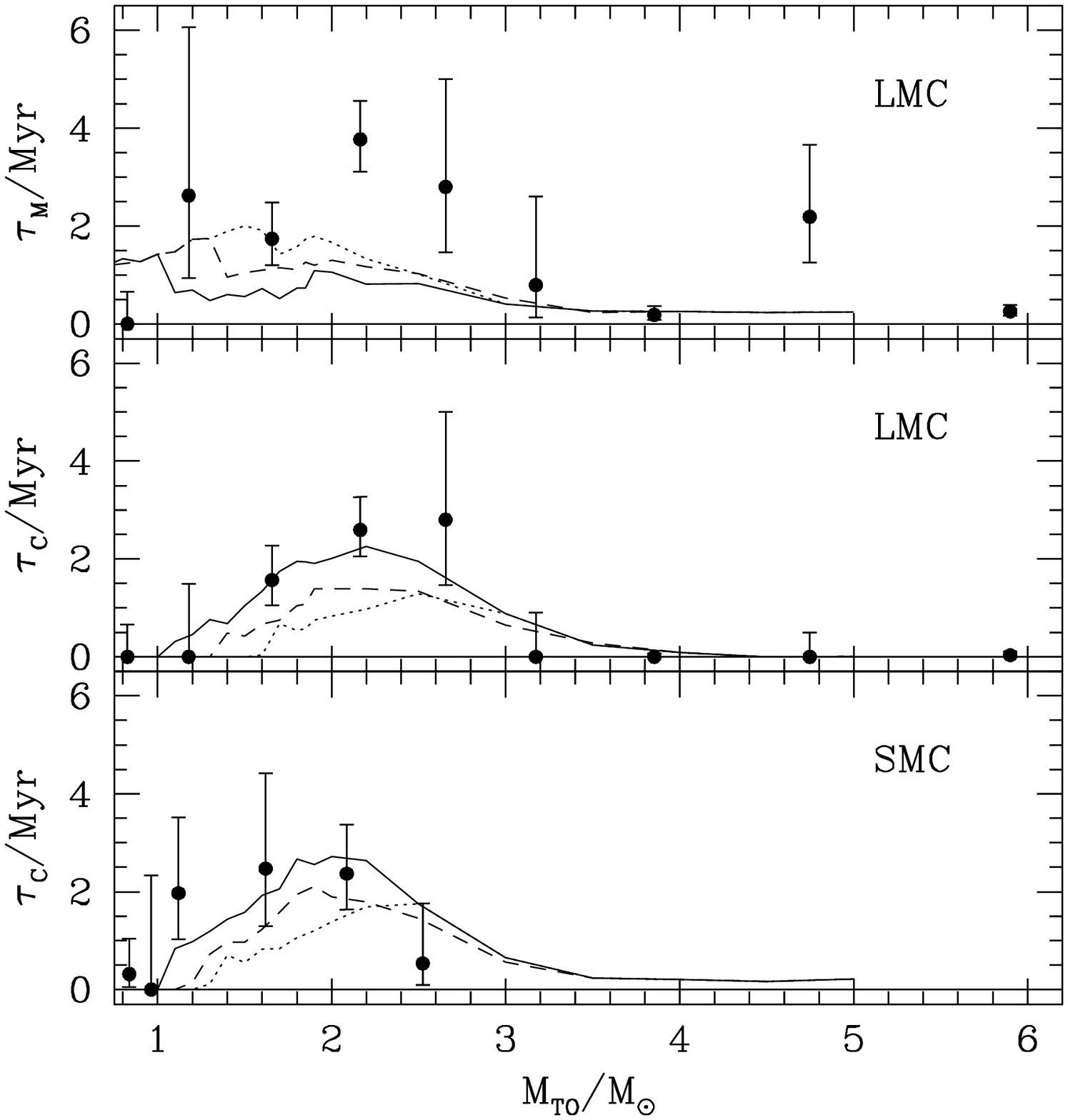}}
\end{minipage} 
\caption{
Left panels: The observed C star luminosity functions (CSLF) in the
Magellanic Clouds (Groenewegen 2002; shaded areas), as compared to the
synthetic ones derived from models with different dredge-up
parameters, assuming Vassiliadis \& Wood's (1993) mass loss
prescription and pulsation on the FM.  Right panels: The observed C-
and M-type lifetimes in the Magellanic Clouds (Girardi \& Marigo
2007), as compared to the same set of models.  }
\label{fig_cslf_kvarl05tbdredP0}
\end{figure*}

The most relevant calibrating steps are illustrated through the
sequence of panels from Figs.~\ref{fig_cslf_kvarl05tbdredP0} to
\ref{fig_cslf_calib2006}, where labels indicate the main model
assumptions, namely:
\begin{itemize}
\item Molecular opacities, either fixed for solar-scaled mixtures,  
$\kappa_{\rm fix}$, or variable chemical composition, $\kappa_{\rm
var}$;
\item Dredge-up efficiency $\lambda$;
\item Onset of the third dredge-up, determined by either 
the temperature $T_{\rm b}^{\rm dred}$ parameter, or the classical
$M_{\rm c}^{\rm min}$ parameter.
\item Pulsation mode, either fundamental with $P=P_0$, or first overtone 
with $P=P_1$;
\item Metallicity of the simulated galaxy, either assumed constant
over its entire history, $Z={\rm const}$, or made vary according 
to a specified age-metallicity relation, AMR.
\end{itemize}
   
Figure~\ref{fig_cslf_kvarl05tbdredP0} illustrates the results obtained
with the same prescriptions for the third dredge-up as in our previous
study (Marigo et al. 1999).  The most significant difference between
the two sets of models is the adoption in the present calculations of
variable molecular opacities in place of those for solar-scaled
mixtures.  In both cases mass loss is treated with the aid the
Vassiliadis \& Wood (1993) formalism, always assuming FM pulsation
($P=P_0$).

We notice that the combination $(\log T_{\rm b}^{\rm dred}=6.40,
\lambda=0.5)$ now fails to reproduce the CSLF in the LMC --
instead providing the best fit in Marigo et al. (1999) previous
calibration -- while it well recovers the C-star lifetimes in both
Magellanic Clouds.  A reasonable fit to the CSLF in the LMC is
obtained with the $(\log T_{\rm b}^{\rm dred}=6.45, \lambda=0.5)$
case, but the duration of the C-star phase as a function of $M_{\rm i}$ is
somewhat underestimated.  The $(\log T_{\rm b}^{\rm dred}=6.50,
\lambda=0.5)$ set of models yields poor results in respect to all
observables.  Finally, we should remark that all combinations of
dredge-up parameters do not reproduce the CSLF in the SMC, and they
substantially under-estimate the duration of the M-star phase in the
$1.8-3.0 M_{\odot}$ range.

\begin{figure*}[!tbp]  
\begin{minipage}{0.47\hsize}
	\resizebox{\hsize}{!}{\includegraphics{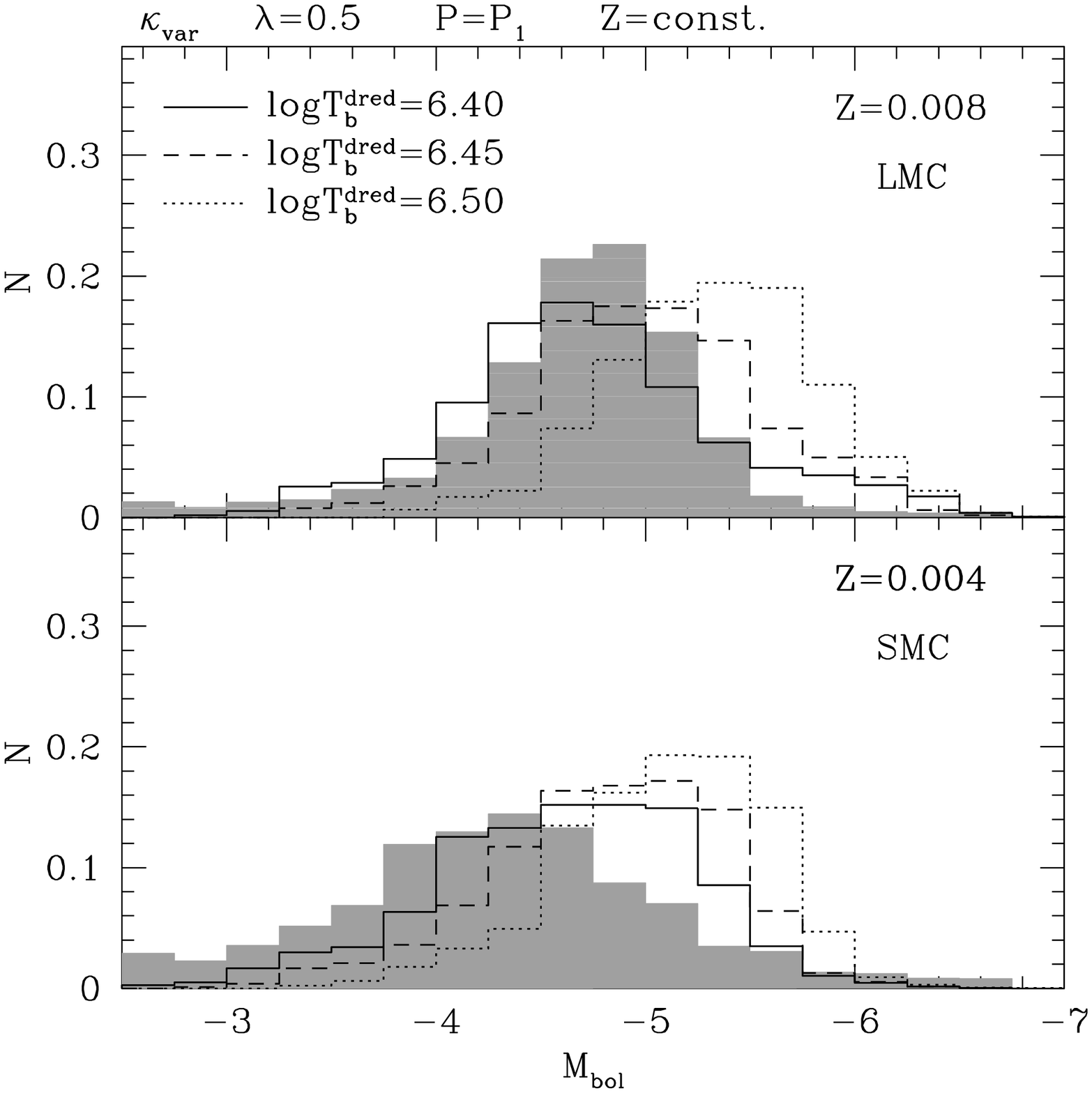}}
\end{minipage} 
\hfill
\begin{minipage}{0.47\hsize}
	\resizebox{\hsize}{!}{\includegraphics{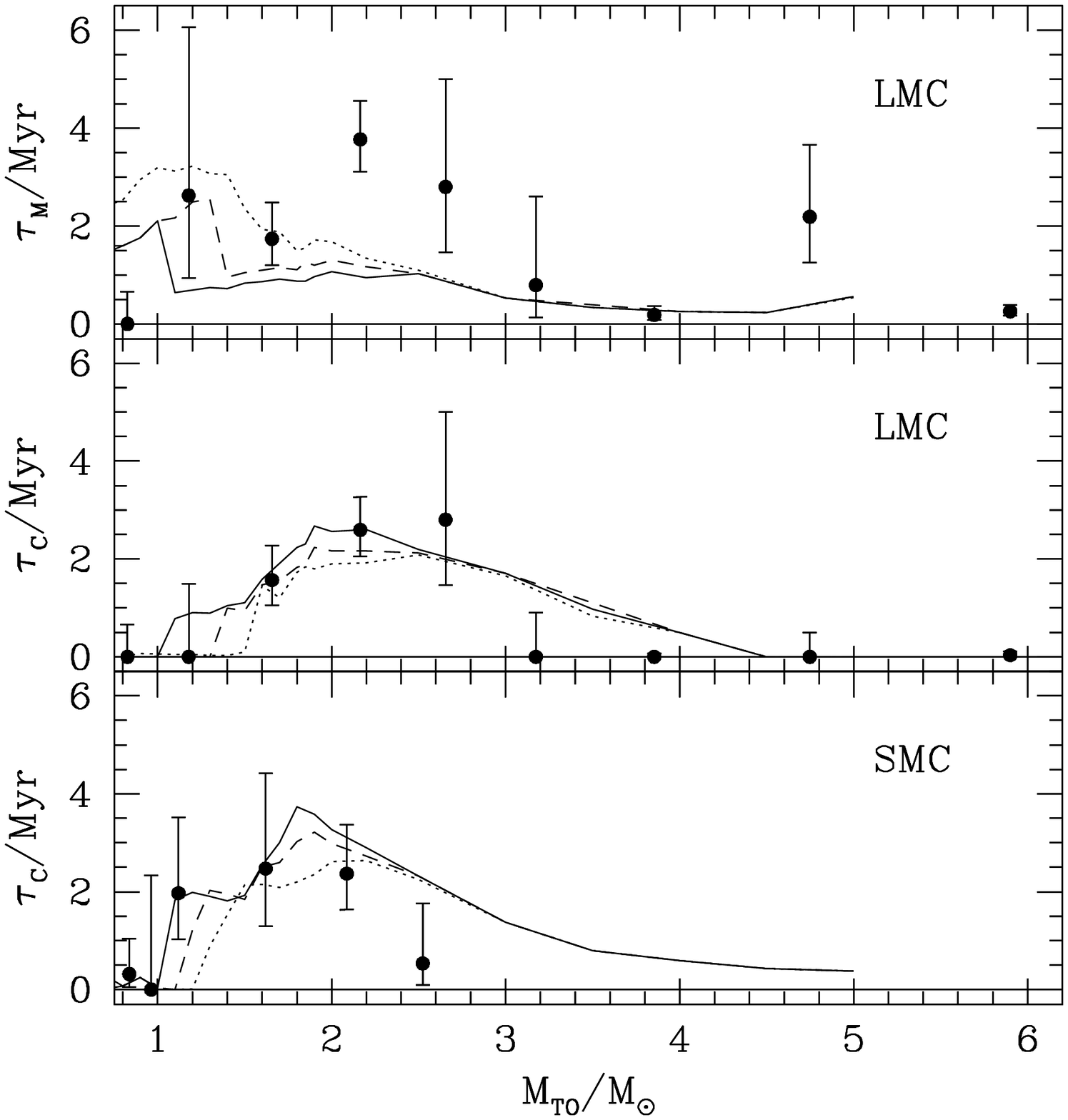}}
\end{minipage} 
\caption{
The same as Fig.~\ref{fig_cslf_kvarl05tbdredP0}, but for models
assuming pulsation on the FOM.  }
\label{fig_cslf_kvarl05tbdredP1}
\end{figure*}

\begin{figure*}[!tbp]  
\begin{minipage}{0.47\hsize}
	\resizebox{\hsize}{!}{\includegraphics{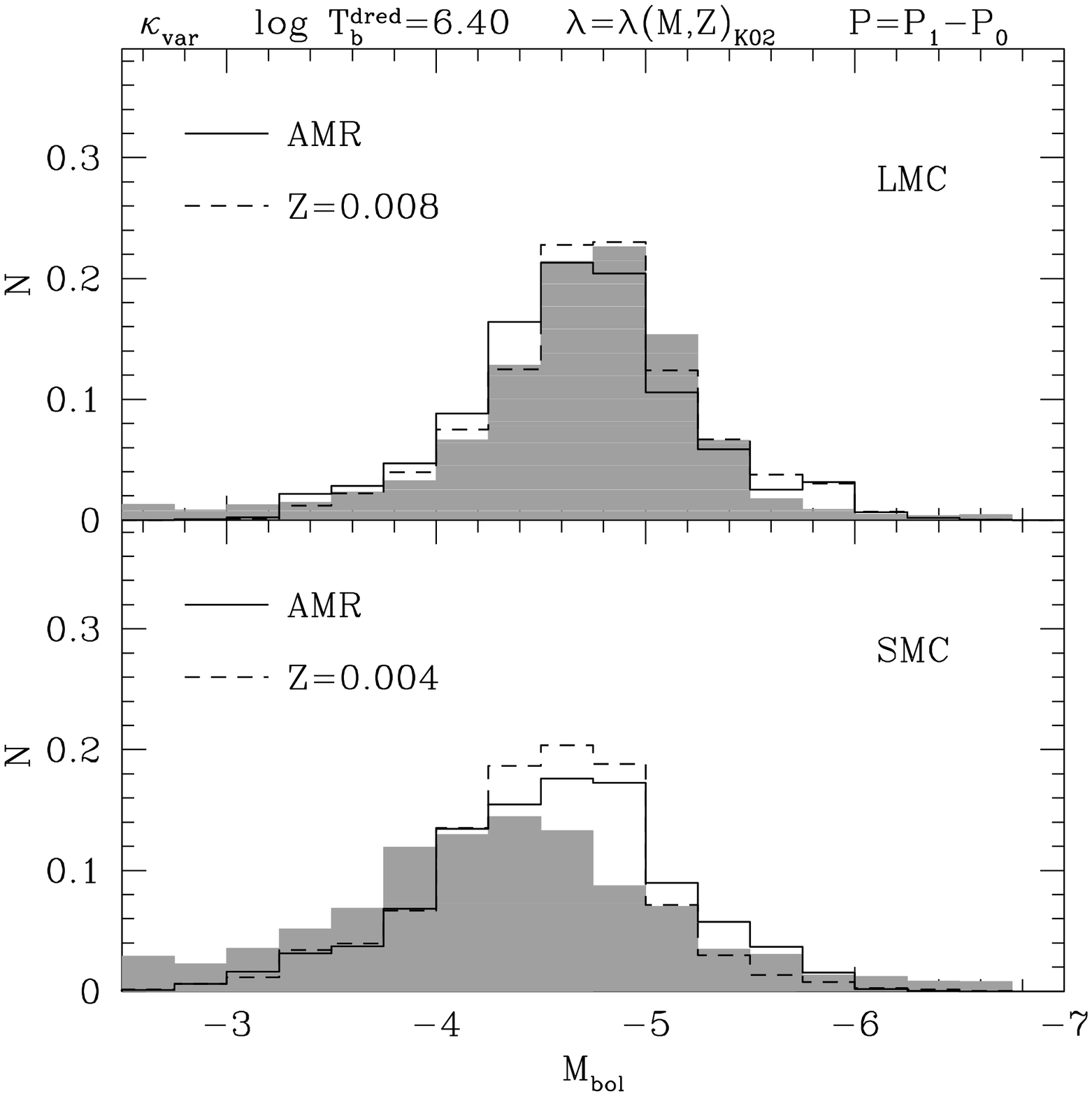}}
\end{minipage} 
\hfill
\begin{minipage}{0.47\hsize}
	\resizebox{\hsize}{!}{\includegraphics{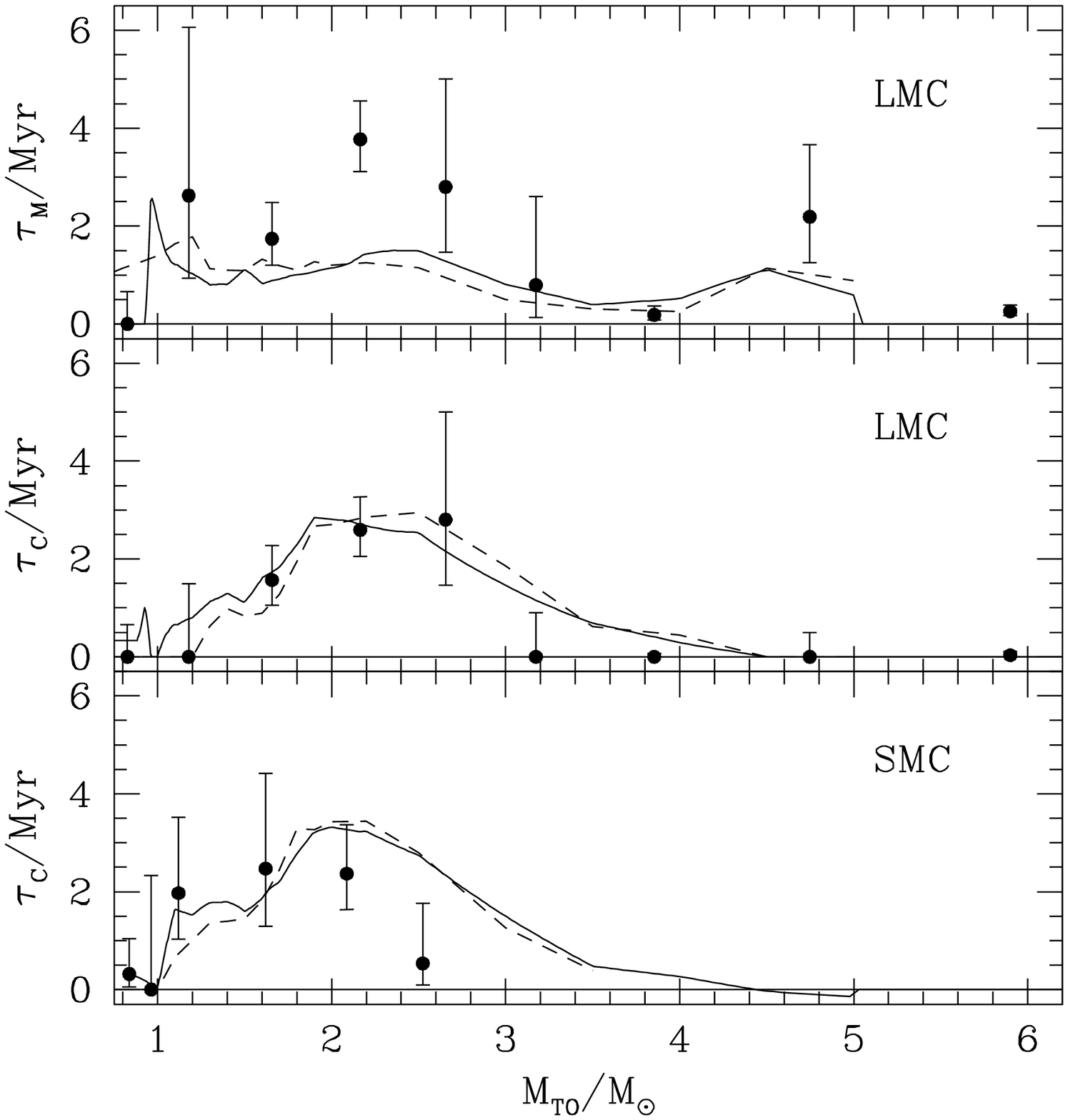}}
\end{minipage} 
\caption{
The same as Fig.~\ref{fig_cslf_kvarl05tbdredP0}, but for models
assuming a variable $\lambda$ (K02), mode switching between FM and
FOM, and an AMR. See the text for details.}
\label{fig_cslf_kvarlvartb642P0P1}
\end{figure*}

\begin{figure*}[!tbp]  
\begin{minipage}{0.47\hsize}
	\resizebox{\hsize}{!}{\includegraphics{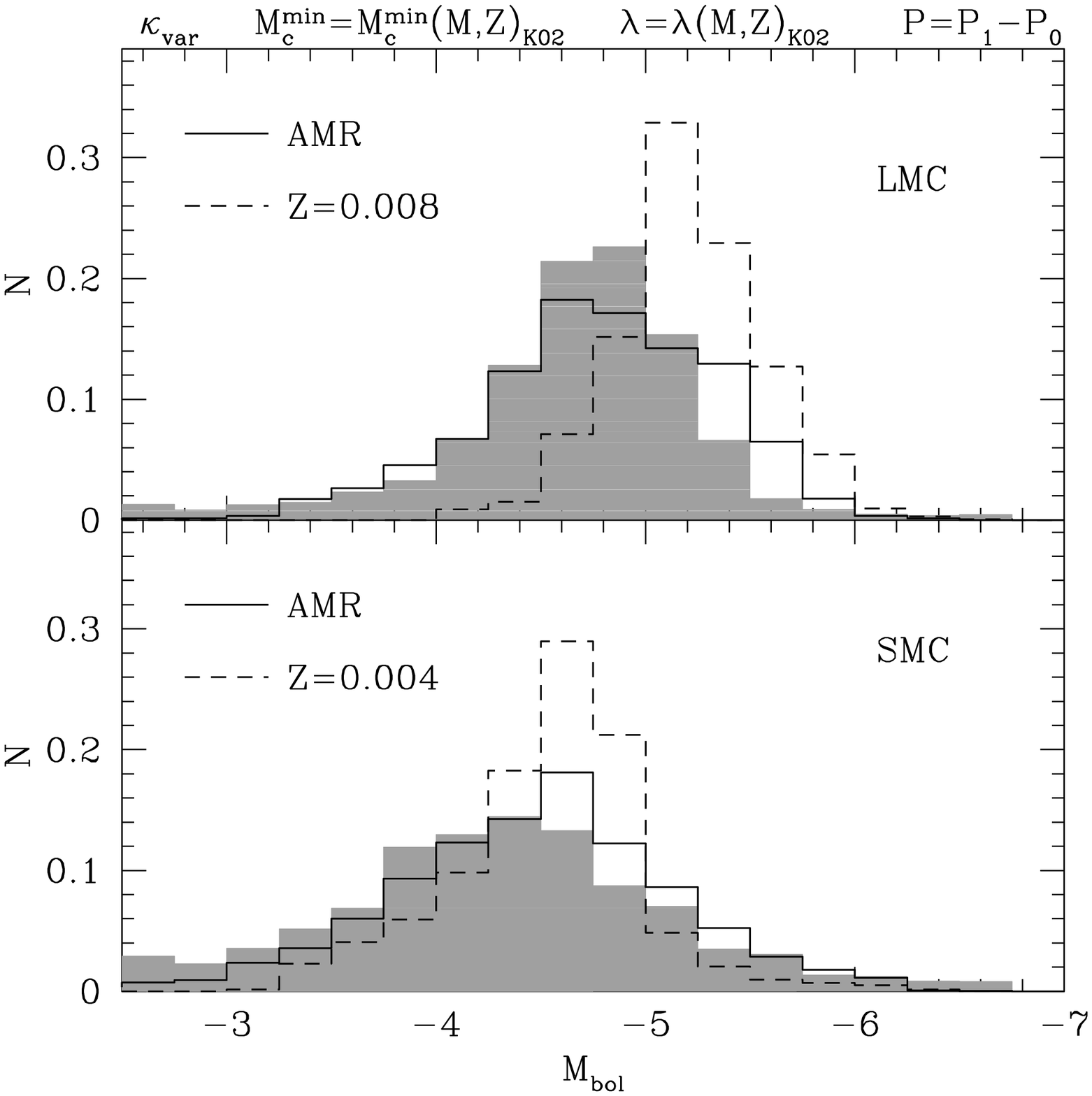}}
\end{minipage} 
\hfill
\begin{minipage}{0.47\hsize}
	\resizebox{\hsize}{!}{\includegraphics{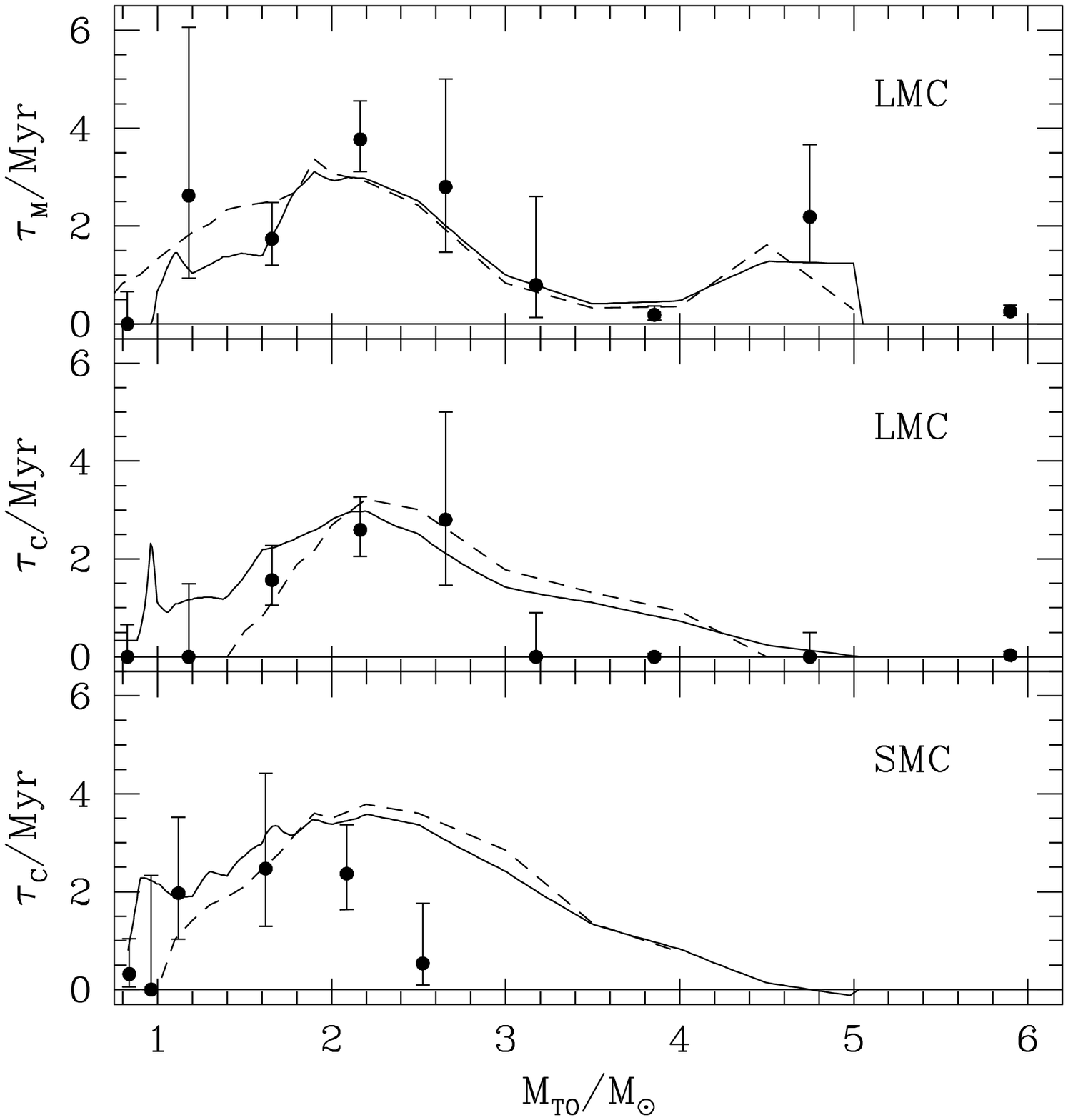}}
\end{minipage} 
\caption{
The same as Fig.~\ref{fig_cslf_kvarlvartb642P0P1}, but for models
assuming the $M_{\rm c}^{\rm min}(M,Z)$ formalism suggested by K02. } 
\label{fig_cslf_kvarlvarmcP0P1}
\end{figure*}
  
Figure~\ref{fig_cslf_kvarl05tbdredP1} is based on models calculated
with exactly the same input parameters of
Fig.~\ref{fig_cslf_kvarl05tbdredP0}, except that pulsation is assumed
to occur in the FOM. With the assumption $P=P_1$, the predicted CSLFs
tend to populate brighter luminosities, as a consequence of the fact
that shorter periods correspond to lower mass-loss rates,
hence longer TP-AGB lifetimes.  From the comparison with the observed
data it turns out that (i) none of the $(T_{\rm b}^{\rm dred},
\lambda)$ combinations leads to a satisfactory reproduction of the
CSLFs; (ii) the C-star lifetimes are relatively well recovered; (iii)
the M-star lifetimes in the LMC are still notably underestimated in
the $1.8-3.0 M_{\odot}$ range.

Figure~\ref{fig_cslf_kvarlvartb642P0P1} introduces a few important
changes, namely (i) the assumption of constant $\lambda$ is relaxed to
account for a more realistic dependence on mass and metallicity
according to K02 recipe; (ii) pulsation periods are calculated
following the scheme indicated in Sect.~\ref{ssec_pulsation}, that
predicts the mode switching from FOM ($P=P_1$) to FM ($P=P_0$); and
(iii) the inclusion of an age-metallicity relation in the galaxy
simulations instead of constant metallicity.  As a matter of fact, the
use of a $\lambda(M,Z)$ law significantly improves the results for the
CSLF in the LMC, while a sizeable overproduction of luminous C stars
affects the CSLF in the SMC.  Moreover, the problem with the too
short  M-star lifetimes still remains.

The way out to this latter discrepancy is obtained when we adopt the
$M_{\rm c}^{\rm min}(M,Z)$ formalism suggested by K02, in place of the
temperature parameter $T_{\rm b}^{\rm dred}$ to determine the onset of
the third dredge-up. This effect is illustrated in
Fig.~\ref{fig_cslf_kvarlvarmcP0P1}.  The reason is that while the
$T_{\rm b}^{\rm dred}$ criterion predicts that models with $M_{\rm i}
\ga 2 M_{\odot}$ experience the mixing events already from the first
thermal pulse, with the $M_{\rm c}^{\rm min}(M,Z)$ scheme the onset of
the dredge-up is delayed, then allowing for a longer duration of the
M-type phase for these models.

We should also note that the theoretical CSLFs obtained with $Z={\rm
const.}$ notably differ from the observed ones in both Magellanic
Clouds. The results gets better when assuming a proper AMR for the
host galaxy, but the comparison with the observed data is still not
completely satisfactory.

Finally, the best fit to the CSLFs, the M-type, and the C-type
lifetimes in the Magellanic Clouds is reached with the aid of the
scheme for the third dredge-up presented in Sect.~\ref{ssec_3dup}.
The original K02 formalism is partly modified in order to (i) correct
for convective overshooting, (ii) anticipate the occurrence of the
third dredge-up, (iii) and make it more efficient at lower
metallicities.  The results are displayed in
Fig.~\ref{fig_cslf_calib2006}.

\end{appendix}
\end{document}